\documentclass[10pt,a4paper]{article}
\usepackage[utf8]{inputenc}
\usepackage{graphicx}
\usepackage{amsmath}
\usepackage{amsfonts}
\usepackage{amssymb}
\usepackage{multirow}
\usepackage{multicol}
\usepackage[font=small,labelfont=bf]{caption}
\usepackage{xcolor}
\usepackage{geometry}
\usepackage{hyperref}
\usepackage{pdflscape}
\usepackage{epstopdf}

\usepackage{ulem}
\usepackage{xcolor}

\begin{document}

\begin{titlepage}
\begin{flushright}
LU TP 18-04\\
17th April 2018
\end{flushright}
\vfill
\begin{center}
{\Large\bf Analytic representations of $m_K$, $F_K$, $m_\eta$ and $F_\eta$ in two loop $SU(3)$ chiral perturbation theory}

\vfill
{\bf B. Ananthanarayan$^a$, Johan Bijnens$^b$, Samuel Friot$^{c,d}$ and Shayan Ghosh$^a$}\\[1cm]
{$^a$ Centre for High Energy Physics \\Indian Institute of Science, Bangalore-560012, Karnataka, India}\\[0.5cm]
{$^b$ Department of Astronomy and Theoretical Physics\\ Lund University,
S\"olvegatan 14A, SE 223-62 Lund, Sweden} \\[0.5cm]
{$^c$ Institut de Physique Nucl\'eaire d’Orsay \\
Universit\'e Paris-Sud 11, IN2P3-CNRS, F-91405 Orsay Cedex, France}\\[0.5cm]
{$^d$ Institut de Physique Nucléaire de Lyon \\ Université Lyon 1, IN2P3-CNRS, F-69622 Villeurbanne Cedex, France}
\end{center}
\vfill

\begin{abstract}

In this work, we consider expressions for the masses and decay constants of the pseudoscalar mesons in $SU(3)$ chiral perturbation theory. These involve sunset diagrams and their derivatives evaluated at $p^2=m_P^2$ ($P=\pi, K, \eta$). Recalling that there are three mass scales in this theory, $m_\pi$, $m_K$ and $m_\eta$, there are instances when the finite part of the sunset diagrams do not admit an expression in terms of elementary functions, and have therefore been evaluated numerically in the past. In a recent publication, an expansion in the external momentum was performed to obtain approximate analytic expressions for $m_\pi$ and $F_\pi$, the pion mass and decay constant. We provide fully analytic exact expressions for $m_K$ and $m_\eta$, the kaon and eta masses, and $F_K$ and $F_\eta$, the kaon and eta decay constants. These expressions, calculated using Mellin-Barnes methods, are in the form of double series in terms of two mass ratios. A numerical analysis of the results to evaluate the relative size of contributions coming from loops, chiral logarithms as well as phenomenological low-energy constants is presented. We also present a set of approximate analytic expressions for $m_K$, $F_K$, $m_\eta$ and $F_\eta$ that facilitate comparisons with lattice results. Finally, we show how exact analytic expressions for $m_\pi$ and $F_\pi$ may be obtained, the latter having been used in conjunction with the results for $F_K$ to produce a recently published analytic representation of $F_K/F_\pi$.
\end{abstract}
\vfill
\vfill
\end{titlepage}

\section{Introduction}

In a recent publication, the important ratio $F_K/F_\pi$ was evaluated in a scheme that allows for the derivation of compact analytic approximations in two loop chiral perturbation theory (ChPT) \cite{Ananthanarayan:2017qmx}, based on the Mellin-Barnes (MB) approach detailed in \cite{Ananthanarayan:2016pos}. In a prior work, a different scheme was employed to obtain analytic approximations of $m_\pi$ and $F_\pi$ in $SU(3)$ ChPT at two-loops \cite{Ananthanarayan:2017yhz}. Recall that ChPT is an effective field theory for the pseudo-scalar octet degrees of freedom, namely the pions, kaons and eta. At one-loop order, this theory was elucidated in \cite{Gasser:1983yg, Gasser:1984gg}. At two-loop order, the $SU(2)$ theory with just the pion degrees of freedom was worked out in \cite{Bijnens:1997vq}, while the significantly more complicated $SU(3)$ theory has been described in \cite{Bijnens:2006zp}. For many observables and processes of interest in the $SU(2)$ theory, there is a single mass scale in the problem when isospin violation and electromagnetic corrections are neglected, namely the pion mass. At the two-loop order, integrals that arise in this context have been discussed in \cite{Gasser:1998qt}. In the $SU(3)$ theory, all three masses of the pseudoscalar mesons may appear in quantities of interest. Of the relevant integrals in the latter case, the self-energy diagram, which is known as the sunset, may be represented as:
\begin{align}
H_{\{\alpha,\beta,\gamma\}}^d \{m_1,m_2,m_3; p^2\} = \frac{1}{i^2} \int \frac{d^dq}{(2\pi)^d} \frac{d^dr}{(2\pi)^d} \frac{1}{[q^2-m_1^2]^{\alpha} [r^2-m_2^2]^{\beta} [(q+r-p)^2-m_3^2]^{\gamma}} \label{Eq:SunsetDef}
\end{align}
In our conventions for dimensional regularisation, $d=4-2\epsilon$.

\begin{figure}[hbtp]
\centering
\includegraphics[scale=0.6]{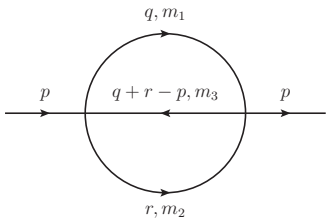}
\caption{Sunset diagram}
\label{sunset}
\end{figure}

Tarasov established that integration by parts allows one to express all sunset integrals using a minimal set of four master integrals (MI) \cite{Tarasov:1997kx}. For some configurations, such as the one mass scale case, the representation of the sunset in terms of MI may require fewer than the full set of four. For quantities of interest in $SU(3)$ ChPT, such as the masses and decay constants of the pion, kaon and eta (denoted by $m_\pi$, $F_\pi$, $m_K$, $F_K$, $m_\eta$ and $F_\eta$, respectively), several variations of the basic sunset diagram of Eq.(\ref{Eq:SunsetDef}) appear in their expressions. These include $H_1$, $H_{21}$ and $H_{22}$, which are the scalar integrals appearing from the Passarino-Veltman decomposition of the tensor sunset integrals:
\begin{align}
	& H_{\mu}^d = p_{\mu} H_{1} \nonumber \\
	& H_{\mu \nu}^d = p_{\mu} p_{\nu} H_{21} + g_{\mu \nu} H_{22}
\end{align}
where
\begin{align}
	H_{\mu}^d \{m_1,m_2,m_3; p^2\} = \frac{1}{i^2} \int \frac{d^dq}{(2\pi)^d} \frac{d^dr}{(2\pi)^d} \frac{q_{\mu}}{[q^2-m_1^2] [r^2-m_2^2] [(q+r-p)^2-m_3^2] } \nonumber \\
	\nonumber \\
	H_{\mu \nu}^d \{m_1,m_2,m_3; p^2\} = \frac{1}{i^2} \int \frac{d^dq}{(2\pi)^d} \frac{d^dr}{(2\pi)^d} \frac{q_{\mu} q_{\nu}}{[q^2-m_1^2] [r^2-m_2^2] [(q+r-p)^2-m_3^2] }
\end{align}

These may be expressed in terms of the MI. Similarly, while the meson masses require the evaluation of only basic sunset integrals, the decay constants also need calculation of the derivatives of the sunsets with respect to the square of the external momentum. However, it is possible to discuss both the mass and decay constant on an equal footing by reducing the task to evaluating just the MI.

It is of interest to obtain representations of the $m_P$ and $F_P$ that do not require numerical evaluation of the higher order loop integrals. Such analytic approaches allow for more widespread use of these expressions, and facilitate cross-disciplinary studies. An example of the latter would be comparisons with lattice simulations as the quark masses are varied in order to obtain insights into the behaviour of these quantities. In \cite{Ecker:2010nc, Ecker:2013pba}, for example, an approximation for $F_K/F_\pi$ was obtained by means of a large-$N$ approach at the Lagrangian level, and the resulting expression was used to fit lattice data to extract values of several ChPT parameters. In \cite{Kaiser:2007kf}, $m_\pi$ was treated by taking an approximation at the level of the loop integral \cite{Kaiser:2007kf}. In \cite{Ananthanarayan:2017yhz}, some of the present authors extended the former work to the case of $F_\pi$, and were able to integrate out the s-quark from the expressions of the pion mass and decay constant, in the same way as the $SU(3)$ ChPT reduces to the $SU(2)$ version, reproducing known results in the chiral limit, as well as evaluating the departure from the chiral limit to leading order in the light quark mass.

The goal of this paper is in the same spirit of furthering studies in the area of analytic approaches to observables and other quantities of interest. In particular, the aims of the current work are:
\begin{itemize}
\item To extend the work of \cite{Kaiser:2007kf, Ananthanarayan:2017yhz} to the case of the kaon and eta, and provide approximate analytic expressions for $m_K$, $m_\eta$, $F_K$ and $F_\eta$ that are easily amenable to fitting with lattice data.
\item To provide exact (non-approximate) two loop analytic expressions for all the pseusoscalar meson masses and decay constants.
\item To perform a first order study of the numerics of $m_K$, $m_\eta$, $F_K$ and $F_\eta$ to determine the relative contributions to them of their different components, as well as their dependence on the values of parameters such as the low energy constants of ChPT.
\end{itemize}
Although this work is a sequel to \cite{Ananthanarayan:2017yhz}, the approach taken here is completely different and novel. In the aforementioned work, as well as in \cite{Kaiser:2007kf}, the three mass scale sunset integrals were approximated by taking an expansion in the external momentum $p^2=m_\pi^2$ around zero. In the case of the kaon and eta, however, such an expansion around $p^2=m_K^2$ or $p^2=m_\eta^2$ results in a poorly converging series due to the presence of the much smaller $m_\pi^2$ in the propagator. An expansion about the propagator mass $m_\pi^2 = 0$ is also not feasible as this gives rise to an infrared divergence. In this work, therefore, we turn to the MB approach to evaluate the three-mass scale sunset integrals.

The analytic evaluation of a sunset integral depends on its mass configuration. For special cases, with upto two distinct mass scales, closed form results are available \cite{Berends:1997vk,Czyz:2002re,Davydychev:1992mt} \footnote{By complete results, we  refer only to the finite $\mathcal{O}(\epsilon^0)$ term obtained using dimensional regularisation. The $\mathcal{O}(\epsilon^{-1})$ and $\mathcal{O}(\epsilon^{-2})$ terms are known exactly for all mass configurations \cite{Davydychev:1992mt}.}. With two mass scales not falling into the threshold or pseudo-threshold configurations \cite{Ananthanarayan:2016pos}, or with three distinct mass scales, the sunsets cannot be written in terms of elementary functions. In fact, the sunset diagram with three different masses and arbitrary $p^2$ cannot even be expressed entirely in terms of multiple polylogarithms. In \cite{Berends:1993ee}, for non-zero $\epsilon$, expressions in terms of Lauricella functions are given but, as emphasized in \cite{Adams:2016sob}, none of the present methods allows for an expansion of the Lauricella functions in terms of multiple polylogarithms. Indeed, it seems that, as shown in \cite{Adams:2015gva}, the sunset diagram requires the introduction of yet another generalisation of the polylogarithms, the so-called elliptic polylogarithms (see also \cite{Ablinger:2017bjx} for more details on elliptic integrals in the context of sunset integrals). In this work, we adopt a more utilitarian approach to get the analytical expressions needed for our analysis. We keep to the spirit of \cite{Berends:1993ee} where, once the $\epsilon\rightarrow 0$ limit of the Lauricella functions is taken, one stays with triple series in powers and logarithms of the mass ratios that one may truncate to achieve a desired level of accuracy. Note, however, that the expressions given in \cite{Berends:1993ee} cannot be used to obtain an analytic expression of the sunset valid in the context of ChPT, since the triple series given there do not converge for the physical values of the pseudo-scalar meson masses. In this work, we therefore present the full analytic expressions valid for the kaon and eta masses and decay constants in terms of double infinite series involving two mass ratios, thus completing the programme first initiated in \cite{Post:1996gg} of evaluating the three mass scale sunset diagrams in $SU(3)$ ChPT \footnote{Some years ago, there had been an attempt to pursue such a programme \cite{David1}. However, the investigations were not completed and publications did not result \cite{David2}.}. Here, we present only the results, and the complete derivation will be given in an upcoming paper \cite{Ananthanarayan:2018}. An overview of the method used is given in \cite{Ananthanarayan:2016pos}, and detailed descriptions can be found in \cite{Friot:2011ic,Aguilar:2008qj}.

One of the possible applications of fully analytic representations of the quantities considered here is their use to obtain different analytic approximations that may easily be fitted with data coming from various lattice simulations in conjunction with various lattice data. In addition to the full results, we therefore also present a set of analytic approximations that may easily be fitted with data from lattice simulations. These approximations are obtained by suitably truncating the infinite series so that their omitted tails are numerically smaller than a chosen percentage of the central value obtained from the exact expression\footnote{By exact expression we mean the partial sum where a big number of terms is retained such that it is assured that by adding more terms the corresponding numerical result stays stable within the standard numerical precision of \texttt{Mathematica}.} for the inputs being considered. And as these inputs will depend on the precise lattice set being used, the approximations will need to be changed accordingly. Therefore, we present along with this paper, a set of supplementary \texttt{Mathematica} files that automates the task of finding a suitable truncation for the sunsets, given a set of (lattice) input masses and a permissible error threshold value. The lattice expressions given in Section~\ref{Sec:LatticeFits} are suitable for fits with the data given in \cite{Durr:2016ulb}, and an illustrative fit with these expressions is presented in \cite{Ananthanarayan:2017qmx}.

The structure of this paper is as follows. In Section~\ref{Sec:MI}, we present our notation and a short discussion on convergence properties of our series representations of the sunset integrals. In Section~\ref{Sec:KaonMass}, the expression for the kaon mass is given, in Section~\ref{Sec:KaonDecay} the same is given for the kaon decay constant, in Section~\ref{Sec:EtaMass} we give the expression for the eta mass, and in Section~\ref{Sec:EtaDecay} we give the expression for the eta decay constant. In these sections, the expressions presented are simplified using the Gell-Mann-Okubo (GMO) formula. The same expressions, in which the eta mass has been retained, are presented in Appendix~\ref{Sec:NonGMOExpr}. Simplified analytic results for each of the two sets of four master integrals appearing in the expressions for the kaon and eta masses and decay constants, obtained from the ancillary \texttt{Mathematica} code, are discussed in Section~\ref{Sec:NumAnalysis}, while full results for these master integrals are given in Appendix~\ref{Sec:SunsetResults}. Numerical implications of the expressions presented in the paper are also shown in Section~\ref{Sec:NumAnalysis}. This work closes by presenting a set of compact expressions for each of $m_K^2$, $m_\eta^2$, $F_K$ and $F_\eta$, which may be used for easy fitting with lattice data, in Section~\ref{Sec:LatticeFits}, which is followed by a detailed summary and conclusion section. In Appendix~\ref{Sec:PionSunsets}, we explain how exact expressions may be obtained for $m_\pi$ and $F_\pi$, and also provide a set of truncated three mass scale sunset results that may be used to produce approximate analytic expressions for $m_\pi$, $F_\pi$ and $F_K/F_\pi$.

\section{Sunset Master Integrals: notation and convergence properties of the series representations \label{Sec:MI}}

The four three-mass-scale sunset master integrals that arise in the kaon mass and decay constant expressions are:
\begin{align}
\nonumber H_{\{1,1,1\}}^d\{ m_K, m_{\pi},m_{\eta}; p^2=m_K^2\},\\
\nonumber H_{\{2,1,1\}}^d\{ m_K, m_{\pi},m_{\eta}; p^2=m_K^2\},\\
\nonumber H_{\{1,2,1\}}^d\{ m_K, m_{\pi},m_{\eta}; p^2=m_K^2\}, \\
H_{\{1,1,2\}}^d\{ m_K, m_{\pi},m_{\eta}; p^2=m_K^2\}.
\end{align}
and the three independent three-mass-scale sunset master integrals that arise in the eta mass and decay constant expressions are:
\begin{align}
\nonumber H_{\{1,1,1\}}^d\{ m_\pi, m_K, m_K; p^2=m_\eta^2 \},\\
\nonumber H_{\{2,1,1\}}^d\{ m_\pi, m_K, m_K; p^2=m_\eta^2\},\\
H_{\{1,2,1\}}^d\{ m_\pi, m_K, m_K; p^2=m_\eta^2\}.
\end{align}

In ChPT, the renormalisation is normally done using a modified form of the $\overline{MS}$ scheme, and involves multiplying the sunset integral with the factor $(\mu^{2}_{\chi})^{4-d}$, where:
\begin{align}
	\mu^2_{\chi} \equiv \mu^2 \frac{e^{\gamma_E - 1}}{4\pi}
\end{align}

We therefore define:
\begin{align}
	H^{\chi} \equiv   (\mu^2_{\chi})^{4-d} H^d
\end{align}
which is the sunset integral suitably renormalised. In this paper, we denote the sunset integrals normalised using the $\overline{MS}_{\chi}$ scheme by $H^{\chi}$, to differentiate it from the unrenormalised sunset integral $H^d$ defined in Eq.(\ref{Eq:SunsetDef}). This renormalisation introduces in each of these integrals terms containing chiral logarithms:
\begin{align}
	l_{P}^{r} = \frac{1}{2(4\pi)^2}\log \left[ \frac{m_P^2}{\mu^2} \right], \; \; \; \; \; \; \; \; \; \; P = \pi, K, \eta .
\end{align}
We denote the chiral log terms of a sunset integral using a $\log$ superscript, i.e. $H^{\log}$, 
and the rest of the integral by a bar, i.e. $\overline{H}$. 
Therefore we have, for instance:
\begin{align}
	H^{\chi} \equiv \overline{H}^{\chi} + H^{\chi,\log}
\end{align} 
We also adopt the notation of \cite{Kaiser:2007kf} and denote the sunset integrals in this paper by means of the abbreviation:
\begin{align}
	H_{aP \, bQ \, cR} \equiv H_{\{a,b,c\}} \{ m_P, m_Q, m_R; p^2 = m_{K}^2 \}
\end{align}
where we normally omit the numerical index $a,b,c$ on the LHS of the above when their values are unity, and with either the $\log$ superscript or bar over the $H$, as well as either a $\chi$ or a $d$ superscript on it, as appropriate.

The $H^{\chi,\log}$ for the master integrals considered in this paper are given by:
\begin{align}
	& H^{\chi,log}_{P \, Q \, R} = 4 m_P^2 (l^r_P)^2 + 4 m_Q^2 (l^r_Q)^2 + 4 m_R^2 (l^r_R)^2 - \frac{m_P^2}{8\pi^2} l^r_P - \frac{m_Q^2}{8\pi^2} l^r_Q -\frac{m_R^2}{8\pi^2} l^r_R + \frac{s}{16\pi^2} l^r_{s} \nonumber \\
	& H^{\chi,log}_{2P \, Q \, R} = 4 (l^r_P)^2 + \frac{1}{8\pi^2} l^r_P \nonumber \\
	& H^{\chi,log}_{P \, 2Q \, R} = 4 (l^r_Q)^2 + \frac{1}{8\pi^2} l^r_Q \nonumber \\
	& H^{\chi,log}_{P \, Q \, 2R} = 4 (l^r_R)^2 + \frac{1}{8\pi^2} l^r_R
\end{align}
where $s=p^2$, and $l^r_s = l^r_K$ or $l^r_\eta$ as the case may be.  

\begin{figure}
\centering

\begin{minipage}{0.45\textwidth}
\includegraphics[scale=0.5]{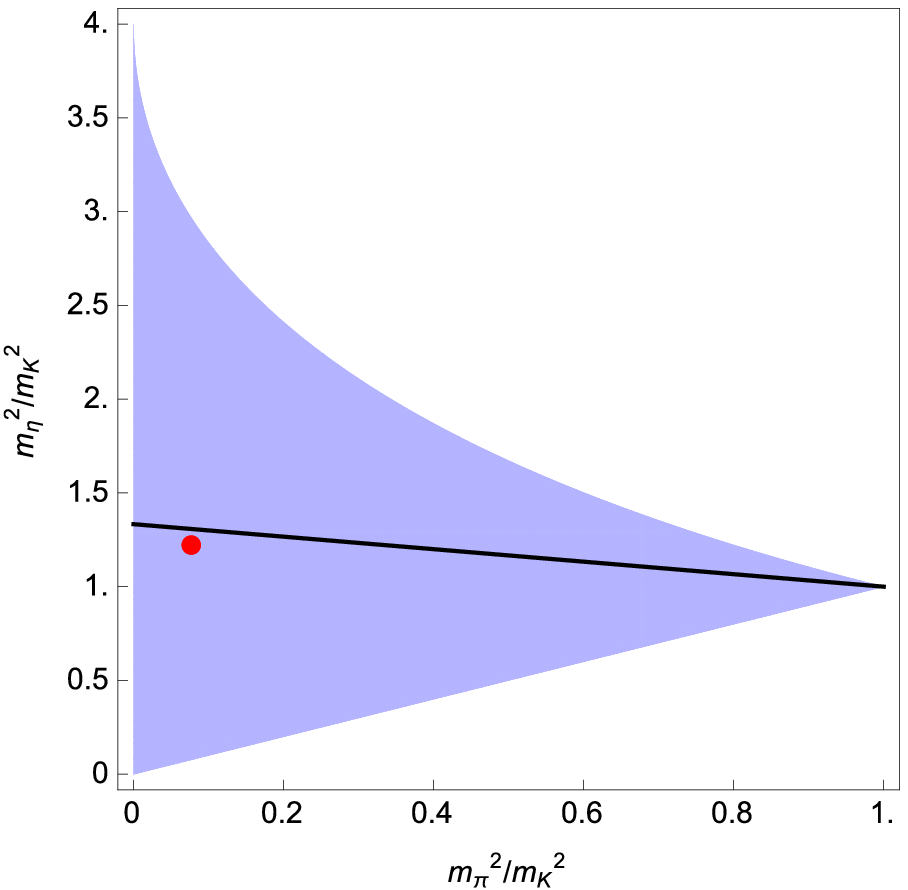} 
\end{minipage}
~~
\begin{minipage}{0.45\textwidth}
\includegraphics[scale=0.5]{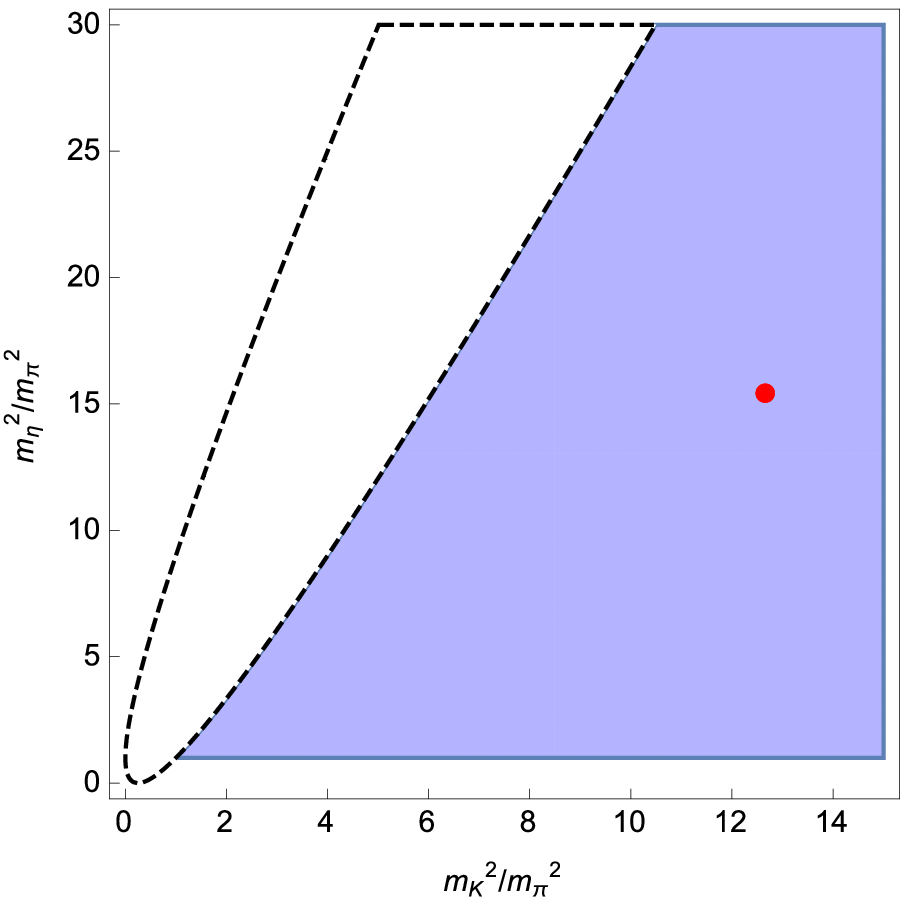}
\end{minipage}

\caption{Region of convergence for results presented in Appendix~\ref{Sec:SunsetResults} (blue area).}

\label{Fig:RegOfConv}
\end{figure}

The full expressions for these master integrals are given in Appendix \ref{Sec:SunsetResults} in the form of linear combinations of independent terms, single infinite series, and double infinite series. These series do not converge for all values of the masses. Rather, they converge for values of the masses that satisfy the following set of inequalities: $(m_{\pi} < m_{\eta}) \bigwedge (m_{\pi} + m_{\eta} < 2m_{K})$, which is graphically described by the blue areas in Figure~\ref{Fig:RegOfConv}, plotted with two different sets of mass ratio axes. The black line in the left panel denotes mass values that obey the GMO formula, while the red dot in both panels marks the physical values of the meson masses.


\section{The pseudoscalar meson masses and decay constants} \label{Sec:MainResults}

The expressions for the pseudo-scalar meson masses is given up to two loops in \cite{Amoros:1999dp} as:
\begin{align}
	m^2_{P} = m^{2}_{P0} + \left( m^{2}_{P} \right)^{(4)} + \left( m^{2}_{P} \right)^{(6)}_{CT} + \left( m^{2}_{P} \right)^{(6)}_{loop} + \mathcal{O}(p^8) \label{NNLOMass}
\end{align}
where $P$ is the particle in question. The model independent $\mathcal{O}(p^6)$ contribution can be subdivided as:
\begin{align}
	F_{\pi}^4 \left( m^{2}_{P} \right)^{(6)}_{loop} = c_{sunset}^{P} + c_{log \times log}^{P} + c_{log}^{P} + c_{log \times L_i}^{P} + c_{L_i}^{P} + c_{L_i \times L_j}^{P}
\end{align}
where the $c_{log}^{P}$ represents the terms containing the chiral logarithms, $c_{log \times log}^{P}$ are the bilinear chiral log terms, $c_{L_i}^{P}$ are those terms proportional to the low energy constants $L_i$,  $c_{L_i \times L_j}^{P}$ are terms bilinear in the LECs, and $c_{log \times L_i}^{P}$ are those terms that contain a product of the low energy constants and a chiral logarithm.

The expressions and breakup for their decay constants is similar:
\begin{align}
	\frac{F_P}{F_0} = 1 + F_P^{(4)} + \left( F_P \right)^{(6)}_{CT} + \left( F_P \right)^{(6)}_{loop} + \mathcal{O}(p^8) \label{NNLODecay}
\end{align}
where:
\begin{align}
	F_{\pi}^4 \left( F_P \right)^{(6)}_{loop} = d_{sunset}^{P} + d_{log \times log}^{P} + d_{log}^{P} + d_{log \times L_i}^{P} + d_{L_i}^{P} + d_{L_i \times L_j}^{P}
\end{align}

In the following sections, we give explicit expressions for each of the terms above for the kaon mass and decay constant. The expressions have been simplified by the use of the GMO relation to rewrite the eta mass in terms of the pion and kaon masses, except in the chiral logarithms, in which the eta mass is understood to mean $ m_{\eta0}^2 = (4 m_{K0}^2-m_{\pi0}^2)/3$. The full expressions, not simplified by use of the GMO relation, are given in Appendix~\ref{Sec:NonGMOExpr} .

\subsection{The kaon mass \label{Sec:KaonMass}}

In this section, we use the expression for the kaon mass to two-loops given in Eqs.(A.5)-(A.7) of \cite{Amoros:1999dp}, and rewrite the linear log terms, the bilinear log terms, and the terms involving the sunset integrals (e.g. $H^F$, $H_1^F$) in Eq.(A.7) of \cite{Amoros:1999dp} by applying Tarasov's relations \cite{Tarasov:1997kx}.

The kaon mass is given in \cite{Amoros:1999dp} as:
\begin{align}
	m^2_{K} = m^{2}_{K0} + \left( m^{2}_{K} \right)^{(4)} + \left( m^{2}_{K} \right)^{(6)}_{CT} + \left( m^{2}_{K} \right)^{(6)}_{loop} + \mathcal{O}(p^8) 
\end{align}

The expressions given here have been simplified using the GMO relation. See Appendix~\ref{Sec:NonGMOExpr} for the full results. The full form of the components above are given by:
\begin{align}
	m^{2}_{K} = B_0 \left( m_s + \hat{m} \right) \label{cTreeMass}
\end{align}
\begin{align}
	\frac{F_{\pi}^2}{m_K^2}	\left( m^{2}_{K} \right)^{(4)} = 8(m_{\pi}^2 + 2 m_K^2)(2 L_6^r - L_4^r) + 8 m_K^2 (2L_8^r - L_5^r) + \frac{2}{9} \left(4 m_{K}^2-m_{\pi}^2\right) l^r_{\eta} \label{KaonMassNLOContrib}
\end{align}
\begin{align}
	 \frac{F_{\pi}^2}{m_K^2} \left( m^{2}_{K} \right)^{(6)}_{CT} =& -32 m_{K}^4 C^r_{12} - 32 m_{K}^2 \left(2 m_{K}^2+m_{\pi}^2\right) C^r_{13} - 16 \left(2 m_{K}^4-2 m_{K}^2 m_{\pi}^2+m_{\pi}^4\right) C^r_{14}  \nonumber \\
	 & -16 m_{K}^2 \left(2 m_{K}^2+m_{\pi}^2\right) C^r_{15} -16 \left(4 m_{K}^4-4 m_{K}^2 m_{\pi}^2+3 m_{\pi}^4\right) C^r_{16} \nonumber \\
	 & +16 m_{\pi}^2 \left(m_{\pi}^2-2 m_{K}^2\right) C^r_{17} + 48 \left(2 m_{K}^4-2 m_{K}^2 m_{\pi}^2+m_{\pi}^4\right) C^r_{19} \nonumber \\
	 & +16 \left(8 m_{K}^4-2 m_{K}^2 m_{\pi}^2+3 m_{\pi}^4\right) C^r_{20} + 48 \left(2 m_{K}^2+m_{\pi}^2\right)^2  C^r_{21} \nonumber \\
	 & +32 m_{K}^4 C^r_{31} + 32 m_{K}^2 \left(2 m_{K}^2+m_{\pi}^2\right) C^r_{32} \label{cCT}
\end{align}
and
\begin{align}
	F_{\pi}^4 \left( m^{2}_{K} \right)^{(6)}_{loop} = c^{K}_{L_i} + c^{K}_{L_i \times L_j} + c^{K}_{log \times L_i} + c^{K}_{log} + c^{K}_{log \times log} + c^{K}_{sunset}
\end{align}
where
\begin{align}
	27 (16 \pi^2)  c^{K}_{L_i} =& 108 m_K^6 L_1^r + 6 m^2_K \left(61 m_K^4-8 m_K^2 m_{\pi}^2+28 m_{\pi}^4\right) L^r_2 + m^2_K \left(89 m_K^4 - 4 m_K^2 m_{\pi}^2 + 41 m_{\pi}^4 \right) L_3^r \nonumber \\
	& -32 m^2_K \left( m_K^2 - m_{\pi}^2\right)^2 \left( L^r_5 -12 L^r_7 -6 L^r_8 \right)
\end{align}

\begin{align}
	c^{K}_{L_i \times L_j} =& -128 \left(4 m_K^6 + 4 m_K^4 m_{\pi}^2 + m_K^2 m_{\pi}^4\right) (L_4^r)^2 - 128 \left(3 m_K^6 + 2 m_K^4 m_{\pi}^2 + m_K^2 m_{\pi}^4\right) L_4^r L_5^r \nonumber \\	
	& + 512 \left(4 m_K^6 + 4 m_K^4 m_{\pi}^2 + m_K^2 m_{\pi}^4\right) L_4^r L_6^r + 128 \left(8 m_K^6 + 3 m_K^4 m_{\pi}^2 + m_K^2 m_{\pi}^4\right) L_4^r L_8^r \nonumber \\
	& - 64 \left(m_K^6 + m_K^4 m_{\pi}^2\right)(L_5^r)^2 + 256 \left(3 m_K^6 + 2 m_K^4 m_{\pi}^2 + m_K^2 m_{\pi}^4\right) L_5^r L_6^r \nonumber \\
	& + 128 \left(3 m_K^6 + m_K^4 m_{\pi}^2\right) L_5^r L_8^r -512 \left(4 m_K^6 + 4 m_K^4 m_{\pi}^2 + m_K^2 m_{\pi}^4\right) (L_6^r)^2 \nonumber \\
	& - 256 \left(8 m_K^6 + 3 m_K^4 m_{\pi}^2 + m_K^2 m_{\pi}^4\right) L_6^r L_8^r - 512 m_K^6 (L_8^r)^2 \label{cLiLj}
\end{align}

\begin{align}
	c^{K}_{log \times L_i} =& 2 m_K^2 m_{\pi}^2 \bigg\{ -3 m_{\pi}^2 (16 L^r_{1}+4 L^r_{2}+5 L^r_{3})+4 \left(8 m_K^2+17 m_{\pi}^2\right) L^r_{4} + 4 \left(4 m_K^2+3 m_{\pi}^2\right) (L^r_{5}-2 L^r_{8})  \nonumber \\
	& \quad -8 \left(8 m_K^2+11 m_{\pi}^2\right) L^r_{6} \bigg\} l_{\pi}^r \nonumber \\
	& - 4 m_K^4 \bigg\{ m_{K}^2 (36 L^r_{1}+18 L^r_{2}+15 L^r_{3}-16 L^r_{5}+32 L^r_{8})-4 \left(10 m_{K}^2+m_{\pi}^2\right) L^r_{4} + 8  \left(8 m_{K}^2+m_{\pi}^2\right) L^r_{6} \bigg\} l_K^r \nonumber \\
	& - \frac{2}{9} m_K^2 \bigg\{ \left(4 m_K^2-m_{\pi}^2\right)^2 (16 L^r_{1}+4 L^r_{2}+7 L^r_{3})-12 L^r_{4} \left(32 m_K^4-12 m_K^2 m_{\pi}^2+m_{\pi}^4\right) \nonumber \\
	& \quad -4 \left(32 m_K^4-2 m_K^2 m_{\pi}^2-3 m_{\pi}^4\right) L^r_{5} + 8 \left(64 m_K^4-20 m_K^2 m_{\pi}^2+m_{\pi}^4\right) L^r_{6} + 96 m_{\pi}^2 \left(m_K^2-m_{\pi}^2\right) L^r_{7} \nonumber \\
	& \quad +8 \left(32 m_K^4-6 m_K^2 m_{\pi}^2-5 m_{\pi}^4\right) L^r_{8} \bigg\} l_{\eta}^r
\end{align}

\begin{align}
	\left(-16\pi^2\right) c^{K}_{log} =& \left( \frac{11}{4} m_{K}^4 m_{\pi}^2 + \frac{455}{144} m_{K}^2 m_{\pi}^4 \right) l^r_{\pi} + \left( \frac{148}{27} m_{K}^6 - \frac{5}{4} m_{K}^4 m_{\pi}^2 - \frac{13}{432} m_{K}^2 m_{\pi}^4 \right)  l^r_{\eta} \nonumber \\
 & + \left( \frac{41}{18} m_{K}^4 m_{\pi}^2 + \frac{487}{72} m_{K}^6 \right) l^r_{K} \label{cBarlog}
\end{align}

\begin{align}
	c^{K}_{log \times log} =& \left(-\frac{11}{81} m_{K}^6 - \frac{47}{81} m_{K}^4 m_{\pi}^2 + \frac{1279}{1296} m_{K}^2 m_{\pi}^4 - \frac{5}{24} m_{\pi}^6 \right) (l^r_{\eta})^2 \nonumber \\
	& + \left( \frac{14}{9} m_{K}^6 + \frac{19}{18} m_{K}^4 m_{\pi}^2 -\frac{1}{4} m_{K}^2 m_{\pi}^4 \right) l^r_{\eta} l^r_{K} + \left(\frac{4}{9} m_{K}^4 m_{\pi}^2 + \frac{43}{6} m_{K}^6 \right) (l^r_{K})^2 \nonumber \\
& - \left(\frac{3}{2} m_{K}^4 m_{\pi}^2 - \frac{1}{4} m_{K}^2 m_{\pi}^4 \right) l^r_{\pi} l^r_{K} + \left(\frac{1}{2} m_{K}^4 m_{\pi}^2 + \frac{169}{48} m_{K}^2 m_{\pi}^4 - \frac{5}{24} m_{\pi}^6 \right) (l^r_{\pi})^2  \nonumber \\
& - \left(\frac{55}{18} m_{K}^4 m_{\pi}^2 + \frac{97}{72} m_{K}^2 m_{\pi}^4 - \frac{5}{12} m_{\pi}^6 \right) l^r_{\pi}l^r_{\eta}  \label{cBarloglog} 
\end{align}

The expressions of Eq.(\ref{cBarlog})-(\ref{cBarloglog}) above are a combination of the linear and bilinear chiral logarithms arising from the evaluation of the sunset integrals, those stemming directly from the $\mathcal{O}(p^6)$ kaon mass expression, as well as contributions arising from the $\mathcal{O}(p^4)$ term due to application of the GMO relation. Similarly, the $c^{K}_{L_i}$and $c^{K}_{log \times L_i}$ components are also made up of terms taken directly from the $\mathcal{O}(p^6)$ kaon mass expression, and contributions arising from the $\mathcal{O}(p^4)$ term due to application of the GMO relation.

The $c^{K}_{sunset}$ term itself has the following contributions to it, where the terms of the first line are a combination of terms from the kaon mass expression as well as from the single mass sunset integrals:
\begin{align}
	c^{K}_{sunset} = \frac{1}{\left(16 \pi ^2\right)^2} & \bigg\{ \left(\frac{767}{108}+\frac{427 \pi^2}{1296}\right) m_{K}^4 m_{\pi}^2 - \left(\frac{12307}{3456}+\frac{275 \pi ^2}{648}\right) m_{K}^6 - \left(\frac{571}{288}+\frac{59 \pi ^2}{216}\right) m_{K}^2 m_{\pi}^4 \nonumber \\
	& - \left(\frac{49}{72}+\frac{\pi ^2}{48}\right) m_{\pi}^6 \bigg\} + c^{K}_{K \pi \pi} + c^{K}_{K \eta \eta} + c^{K}_{K \pi \eta} 
\end{align}
where
\begin{align}
	c^{K}_{K \pi \pi} &= \left(-\frac{3}{32}  m_{K}^4 + \frac{9}{16} m_{K}^2 m_{\pi}^2 + \frac{9}{32} m_{\pi}^4 \right) \overline{H}^{\chi}_{K \pi \pi} + \left(\frac{3}{8} m_{K}^6 - \frac{3}{8} m_{K}^2 m_{\pi}^4 \right) \overline{H}^{\chi}_{2K \pi \pi} \label{cBarkpp}
\end{align}

\begin{align}
	c^{K}_{K \eta \eta} &= \left(\frac{289}{288} m_{K}^4 -\frac{41}{48} m_{K}^2 m_{\pi}^2 + \frac{5}{32} m_{\pi}^4 \right) \overline{H}^{\chi}_{K \eta \eta} + \left(-\frac{73}{72} m_{K}^6 + \frac{11}{9} m_{K}^4 m_{\pi}^2 - \frac{5}{24} m_{K}^2 m_{\pi}^4 \right) \overline{H}^{\chi}_{2K \eta \eta}
\end{align}

\begin{align}
	c^{K}_{K \pi \eta} &= \left( \frac{17}{16} m_{K}^4 - \frac{17}{24} m_{K}^2 m_{\pi}^2 + \frac{7}{48} m_{\pi}^4 \right) \overline{H}^{\chi}_{K \pi \eta} - \left( m_{K}^4 m_{\pi}^2 - \frac{5}{4} m_{K}^2 m_{\pi}^4 + \frac{1}{4} m_{\pi}^6 \right) \overline{H}^{\chi}_{K 2\pi \eta} \nonumber \\
	& - \left( \frac{1}{3} m_{K}^6 - \frac{7}{36} m_{K}^4 m_{\pi}^2 - \frac{7}{36} m_{K}^2 m_{\pi}^4 + \frac{1}{18} m_{\pi}^6 \right) \overline{H}^{\chi}_{K \pi 2\eta}
\end{align}

The terms $c^{K}_{\pi \pi K}, c^{K}_{\eta \eta K}, c^{K}_{K \pi \eta}$ are the result of applying Tarasov's relations to the variety of sunset integrals appearing in Eq.(A.7) of \cite{Amoros:1999dp} and rewriting them in terms of the master integrals given in Appendix~\ref{Sec:SunsetResults}.

\subsection{The kaon decay constant \label{Sec:KaonDecay}}

The treatment of the kaon decay constant is similar to that of the kaon mass in the previous section, except that the expression for the kaon decay constant also involves derivatives of the sunsets with respect to the external momentum. The kaon decay constant to two-loops is given in Eqs.(A.15)-(A.17) of \cite{Amoros:1999dp} as:
\begin{align}
	\frac{F_K}{F_0} = 1 + F_K^{(4)} + \left( F_K \right)^{(6)}_{CT} + \left( F_K \right)^{(6)}_{loop} + \mathcal{O}(p^8) 
\end{align}
where:
\begin{align}
	F_{\pi}^2 F_K^{(4)} = 4\left(2 m_{K}^2+m_{\pi}^2\right) L_4^r  + 4 m_{K}^2 L_5^r - \frac{3}{4} m_{\pi}^2 l_{\pi}^r - \frac{3}{2} m_{K}^2 l^r_{K} -\frac{1}{4} \left(4 m_{K}^2-m_{\pi}^2\right) l^r_{\eta} \label{KaonDecayNLOContrib}
\end{align}
\begin{align}
	F_{\pi}^4 \left( F_K \right)^{(6)}_{CT} &= 8 \left(2 m_{K}^4-2 m_{K}^2 m_{\pi}^2+m_{\pi}^4\right) C^r_{14} + 8 m_{K}^2 \left(2 m_{K}^2+m_{\pi}^2\right) C^r_{15} \nonumber \\
	& + 8 \left(4 m_{K}^4-4 m_{K}^2 m_{\pi}^2+3 m_{\pi}^4\right) C^r_{16} + 8 m_{\pi}^2 \left(2 m_{K}^2-m_{\pi}^2\right) C^r_{17} \label{dCT}
\end{align}
and
\begin{align}
	F_{\pi}^4 \left( F_K \right)^{(6)}_{loop} = d^K_{L_i} + d^K_{L_i \times L_j} + d^K_{log \times L_i} + d^K_{log} + d^K_{log \times log} + d^K_{sunset}
\end{align}
where:
\begin{align}
	- 54(16\pi^2) d^K_{L_i} =& 108 m_{K}^4 L_1^r + 6 \left(61 m_{K}^4 - 8 m_{K}^2 m_{\pi}^2 + 28 m_{\pi}^4 \right) L_2^r + \left(89 m_{K}^4 - 4 m_{K}^2 m_{\pi}^2 + 41 m_{\pi}^4 \right) L_3^r \nonumber \\
	& - 72 \left(m_{K}^2-m_{\pi}^2\right)^2 \left( L^r_{5} - 12 L^r_{7} - 6 L^r_{8} \right)
\end{align}

\begin{align}
	d^K_{L_i \times L_j} =& 56 \left(4 m_{K}^4 + 4 m_{K}^2 m_{\pi}^2 + m_{\pi}^4 \right) (L_{4}^r)^2 + 16 \left(10 m_{K}^4 + 7 m_{K}^2 m_{\pi}^2 + 4 m_{\pi}^4\right) L_4^r L_5^r \nonumber \\
	& -64 \left(4 m_{K}^4 + 4 m_{K}^2 m_{\pi}^2 + m_{\pi}^4\right) L_4^r L_6^r - 64 \left(2 m_{K}^4 + m_{\pi}^4\right) L_4^r L_8^r \nonumber \\
	& + 8 \left(3 m_{K}^4 + 4 m_{K}^2 m_{\pi}^2\right) (L_{5}^r)^2 -64  \left(2 m_{K}^4 + m_{K}^2 m_{\pi}^2\right) L_5^r L_6^r -64 m_{K}^4 L_5^r L_8^r \label{dBarLiLj}
\end{align}

\begin{align}
	d^K_{log \times L_i} =& \left\{ 48 m_{\pi}^4 L^r_{1} + 12 m_{\pi}^4 L^r_{2} + 15 m_{\pi}^4 L^r_{3} - \left(38 m_{K}^2 m_{\pi}^2+47 m_{\pi}^4\right) L^r_{4} - \left(19 m_{K}^2 m_{\pi}^2+6 m_{\pi}^4 \right) L^r_{5} \right\} l^r_{\pi} \nonumber \\
	& + \left\{ 72 m_{K}^4 L^r_{1} + 36 m_{K}^4 L^r_{2} + 30 m_{K}^4 L^r_{3} - 2 m_{K}^2 \left(30 m_{K}^2+7 m_{\pi}^2\right) L^r_{4} - 2 m_{K}^2 \left(7 m_{K}^2+6 m_{\pi}^2\right) L^r_{5} \right\} l^r_{K} \nonumber \\
	& + \bigg\{ \frac{1}{9} \left(4 m_K^2-m_{\pi}^2\right)^2 \left( 16 L^r_{1} + 4 L^r_{2} +7 L^r_{3} \right) - \frac{1}{3} \left(4 m_K^2-m_{\pi}^2\right) \left(22 m_K^2 - m_{\pi}^2\right) L^r_{4}  \nonumber \\
	& \quad - \frac{1}{3} \left(4 m_K^4 + 37 m_K^2 m_{\pi}^2 - 14 m_{\pi}^4\right) L^r_{5} - 16 \left(m_K^2-m_{\pi}^2\right)^2 \left( 2 L^r_{7} + L^r_{8} \right) \bigg\} l^r_{\eta}
\end{align}

\begin{align}
 	\left(16\pi^2\right) d^K_{log} =& \left(\frac{3}{8} m_{K}^2 m_{\pi}^2 + \frac{53}{32} m_{\pi}^4 \right) l^r_{\pi} + \left( \frac{19}{9} m_{K}^4 - \frac{65}{72} m_{K}^2 m_{\pi}^2 + \frac{3}{32} m_{\pi}^4 \right) l^r_{\eta} \nonumber \\
 	& + \left(\frac{245}{48} m_{K}^4 + \frac{173}{72} m_{K}^2 m_{\pi}^2 \right) l^r_{K} \label{dBarlog}
\end{align}

\begin{align}
	d^K_{log \times log} =& \left(\frac{5}{16} \frac{m_{\pi}^6}{m_{K}^2} +\frac{2}{3} m_{K}^2 m_{\pi}^2 - \frac{5}{48} m_{\pi}^4 \right) (l^r_{\pi})^2 - \left(\frac{5}{8} \frac{m_{\pi}^6}{m_{K}^2} - \frac{25}{6} m_{K}^2 m_{\pi}^2 - \frac{47}{24} m_{\pi}^4 \right) l^r_{\eta} l^r_{\pi} \nonumber \\
	& + \left(\frac{31}{9} m_{K}^4 + \frac{5}{16} \frac{m_{\pi}^6}{m_{K}^2} - \frac{11}{18} m_{K}^2 m_{\pi}^2 - \frac{21}{16} m_{\pi}^4 \right) (l^r_{\eta})^2 + \left(\frac{155}{72} m_{K}^4 + \frac{11}{36} m_{K}^2 m_{\pi}^2 \right) (l^r_{K})^2 \nonumber \\
	& - \left(\frac{91}{18} m_{K}^4 + \frac{53}{72} m_{K}^2 m_{\pi}^2 - \frac{3}{8} m_{\pi}^4 \right) l^r_{\eta} l^r_{K} + \left(\frac{51}{8} m_{K}^2 m_{\pi}^2 - \frac{3}{8} m_{\pi}^4 \right) l^r_{K} l^r_{\pi} \label{dBarloglog}
\end{align}

The linear and bilinear chiral log terms given in Eq.(\ref{dBarlog}) and Eq.(\ref{dBarloglog}) are a combination of the terms coming directly from the $\mathcal{O}(p^6)$ kaon decay constant expression, the chiral logs arising from the sunset integrals, and contributions stemming from the $\mathcal{O}(p^4)$ term due to application of the GMO relation. $d^K_{L_i}$ and $d^K_{log \times L_i}$ are similarly made up of terms taken directly from the $\mathcal{O}(p^6)$ expression, and contributions arising from the $\mathcal{O}(p^4)$ term due to application of the GMO relation.

As in the case of the kaon mass, we break up the sunset contribution as follows, in which the first line contains contributions from the single mass sunsets, as well as terms arising from the free terms (i.e. not containing a chiral logarithm or a low energy constant) from the expression for the $\mathcal{O}(p^6)$ contribution to the kaon decay constant:
\begin{align}
	d^K_{sunset} = \frac{1}{\left( 16 \pi ^2\right)^2} & \bigg\{ \left(\frac{17671}{2304}+\frac{1195 \pi ^2}{2592}\right) m_{K}^4 + \left(\frac{49}{48}+\frac{\pi ^2}{32}\right) \frac{m_{\pi}^6}{m_{K}^2}-\left(\frac{1625}{144}+\frac{689 \pi ^2}{1296}\right) m_{K}^2 m_{\pi}^2 \nonumber \\
	& + \left(\frac{2153}{576}+\frac{151 \pi ^2}{432}\right) m_{\pi}^4 \bigg\} + d^K_{K \pi \pi} + d^K_{K \eta \eta} + d^K_{K \pi \eta} 
\end{align}
where
\begin{align}
	d^K_{K \pi \pi} &=  -\left(\frac{27 m_{\pi}^4}{64 m_{K}^2}+\frac{m_{K}^2}{64}+\frac{9 m_{\pi}^2}{16}\right) \overline{H}^{\chi}_{K \pi \pi}  + \left(\frac{m_{K}^4}{16}+\frac{m_{K}^2 m_{\pi}^2}{8}+\frac{9 m_{\pi}^4}{16}\right) \overline{H}^{\chi}_{2K \pi \pi} \label{dBarkpp}
\end{align}

\begin{align}
	d^K_{K \eta \eta} &= - \left(\frac{15 m_{\pi}^4}{64 m_{K}^2}+\frac{1189 m_{K}^2}{576}-\frac{65 m_{\pi}^2}{48}\right) \overline{H}^{\chi}_{K \eta \eta} + \left(\frac{143 m_{K}^4}{48}-\frac{139 m_{K}^2 m_{\pi}^2}{72}+\frac{5 m_{\pi}^4}{16}\right) \overline{H}^{\chi}_{2K \eta \eta}
\end{align}

\begin{align}
	d^K_{K \pi \eta} &= \left( - \frac{7}{32} \frac{m_{\pi}^4}{m_{K}^2} + \frac{5}{96} m_{K}^2 + \frac{7}{6} m_{\pi}^2 \right) \overline{H}^{\chi}_{K \pi \eta} + \left( \frac{3}{8} \frac{m_{\pi}^6}{m_{K}^2} + \frac{1}{4} m_{K}^2 m_{\pi}^2 - \frac{15}{8} m_{\pi}^4 \right) \overline{H}^{\chi}_{K 2\pi \eta} \nonumber \\
	& - \left( \frac{11}{18} m_{K}^4 - \frac{1}{12} \frac{m_{\pi}^6}{m_{K}^2} + \frac{41}{72} m_{K}^2 m_{\pi}^2 + \frac{11}{72}m_{\pi}^4 \right) \overline{H}^{\chi}_{K \pi 2\eta}  - \left( \frac{1}{2} m_{K}^4 \right) \overline{H}^{\chi}_{2K \pi \eta}
\end{align}

The terms $d^{K}_{\pi \pi K}$, $d^K_{\eta \eta K}$ and $d^K_{K \pi \eta}$ are the result of applying Tarasov's relations to the sunset integrals appearing in Eq.(A.17) of \cite{Amoros:1999dp} and rewriting them in terms of the sunset diagram master integrals given in Appendix~\ref{Sec:SunsetResults}.

\subsection{The eta mass \label{Sec:EtaMass}}

The eta mass is given in \cite{Amoros:1999dp} as:
\begin{align}
	m^2_{\eta} = m^{2}_{\eta 0} + \left( m^{2}_{\eta} \right)^{(4)} + \left( m^{2}_{\eta} \right)^{(6)}_{CT} + \left( m^{2}_{\eta} \right)^{(6)}_{loop} + \mathcal{O}(p^8)
\end{align}
where
\begin{align}
	m^{2}_{\eta} = \frac{2}{3} B_0 \left(2 m_s + \hat{m} \right)
\end{align}
and
\begin{align}
	F_{\pi}^2 \left( m_{\eta}^{2} \right)^{(4)} =& \frac{8}{9}(3 L^r_{4}-L^r_{5}-6 L^r_{6}+48 L^r_{7}+18 L^r_{8})  m_{\pi}^4  -\frac{16}{9} (3 L^r_{4}-4 L^r_{5}-6 L^r_{6}+48 L^r_{7}+24 L^r_{8}) m_K^2 m_{\pi}^2 \nonumber \\
	&  -\frac{64}{9} (3 L^r_{4}+2 L^r_{5}-6 L^r_{6}-6 L^r_{7}-6 L^r_{8}) m_K^4 + \left(\frac{8}{3} l^r_{K} - \frac{64}{27} l^r_{\eta} \right) m_K^4  - \left(\frac{7}{27} l^r_{\eta} + l^r_{\pi}\right) m_{\pi}^4 \nonumber \\
	& + \frac{44}{27} l^r_{\eta} m_K^2 m_{\pi}^2 \label{EqMeP4}
\end{align}

The $\mathcal{O}(p^6)$ counter-term contribution is given by:
\begin{align}
	 F_{\pi}^4 \left( m_{\eta}^{2} \right)^{(6)}_{CT} =& -\frac{256}{27} m_{K}^6 ( 8 C_{12}^r + 12 C_{13}^r + 6 C_{14}^r + 6 C_{15}^r + 9 C_{16}^r + 6 C_{17}^r + 6 C_{18}^r - 27 C_{19}^r - 27 C_{20}^r \nonumber \\
	 & - 27 C_{21}^r - 18 C_{31}^r - 18 C_{32}^r - 18 C_{33}^r ) + \frac{16}{27} m_{\pi}^6 ( 2 C_{12}^r - 6 C_{13}^r + 9 C_{14}^r - 3 C_{15}^r + 27 C_{16}^r \nonumber \\
	 & + 9 C_{17}^r + 24 C_{18}^r - 27 C_{19}^r + 27 C_{20}^r - 27 C_{21}^r - 18 C_{31}^r + 54 C_{32}^r) -\frac{32}{9} m_{K}^2 m_{\pi}^4 ( 4 C_{12}^r \nonumber \\
	 & - 6 C_{13}^r + 10 C_{14}^r - 3 C_{15}^r + 24 C_{16}^r +10 C_{17}^r + 24 C_{18}^r - 54 C_{19}^r - 18 C_{20}^r -36 C_{31}^r + 6 C_{32}^r \nonumber \\
	 & - 48 C_{33}^r)+\frac{64}{9} m_{K}^4 m_{\pi}^2 (8 C_{12}^r + 10 C_{14}^r + 15 C_{16}^r + 10 C_{17}^r + 18 C_{18}^r - 54 C_{19}^r - 27 C_{20}^r \nonumber \\
	 & + 27 C_{21}^r - 36 C_{31}^r - 12 C_{32}^r - 48 C_{33}^r )
\end{align}
and the model independent $\mathcal{O}(p^6)$ contribution can be subdivided as:
\begin{align}
	F_\pi^4 \left( m^{2}_{\eta} \right)^{(6)}_{loop} = c_{sunset}^{\eta} + c_{log \times log}^{\eta} + c_{log}^{\eta} + c_{log \times L_i}^{\eta} + c_{L_i}^{\eta} + c_{L_i \times L_j}^{\eta}
\end{align}
where $c_{log}^{\eta}$ represents the terms containing the chiral logarithms:
\begin{align}
	(16 \pi^2) c_{log}^{\eta} =& \left(\frac{41}{324} l^r_{\eta} + \frac{961}{108} l^r_K - 3 l^r_{\pi} \right) m_K^4 m_{\pi}^2 + \left(\frac{371}{486} l^r_{\eta} - 3 l^r_K +\frac{61}{27} l^r_{\pi} \right) m_K^2 m_{\pi}^4 -\left(\frac{1093}{729} l^r_{\eta} + \frac{577}{27} l^r_K \right) m_K^6 \nonumber \\
	& - \left(\frac{2045}{11664} l^r_{\eta} + \frac{931}{432} l^r_{\pi} \right) m_{\pi}^6
\end{align}
The $c_{log \times log}^{\eta}$ term refers to the collection of bilinear chiral log terms:
\begin{align}
	c_{log \times log}^{\eta} &= \left(-\frac{2713}{108} (l^r_{\eta})^2 +\frac{473}{54} l^r_{\eta} l^r_{K} + \frac{256}{27} l^r_{\eta} l^r_{\pi} - \frac{133}{18} (l^r_{K})^2 - \frac{55}{6} l^r_{K} l^r_{\pi} - \frac{3}{4}(l^r_{\pi})^2 \right) m_K^4 m_{\pi}^2 \nonumber \\
	& + \left(\frac{1367}{162} (l^r_{\eta})^2 - \frac{31}{27} l^r_{\eta} l^r_{K} - \frac{172}{27} l^r_{\eta} l^r_{\pi} + \frac{10}{3} (l^r_{K})^2 - 3 l^r_{K} l^r_{\pi} + \frac{5}{2} (l^r_{\pi})^2 \right) m_K^2 m_{\pi}^4 \nonumber \\
	& + \left(\frac{6185}{243}(l^r_{\eta})^2 - \frac{118}{9} l^r_{\eta} l^r_{K} + \frac{103}{9} (l^r_{K})^2 \right) m_K^6 + \left(-\frac{911}{972} (l^r_{\eta})^2 + \frac{7}{9} l^r_{\eta} l^r_{\pi} + \frac{65}{12} (l^r_{\pi})^2 \right) m_{\pi}^6 
\end{align}
and $c_{L_i}^{\eta}$ are those terms proportional to the low energy constants $L_i$:
\begin{align}
	\left(16 \pi ^2\right) c_{L_i}^{\eta} &=  \frac{1}{27} \left(256 L^r_{1}+544 L^r_{2}+152 L^r_{3}+\frac{256}{3} L^r_{5} - 1024 L^r_{7}-512 L^r_{8}\right) m_K^6 \nonumber \\
	& + \frac{1}{9} \left(-64 L^r_{1}-88 L^r_{2}-34 L^r_{3}-\frac{208}{3}L^r_{5} + 832 L^r_{7}+416 L^r_{8}\right) m_K^4 m_{\pi}^2 \nonumber \\
	& +\frac{1}{9} \left(16 L^r_{1}+88 L^r_{2}+32 L^r_{3}+\frac{160}{3}L^r_{5} - 640 L^r_{7}-320 L^r_{8}\right)  m_K^2 m_{\pi}^4 \nonumber \\
	& + \frac{1}{27} \left(-4 L^r_{1}-58 L^r_{2}-20 L^r_{3}-\frac{112}{3} L^r_{5} + 448 L^r_{7}+224 L^r_{8}\right) m_{\pi}^6 \label{cLi}
\end{align}
while bilinears in the LECs are given by $c_{L_i \times L_j}^{\eta}$:

\begin{align}
	c_{L_i \times L_j}^{\eta} =& -\frac{128}{9} \bigg(36 (L^r_{4})^2+15 L^r_{4} L^r_{5}-144 L^r_{4} L^r_{6}+144 L^r_{4} L^r_{7}+42 L^r_{4} L^r_{8}+12 (L^r_{5})^2-30 L^r_{5} L^r_{6} \nonumber \\
	& \qquad -48 L^r_{5} L^r_{7}  -32 L^r_{5} L^r_{8}+144 (L^r_{6})^2-288 L^r_{6} L^r_{7}-84 L^r_{6} L^r_{8}-96 L^r_{7} L^r_{8}-48 (L^r_{8})^2 \bigg)  m_K^4 m_{\pi}^2 \nonumber \\
	& -\frac{128}{9} \bigg( 3 L^r_{4} L^r_{5}+6 L^r_{4} L^r_{8}-10 (L^r_{5})^2-6 L^r_{5} L^r_{6}+144 L^r_{5} L^r_{7}+76 L^r_{5} L^r_{8}-12 L^r_{6} L^r_{8} \nonumber \\
	& \qquad -96 L^r_{7} L^r_{8}-48 (L^r_{8})^2 \bigg)  m_K^2 m_{\pi}^4  \nonumber \\
	& -\frac{1024}{27} \bigg( 6 L^r_{4}+L^r_{5}-12 L^r_{6}-6 L^r_{8}) (3 L^r_{4}+2 L^r_{5}-6 L^r_{6}-6 L^r_{7}-6 L^r_{8} \bigg)  m_K^6 \nonumber \\
	& +\frac{128}{27} \bigg( 3 L^r_{4}+5 L^r_{5}-6 L^r_{6}-6 L^r_{8}) (3 L^r_{4}-L^r_{5}-6 L^r_{6}+48 L^r_{7}+18 L^r_{8} \bigg)  m_{\pi}^6 
\end{align}

$c_{log \times L_i}^{\eta}$ are those terms that contain a product of the low energy constants and a chiral log:
\begin{align}
	c_{log \times L_i}^{\eta} &= \bigg\{ \frac{32}{27}(72 L^r_{1}+72 L^r_{2}+36 L^r_{3}-54 L^r_{4}-113 L^r_{5}+156 L^r_{6}+684 L^r_{7}+422 L^r_{8})  l^r_{\eta}  \nonumber \\
	& \quad +\frac{8}{3} (16 L^r_{1}+4 L^r_{2}+7 L^r_{3}-12 L^r_{4}-4 L^r_{5}+8 L^r_{6}+96 L^r_{7}+56 L^r_{8}) l^r_K  \nonumber \\
	& \quad +\frac{256}{9} (3 L^r_{4}+2 L^r_{5}-6 L^r_{6} -6 L^r_{7} - 6 L^r_{8}) l^r_{\pi} \bigg\} m_K^4 m_{\pi}^2 \nonumber \\
	& + \bigg\{ -\frac{16}{27} (36 L^r_{1}+36 L^r_{2}+18 L^r_{3}-27 L^r_{4}-104 L^r_{5}+78 L^r_{6}+720 L^r_{7}+404 L^r_{8})  l^r_{\eta}  \nonumber \\
	& \quad -\frac{16}{9} (72 L^r_{1}+18 L^r_{2}+18 L^r_{3}-87 L^r_{4}+8 L^r_{5}+102 L^r_{6}-312 L^r_{7}-120 L^r_{8}) l^r_{\pi} \nonumber \\
	& \quad -\frac{16}{9} (3 L^r_{4}-L^r_{5}-6 L^r_{6}+48 L^r_{7}+18 L^r_{8}) l^r_K \bigg\} m_K^2 m_{\pi}^4 \nonumber \\
	& + \bigg\{ -\frac{512}{27} (6 L^r_{1}+6 L^r_{2}+3 L^r_{3}-6 L^r_{4}-6 L^r_{5}+16 L^r_{6}+24 L^r_{7}+20 L^r_{8})  l^r_{\eta} \nonumber \\
	& \quad -\frac{32}{9} (48 L^r_{1}+12 L^r_{2}+21 L^r_{3}-60 L^r_{4}-22 L^r_{5}+72 L^r_{6}+48 L^r_{7}+60 L^r_{8}) l^r_K  \bigg\} m_K^6 \nonumber \\
	& + \bigg\{ \frac{8}{27} (6 L^r_{1}+6 L^r_{2}+3 L^r_{3}-6 L^r_{4}-32 L^r_{5}+16 L^r_{6}+240 L^r_{7}+130 L^r_{8}) l^r_{\eta} \nonumber \\
	& \quad + 8 (4 L^r_{1}+L^r_{2}+L^r_{3}-6 L^r_{4}+8 L^r_{6}-48 L^r_{7}-18 L^r_{8}) l^r_{\pi} \bigg\} m_{\pi}^6 
\end{align}

The contributions from the sunset integrals $\overline{c}_{sunset}^{\eta}$ can in turn be expressed as:
\begin{align}
	c^{\eta}_{sunset} = \frac{1}{\left(16 \pi ^2\right)^2}  &\Bigg\{ \left(\frac{8783}{1944}-\frac{115 \pi ^2}{162}\right) m_K^6+\left(\frac{629 \pi ^2}{1296}-\frac{3515}{864}\right) m_K^4 m_{\pi}^2-\left(\frac{1259}{2592}+\frac{77 \pi ^2}{216}\right) m_K^2 m_{\pi}^4 \nonumber \\
	& \quad -\left(\frac{20183}{31104}+\frac{7 \pi ^2}{432}\right) m_{\pi}^6 \Bigg\} + c_{\eta \pi \pi}^{\eta} + c_{\eta K K}^{\eta} + c_{\pi K K}^{\eta}
\end{align}
where the contributions in the square brackets come from a combination of the single mass scale sunsets and the free terms (i.e. those not involving a chiral log, a low energy constant, or arising from a sunset diagram) of $\mathcal{O}(p^6)$, and where:
\begin{align}
c_{\eta \pi \pi}^{\eta} &= \frac{1}{6} m_{\pi}^4 \overline{H}^\chi_{\eta \pi \pi}
\end{align}
\begin{align}
	c_{\eta K K}^{\eta} = \left(\frac{53}{36} m_K^2 m_{\pi}^2 -\frac{1}{24} m_K^4 - \frac{5}{24} m_{\pi}^4 \right) \overline{H}^\chi_{\eta K K} + \left(\frac{146}{27} m_K^6 - \frac{425}{54} m_K^4 m_{\pi}^2 + \frac{74}{27} m_K^2 m_{\pi}^4 - \frac{5}{18} m_{\pi}^6 \right) \overline{H}^\chi_{2\eta K K}
\end{align}
\begin{align}
	\overline{c}_{\pi K K}^{\eta} =& \left(\frac{9}{8} m_{K}^4 - \frac{13}{12} m_{K}^2 m_{\pi}^2 + \frac{23}{24} m_{\pi}^4 \right) \overline{H}^\chi_{\pi K K} + \left(m_{K}^6 - \frac{1}{3} m_{K}^4 m_{\pi}^2 - \frac{2}{3} m_{K}^2 m_{\pi}^4 \right) \overline{H}^\chi_{\pi 2K K} \nonumber \\
	&  + \left(-\frac{3}{2} m_{K}^4 m_{\pi}^2 + \frac{7}{3} m_{K}^2 m_{\pi}^4 - \frac{5}{6} m_{\pi}^6 \right) \overline{H}^\chi_{2\pi K K}
\end{align}

\subsection{The eta decay constant \label{Sec:EtaDecay}}

The eta decay constant is given in \cite{Amoros:1999dp} as:
\begin{align}
	F_{\eta} = F^0 \left( \overline{F}_{\eta}^{(4)} + ( \overline{F}_{\eta}^{(6)} )_{CT} + ( \overline{F}_{\eta}^{(6)} )_{loop} \right) + \mathcal{O}(p^8)
\end{align}
where the $\mathcal{O}(p^4)$ term is:
\begin{align}
	F_{\pi}^2 \overline{F}_{\eta}^{(4)} =& 8\left(L^r_{4}+\frac{2}{3} L^r_{5}\right)  m_K^2 + 4 \left(L^r_{4}-\frac{1}{3}L^r_{5}\right) m_{\pi}^2 - 3 l^r_K m_K^2
	\label{EqFeP4}
\end{align}
and the $\mathcal{O}(p^6)$ counter-term contribution is given by:
\begin{align}
	 F_{\pi}^4 \left( F_{\eta}^{2} \right)^{(6)}_{CT} =& \left(\frac{64}{3} C^r_{14}+\frac{64}{3} C^r_{15}+32 C^r_{16}+\frac{64}{3} C^r_{17}+\frac{64}{3} C^r_{18}\right) m_K^4 \nonumber \\
	 & + \left(-\frac{64}{3} C^r_{14}+\frac{16}{3} C^r_{15}-32 C^r_{16}-\frac{64}{3} C^r_{17}-\frac{128}{3} C^r_{18}\right) m_K^2 m_{\pi}^2 \nonumber \\
	 & + \left(8 C^r_{14}-\frac{8}{3} C^r_{15}+24 C^r_{16}+8 C^r_{17}+\frac{64}{3} C^r_{18}\right) m_{\pi}^4 
\end{align}
and the model independent $\mathcal{O}(p^6)$ contribution can be subdivided as:
\begin{align}
	\left( F_{\eta} \right)^{(6)}_{loop} = d_{sunset}^{\eta} + d_{log \times log}^{\eta} + d_{log}^{\eta} + d_{log \times L_i}^{\eta} + d_{L_i}^{\eta} + d_{L_i \times L_j}^{\eta}
\end{align}
where $d_{log}^{\eta}$ represents the terms containing the chiral logarithms:
\begin{align}
	(16 \pi^2) d_{log}^{\eta} =& \left( \frac{3}{8} l^r_{\pi} + \frac{3}{2} l^r_K - \frac{4363}{1944} l^r_{\eta}  \right) m_K^2 m_{\pi}^2 + \left(\frac{16631}{1944} l^r_{\eta} + \frac{17}{24} l^r_K \right) m_K^4 + \left(\frac{3713}{7776} l^r_{\eta} + \frac{47}{32} l^r_{\pi} \right) m_{\pi}^4
\end{align}
The $d_{log \times log}^{\eta}$ term refers to the collection of bilinear chiral log terms:
\begin{align}
	(4 m_K^2-m_{\pi}^2) d_{log \times log}^{\eta} =& -\frac{1}{4} \left(\frac{23}{6} (l^r_{\eta})^2 - \frac{167}{3} l^r_{\eta} l^r_K + \frac{43}{3} (l^r_K)^2 - 93 l^r_K l^r_{\pi} - \frac{99}{2} (l^r_{\pi})^2 \right) m_K^4 m_{\pi}^2  \nonumber \\
	& + \frac{1}{3} \left(\frac{71}{2} (l^r_{\eta})^2 - 119 l^r_{\eta} l^r_K + \frac{191}{2} (l^r_K)^2 \right) m_K^6  + \frac{1}{8} \bigg( (l^r_{\eta})^2+9 (l^r_{\pi})^2 \bigg) m_{\pi}^6 \nonumber \\
	& - \left( (l^r_{\eta})^2 + l^r_{\eta} l^r_K + (l^r_K)^2+6 l^r_K l^r_{\pi}+\frac{15}{2} (l^r_{\pi})^2 \right) m_K^2 m_{\pi}^4 
\end{align}
and $d_{L_i}^{\eta}$ are those terms proportional to the low energy constants $L_i$:
\begin{align}
	9 \left(16 \pi ^2\right) d_{L_i}^{\eta} &= 8(2 L^r_{1}+2 L^r_{2}+L^r_{3}) m_K^2 m_{\pi}^2 - (2 L^r_{1}+29 L^r_{2}+10 L^r_{3}) m_{\pi}^4 - (32 L^r_{1}+68 L^r_{2}+19 L^r_{3}) m_K^4 \label{dLi}
\end{align}
while bilinears in the LECs are given by $d_{L_i \times L_j}^{\eta}$:
\begin{align}
	d_{L_i \times L_j}^{\eta} =& \left(224 (L^r_{4})^2+192 L^r_{4} L^r_{5}-256 L^r_{4} L^r_{6}-128 L^r_{4} L^r_{8}+\frac{256}{9} (L^r_{5})^2-\frac{512}{3} L^r_{5} L^r_{6}-\frac{256}{3} L^r_{5} L^r_{8}\right)  m_K^4 \nonumber \\
	& + \left(56 (L^r_{4})^2 + 48 L^r_{4} L^r_{5}-64 L^r_{4} L^r_{6}-64 L^r_{4} L^r_{8} - \frac{200}{9} (L^r_{5})^2 + \frac{64}{3} L^r_{5} L^r_{6} + \frac{64}{3} L^r_{5} L^r_{8} \right) m_{\pi}^4 \nonumber \\
	& + \left(224 (L^r_{4})^2+96 L^r_{4} L^r_{5}-256 L^r_{4} L^r_{6}+\frac{448}{9} (L^r_{5})^2 - \frac{128}{3} L^r_{5} L^r_{6} \right) m_K^2 m_{\pi}^2
\end{align}

$d_{log \times L_i}^{\eta}$ are those terms that contain a product of the low energy constants and a chiral log:
\begin{align}
	d_{log \times L_i}^{\eta} &= \left\{ \frac{4}{3} \left(2 L^r_{1}+2 L^r_{2}+L^r_{3}-L^r_{4}-\frac{2}{3} L^r_{5} \right) l^r_{\eta} + 4 \left(12 L^r_{1}+3 L^r_{2}+3 L^r_{3}-11 L^r_{4}+\frac{2}{3} L^r_{5} \right) l^r_{\pi} \right\} m_{\pi}^4 \nonumber \\
	& - \left\{ \frac{32}{3} \left(2 L^r_{1}+2 L^r_{2}+L^r_{3}-L^r_{4}-\frac{2}{3} L^r_{5}\right) l^r_{\eta} + 4 \left(5 L^r_{4}+\frac{13}{3} L^r_{5}\right) l^r_K + \frac{32}{3} (3 L^r_{4}+2 L^r_{5}) l^r_{\pi} \right\} m_K^2 m_{\pi}^2 \nonumber \\
	& + \left\{ \frac{64}{3} \left(2 L^r_{1}+2 L^r_{2}+L^r_{3}-L^r_{4}-\frac{2}{3} L^r_{5} \right) l^r_{\eta} + 4 (16 L^r_{1}+4 L^r_{2}+7 L^r_{3}-18 L^r_{4}-4 L^r_{5})  l^r_K  \right\} m_K^4
\end{align}

The contributions from the sunset integrals $d_{sunset}^{\eta}$ can in turn be expressed as:
\begin{align}
	\left(4 m_K^2-m_{\pi}^2\right) d^{\eta}_{sunset} &= \frac{1}{\left(16 \pi ^2\right)^2} \bigg\{ \left(\frac{65765}{3888}+\frac{59 \pi ^2}{36}\right) m_K^6-\left(\frac{13465}{1728}+\frac{47 \pi ^2}{96}\right) m_K^4 m_{\pi}^2 \nonumber \\
	&  -\left(\frac{3377}{5184}-\frac{3 \pi ^2}{8}\right) m_K^2 m_{\pi}^4+\left(\frac{46099}{62208}-\frac{\pi ^2}{96}\right) m_{\pi}^6 \bigg\} + d_{\eta \pi \pi}^{\eta} + d_{\eta K K}^{\eta} + d_{\pi K K}^{\eta}
\end{align}
where the contributions in the square brackets come from a combination of the single mass scale sunsets and the free terms (i.e. those not involving a chiral log, a low energy constant, or arising from a sunset diagram) of $\mathcal{O}(p^6)$, and where:
\begin{align}
	d_{\eta \pi \pi}^{\eta} &= \left(\frac{1}{3}m_K^2 m_{\pi}^4 - \frac{1}{12} m_{\pi}^6 \right) \overline{H}^\chi_{2\eta \pi \pi}  - \frac{1}{4} m_{\pi}^4 \overline{H}^\chi_{\eta \pi \pi}
\end{align}
\begin{align}
	d_{\eta K K}^{\eta} = \left(-\frac{479}{48} m_K^4 - \frac{17}{12} m_K^2 m_{\pi}^2 + \frac{1}{16} m_{\pi}^4 \right) \overline{H}^\chi_{\eta K K} + \left(\frac{173}{9}  m_K^6 - \frac{23}{12} m_K^4 m_{\pi}^2 - \frac{19}{18} m_K^2 m_{\pi}^4 + \frac{1}{12}m_{\pi}^6 \right) \overline{H}^\chi_{2\eta K K}
\end{align}
\begin{align}
	d_{\pi K K}^{\eta} &= \left(\frac{87}{16} m_K^4 + \frac{1}{4} m_K^2 m_{\pi}^2 + \frac{5}{16} m_{\pi}^4 \right) \overline{H}^\chi_{\pi K K} + \left(\frac{3}{4}  m_K^4 m_{\pi}^2 - 4 m_K^2 m_{\pi}^4 + \frac{1}{4} m_{\pi}^6 \right) \overline{H}^\chi_{2\pi K K} \nonumber \\
	& + \left(-\frac{33}{2} m_K^6 + \frac{5}{2} m_K^4 m_{\pi}^2 - m_K^2 m_{\pi}^4\right) \overline{H}^\chi_{\pi 2K K} 
\end{align}

\section{Approximate Results for the Three Mass Sunsets \label{Sec:ApproxSunsets}}

We now present truncated results which numerically agree to within 1\% of the full results of the master integrals given in Appendix \ref{Sec:SunsetResults} for much of the range of masses we are interested in.

These partial sums have been obtained with the help of an ancillary \texttt{Mathematica} file, called \texttt{truncation.nb}, provided with this paper. In this file, as inputs one can choose the numerical values of the meson masses and the maximum error acceptable due to the truncation. The file gives the partial sums for each of the master integrals accordingly as an output.

The truncation procedure that we use does not follow from a rigorous asymptotic analysis. Our aim here is more to give simplified formulas that may be used in numerical simulations to save CPU time, mainly for interested lattice practitioners. To get the simplified expressions, we use a simple criterion: for a given set of numerical values of the pseudo-scalar masses, in each of the different contributions of Eqs.(\ref{Eq:Hkpe})-(\ref{Eq:Hp2kk}) we keep terms that are bigger than $10^{-p}$, $p\geq1$ being incremented until we achieve the precision goal given by the numerical difference between the corresponding partial sum and the sum of the first hundreds\footnote{1000 terms for single series and 10000 terms for double series.} of terms, the latter being defined as the `exact' value (notice that the very small uncertainties on the pseudo-scalar masses are neglected in this procedure). This way of getting truncations of course implies that for sufficiently different sets of pseudo-scalar masses one gets non-identical simplified expressions of the master integrals. This however does not detract from their numerical utility.

The truncated results presented below have been tested for all the sets of meson masses presented in the lattice study of \cite{Durr:2016ulb}, and for the majority of these mass values, the truncated expressions give results that are accurate upto $1\%$ of the exact value. The numerical implications and accuracy of these approximate results are studied in more detail in Section~\ref{Sec:NumAnalysis}.

\subsection{Truncated kaon sunsets}

\begin{align}
	& \overline{H}^{\chi}_{K \pi \eta} \approx \frac{m_{K}^2}{512\pi ^4} \Bigg\{ \frac{5 \pi^2}{6} -\frac{1}{4} - \frac{7}{4}\left(\frac{m_{\eta}^4}{m_{K}^4}+\frac{m_{\pi}^4}{m_{K}^4}\right) + \left(1-\frac{\pi^2}{2}\right)\left(\frac{m_{\eta}^2}{m_{K}^2}+\frac{m_{\pi}^2}{m_{K}^2}\right) + \frac{1}{2} \frac{m_{\eta}^4}{m_{K}^4} \log\left[\frac{m_{\eta}^2}{m_{K}^2}\right] \nonumber \\
	& \quad +\frac{m_{\pi}^2 m_{\eta}^2}{m_{K}^4} \left(7+\frac{2 \pi^2}{3}-2 \log\left[\frac{m_{\eta}^2}{m_{K}^2}\right]-2 \log\left[\frac{m_{\pi}^2}{m_{K}^2}\right]+\log\left[\frac{m_{\eta}^2}{m_{K}^2}\right] \log\left[\frac{m_{\pi}^2}{m_{K}^2}\right]\right) + \frac{1}{2} \frac{m_{\pi}^4}{m_{K}^4} \log\left[\frac{m_{\pi}^2}{m_{K}^2}\right]  \nonumber \\
	& \quad  -\frac{m_{\pi}^2}{m_{K}^2} \log\left[\frac{m_{\pi}^2}{m_{K}^2}\right]^2-\frac{m_{\eta}^2}{m_{K}^2} \log\left[\frac{m_{\eta}^2}{m_{K}^2}\right]^2 + 
	\frac{8 \pi }{3} \frac{m_{\eta}^3}{m_{K}^3}
	{}_2F_1 \bigg[ \begin{array}{c}
		-\frac{1}{2},\frac{1}{2} \\
		\frac{5}{2} \\
	\end{array}	\bigg| \frac{m_\eta^2}{4 m_{K}^2} \bigg]
	+\frac{1}{36}\frac{m_{\eta}^6}{m_{K}^6}
	{}_3F_2 \bigg[ \begin{array}{c}
		1,1,2 \\
		\frac{5}{2},4 \\
	\end{array}	\bigg| \frac{m_\eta^2}{4 m_K^2} \bigg] \nonumber \\
	& \quad + \frac{1}{6}  \frac{m_{\pi}^2 m_{\eta}^2}{m_K^4}  \left(  \log \left[\frac{m_{\eta}^2}{4 m_K^2}\right] + \log \left[\frac{m_{\pi}^2}{4 m_K^2}\right] \right) \left( \frac{m_{\eta}^2}{m_K^2}
	{}_2F_1 \bigg[ \begin{array}{c}
		1,1 \\
		\frac{5}{2} \\
	\end{array}	\bigg| \frac{m_\eta^2}{4m_K^2} \bigg] 	
	+ \frac{m_{\pi}^2}{m_K^2} 
	{}_2F_1 \bigg[ \begin{array}{c}
		1,1 \\
		\frac{5}{2} \\
	\end{array}	\bigg| \frac{m_\pi^2}{4m_K^2} \bigg] \right)  \nonumber \\
	& \quad - \frac{15\pi}{512} \frac{m_{\pi}^4 m_{\eta}^3}{m_{K}^7} \left(\log \left[\frac{m_{\pi}^2}{16 m_{\eta}^2}\right]+\frac{13}{6}\right) - \frac{1}{20} \frac{m_{\pi}^4 m_{\eta}^4}{m_{K}^8} \left(\frac{37}{15}-\log \left[\frac{m_{\eta}^2}{m_{K}^2}\right]-\log \left[\frac{m_{\pi}^2}{m_{K}^2}\right] \right) \nonumber \\
	& \quad - \frac{\pi}{4} \frac{m_{\pi}^2 m_{\eta}^3}{m_{K}^5} \left(\log \left[\frac{m_{\pi}^2}{16 m_{\eta}^2}\right] + \frac{11}{3}\right) + \frac{1}{12} \frac{m_{\pi}^2 m_{\eta}^2}{m_{K}^4} \left(\frac{m_{\eta}^2}{m_{K}^2}+\frac{m_{\pi}^2}{m_{K}^2}\right) \left(5-8 \gamma-4\psi \left[\frac{5}{2}\right] \right) \nonumber \\
	& \quad + 2 \pi \frac{m_{\pi}^2 m_{\eta}}{m_{K}^3} \left(\log \left[\frac{m_{\pi}^2}{16 m_{\eta}^2}\right]+1\right) + \frac{\pi}{32} \frac{ m_{\pi}^4}{m_{K}^4}  \left(8 \frac{m_{K}}{m_{\eta}} + 3 \frac{m_{\eta}}{m_{K}}\right) \left(\frac{1}{2}-\log \left[\frac{m_{\pi}^2}{16 m_{\eta}^2}\right] \right) \Bigg\}
\label{HkpeApprox}
\end{align}

\begin{align}
& \overline{H}^{\chi}_{2K\pi\eta} \approx \frac{1}{512\pi^4} \Bigg\{ \frac{5\pi^2}{6} -1 - \frac{m_{\eta}^2}{m_{K}^2} \left( 1+\frac{\pi^2}{3}+ \frac{1}{2} \log^2 \left[\frac{m_{K}^2}{m_{\eta}^2}\right] + \log \left[\frac{m_{K}^2}{m_{\eta}^2}\right]  + \text{Li}_2 \left[1-\frac{m_{\pi}^2}{m_{\eta}^2}\right] \right) \nonumber \\
& \quad - \frac{m_{\pi}^2}{m_{K}^2} \left( 1+\frac{\pi^2}{3} - \log \left[\frac{m_{\pi}^2}{m_{K}^2}\right] - \log \left[\frac{m_{K}^2}{m_{\eta}^2}\right]\log \left[\frac{m_{\pi}^2}{m_{K}^2}\right]  - \text{Li}_2 \left[1-\frac{m_{\pi}^2}{m_{\eta}^2}\right] \right) + \frac{2\pi}{3} \frac{m_{\eta}^3}{m_K^3} 
	{}_2F_1 \bigg[ \begin{array}{c}
		\frac{1}{2},\frac{1}{2} \\
		\frac{5}{2} \\
	\end{array}	\bigg| \frac{m_\eta^2}{4m_K^2} \bigg] \nonumber \\
& \quad 
- \frac{1}{4} \frac{m_{\eta }^4}{m_K^4} 
	{}_3F_2 \bigg[ \begin{array}{c}
		1,1,1 \\
		\frac{3}{2},3 \\
	\end{array}	\bigg| \frac{m_\eta^2}{4m_K^2} \bigg] + \frac{1}{2} \frac{m_{\pi}^2 m_{\eta}^2}{m_{K}^4} \left( 4 - \log\left[ \frac{m_{\pi}^2}{m_{K}^2} \right] + \log\left[ \frac{m_{K}^2}{m_{\eta}^2} \right] \right) + \frac{\pi}{2} \frac{m_{\pi}^2 m_{\eta}}{m_{K}^3} \left( 1 + \log \left[ \frac{m_{\pi}^2}{16 m_{\eta}^2} \right] \right) \nonumber \\
& \quad + \frac{1}{60} \frac{m_{\pi}^2 m_{\eta}^6}{m_{K}^8} \left( \frac{4}{5} - \log\left[ \frac{m_{\pi}^2}{m_{K}^2} \right] + \log\left[ \frac{m_{K}^2}{m_{\eta}^2} \right] \right)  + \frac{1}{12} \frac{m_{\pi}^2 m_{\eta}^4}{m_{K}^6} \left( \frac{11}{6} - \log\left[ \frac{m_{\pi}^2}{m_{K}^2} \right] + \log \left[ \frac{m_{K}^2}{m_{\eta}^2} \right] \right) \nonumber \\
& \quad + \frac{\pi}{16} \frac{m_{\pi}^2 m_{\eta}^3}{m_{K}^5} \left( \frac{11}{3} + \log \left[ \frac{m_{\pi}^2}{16 m_{\eta}^2} \right] \right) + \frac{\pi}{16} \frac{m_{\pi}^4}{m_{K}^3 m_{\eta}} \left( \frac{1}{2} - \log \left[ \frac{m_{\pi}^2}{16 m_{\eta}^2} \right] \right)  \Bigg\}
\label{H2kpeApprox}
\end{align}

\begin{align}
& \overline{H}^{\chi}_{K 2\pi \eta} \approx \frac{1}{512\pi ^4} \Bigg\{ - 1 - \frac{\pi^2}{2} - 2\log \left[ \frac{m_{\pi}^2}{m_{K}^2} \right] -\log^2 \left[\frac{m_{\pi}^2}{m_{K}^2} \right] -\frac{m_{\pi}^2}{m_{K}^2} \left(3-\log \left[ \frac{m_{\pi}^2}{m_{K}^2}\right] \right) \nonumber \\
	& \quad + \frac{m_{\eta}^2}{m_{K}^2} \left( 5 + \frac{2 \pi^2}{3} - 2 \log \left[ \frac{m_{\pi}^2}{m_{K}^2} \right] + \log \left[ \frac{m_{K}^2}{m_{\eta}^2}\right] - \log \left[ \frac{m_{K}^2}{m_{\eta}^2}\right]  \log \left[ \frac{m_{\pi}^2}{m_{K}^2}\right] \right) + \frac{1}{12} \frac{m_{\pi}^4}{m_{K}^4}
	{}_3F_2 \bigg[ \begin{array}{c}
		1,1,2 \\
		\frac{5}{2},3 \\
	\end{array}	\bigg| \frac{m_\pi^2}{4m_K^2} \bigg]  \nonumber \\
	& \quad + \frac{1}{6} \frac{m_{\eta}^4}{m_{K}^4}  \left( 2\gamma_E + \log \left[ \frac{m_{\eta}^2}{4 m_{K}^2} \right] + \log \left[ \frac{m_{\pi}^2}{4 m_{K}^2} \right] \right) 
	{}_2F_1 \bigg[ \begin{array}{c}
		1,1 \\
		\frac{5}{2} \\
	\end{array}	\bigg| \frac{m_\eta^2}{4m_K^2} \bigg] + \frac{1}{3} \frac{m_{\eta}^4}{m_{K}^4} \left(\frac{5}{4} - 2 \gamma -\psi\left[\frac{5}{2}\right]\right) \nonumber \\
	& \quad + \frac{1}{3} \frac{m_{\pi}^2 m_{\eta}^2}{m_{K}^4} \left( \log \left[ \frac{m_{\eta}^2}{4 m_{K}^2} \right] + \log \left[ \frac{m_{\pi}^2}{4 m_{K}^2} \right] \right)	{}_3F_2 \bigg[ \begin{array}{c}
		1,1,3 \\
		2,\frac{5}{2} \\
	\end{array}	\bigg| \frac{m_\pi^2}{4m_K^2} \bigg] - \frac{2}{3} \frac{m_{\pi}^2 m_{\eta}^2}{m_{K}^4} \left(\frac{7}{6}+\gamma -\log [4] \right) \nonumber \\
	& \quad - \frac{15 \pi}{256} \frac{m_{\pi}^2 m_{\eta}^3}{m_{K}^5}\left(\log \left[ \frac{m_{\pi}^2}{16 m_{\eta}^2}\right] + \frac{8}{3}\right) - \frac{\pi}{4} \frac{m_{\eta}^3}{m_{K}^3} \left(\log \left[ \frac{m_{\pi}^2}{16 m_{\eta}^2}\right] + \frac{14}{3}\right) - \frac{3 \pi}{32} \frac{m_{\pi}^4 }{m_{K} m_{\eta}^3} \left(\log \left[ \frac{m_{\pi}^2}{16 m_{\eta}^2} \right] + \frac{5}{3}\right) \nonumber \\
		& \quad - \frac{\pi}{2} \frac{m_{\eta}}{m_{K}} \left(\frac{m_{\pi}^2}{m_{\eta}^2} + \frac{3}{8} \frac{m_{\pi}^2}{m_{K}^2}\right) \log \left[\frac{m_{\pi}^2}{16 m_{\eta}^2}\right]   - \frac{1}{10} \frac{m_{\pi}^2 m_{\eta}^4}{m_{K}^6} \left( \frac{59}{30}-\log \left[ \frac{m_{\eta}^2}{m_{K}^2} \right] - \log \left[ \frac{m_{\pi}^2}{m_{K}^2} \right] \right) \nonumber \\
		& \quad - \frac{\pi}{64} \frac{m_{\eta}^5}{m_{K}^5} \left(\log \left[ \frac{m_{\pi}^2}{16 m_{\eta}^2} \right] + \frac{86}{15}\right) + 2 \pi \frac{m_{\eta}}{m_{K}} \left(\log \left[ \frac{m_{\pi}^2}{16 m_{\eta}^2}\right]+2 \right) \Bigg\}
\label{Hk2peApprox}
\end{align}

\begin{align}
	& \overline{H}^{\chi}_{K \pi 2\eta} \approx \frac{1}{512\pi^4} \Bigg\{ - 1 -\frac{\pi^2}{2} + 2 \log \left[ \frac{m_k^2}{m_{\eta}^2} \right] - \log^2 \left[\frac{m_{K}^2}{m_{\eta}^2}\right] + \pi \left(\frac{m_{\eta}^2}{m_{K}^2}\right)^{1/2} \left(4-\frac{m_{\eta}^2}{m_{K}^2}\right)^{1/2} - \frac{\pi m_{\pi}^2}{m_{\eta} m_{K}} \nonumber \\
	& \quad + \frac{m_{\pi}^2}{m_{K}^2} \left( 5 + \frac{2 \pi ^2}{3} + 2 \log \left[\frac{m_{K}^2}{m_{\eta}^2}\right] - \log \left[\frac{m_{\pi}^2}{m_{K}^2}\right] - \log \left[\frac{m_{\pi}^2}{m_{K}^2}\right] \log \left[ \frac{m_{K}^2}{m_{\eta}^2} \right] \right) - \frac{m_{\eta}^2}{m_{K}^2} \left(3 + \log \left[\frac{m_{K}^2}{m_{\eta}^2}\right] \right) \nonumber \\
	& \quad + 2 \pi \frac{m_{\eta}}{m_K}
	{}_2F_1 \bigg[ \begin{array}{c}
		\frac{1}{2},\frac{1}{2} \\
		\frac{3}{2} \\
	\end{array}	\bigg| \frac{m_\eta^2}{4m_K^2} \bigg]
	+ \frac{m_{\eta}^4}{12 m_{K}^4}
	{}_3F_2 \bigg[ \begin{array}{c}
		1,1,2 \\
		\frac{5}{2},3 \\
	\end{array}	\bigg| \frac{m_\eta^2}{4m_K^2} \bigg]
	- \frac{1}{10} \frac{m_{\eta}^4 m_{\pi}^2}{m_{K}^6} \left(\frac{7}{30}+\gamma -\log (4)\right) \nonumber \\
	& \quad -\frac{2}{3} \frac{m_{\pi}^2 m_{\eta}^2}{m_{K}^4} \left(\frac{7}{6}+\gamma -\log (4)\right) 
	- \frac{3 \pi}{8} \frac{m_{\pi}^2 m_{\eta}}{m_{K}^3} \left(\log \left[\frac{m_{\pi}^2}{16 m_{\eta}^2}\right]+3\right) + \frac{\pi m_{\pi}^2}{m_{\eta} m_{K}} \log \left[\frac{m_{\pi}^2}{16 m_{\eta}^2}\right] \nonumber \\
	& \quad  - \frac{5\pi}{128} \frac{m_{\pi}^2 m_{\eta}^3}{m_{K}^5} \left(\log \left[\frac{m_{\pi}^2}{16 m_{\eta}^2}\right] + \frac{13}{3}\right) 
	+ \frac{\pi}{16} \frac{m_{\pi}^4}{m_{\eta}^3 m_{K}}  \left(2 \log \left[\frac{m_{\pi}^2}{16 m_{\eta}^2}\right]+3\right) \nonumber \\
	& \quad + \frac{1}{3} \frac{m_\pi^2}{m_K^2} \frac{m_\eta^2}{m_K^2}
	{}_3F_2 \bigg[ \begin{array}{c}
		1,1,3 \\
		\frac{5}{2},2 \\
	\end{array}	\bigg| \frac{m_\eta^2}{4m_K^2} \bigg] \left( 2\gamma - 1 + \log \left[ \frac{m_\eta^2}{4 m_K^2} \right] + \log \left[ \frac{m_\pi^2}{4 m_K^2} \right] \right) \Bigg\}
\label{Hkp2eApprox}
\end{align}

\subsection{Truncated eta sunsets}

\begin{align}
	& \overline{H}^\chi_{\pi K K} \approx \frac{m_\pi^2}{512\pi^4} \Bigg\{ \frac{\pi ^2}{6}-5-\log^2 \left[ \frac{m_\pi^2}{m_K^2} \right] + 4 \log \left[\frac{m_\pi^2}{m_K^2}\right] + \frac{m_\eta^2}{m_\pi^2}  \left(\log \left[ \frac{m_K^2}{m_\eta^2}\right] + \frac{5}{4}\right) + \frac{m_K^2}{m_\pi^2} \left(6+\frac{\pi ^2}{3}\right) \nonumber \\
	& + \frac{1}{3} \frac{m_\eta^2}{m_K^2} \left(\frac{7}{6}-\log \left[\frac{m_\pi^2}{m_K^2}\right] \right) + \frac{1}{10} \frac{m_\pi^2}{m_K^2} \frac{m_\eta^2}{m_K^2} \left(\frac{37}{30}-\log \left[ \frac{m_\pi^2}{m_K^2}\right] \right) - \frac{1}{18} \frac{m_\eta^2}{m_\pi^2} \frac{m_\eta^2}{m_K^2} {}_3F_2 \bigg[ \begin{array}{c}
		1,1,2 \\
		\frac{5}{2},4 \\
	\end{array}	\bigg| \frac{m_\eta^2}{4 m_K^2} \bigg] \nonumber \\
	& + \frac{1}{3} \frac{m_\pi^2}{m_K^2} \left( \frac{8}{3}-\log [4] 
	-{}_2F_1 \bigg[ \begin{array}{c}
		1,1 \\
		\frac{5}{2} \\
	\end{array}	\bigg| \frac{m_\pi^2}{4 m_K^2} \bigg] \log \left[ \frac{m_\pi^2}{4 m_K^2} \right] \right) \Bigg\}
\label{HpkkApprox} 
\end{align}

\begin{align}
	& \overline{H}^\chi_{2\pi K K} \approx \frac{1}{512\pi^4} \Bigg\{2 \log \left[\frac{m_\pi^2}{m_K^2}\right]-\log ^2\left[\frac{m_\pi^2}{m_K^2}\right] + \frac{1}{3} \frac{m_\eta^2}{m_K^2} \left(\frac{1}{6}-\log \left[\frac{m_\pi^2}{m_K^2} \right] \right) - \frac{1}{30} \frac{m_\eta^4}{m_K^4} \left(\frac{19}{15} + \log \left[\frac{m_\pi^2}{m_K^2}\right] \right) \nonumber \\
	& \quad + \frac{2}{3} \frac{m_\pi^2}{m_K^2}  \Bigg( \frac{8}{3}-\log [4] -\, {}_2F_1 \bigg[ \begin{array}{c}
		1,1 \\
		\frac{5}{2} \\
	\end{array}	\bigg| \frac{m_\pi^2}{4 m_K^2} \bigg] \left(\frac{1}{2} + \log \left[\frac{m_\pi^2}{4 m_K^2}\right] \right) \Bigg) + \frac{1}{5} \frac{m_\pi^2}{m_K^2} \frac{m_\eta^2}{m_K^2} \left(\frac{11}{15}-\log \left[\frac{m_\pi^2}{m_K^2}\right] \right)  \nonumber \\
	& \quad + \frac{1}{10} \frac{m_\pi^4}{m_K^4} \left( \frac{31}{15} - \log [4] -\frac{1}{3} {}_2F_1 \bigg[ \begin{array}{c}
		2,2 \\
		\frac{7}{2} \\
	\end{array}	\bigg| \frac{m_\pi^2}{4 m_K^2} \bigg] \log \left[\frac{m_\pi^2}{4 m_K^2}\right] \right) + \frac{5}{231} \frac{m_\pi^4}{m_K^4} \frac{m_\eta^6}{m_K^{6}}  \left(\frac{757}{2772} - \log \left[ \frac{m_\pi^2}{m_K^2}\right] \right) \nonumber \\
	& \quad - \frac{1}{210} \frac{m_\eta^6}{m_K^6} - \frac{1}{63} \frac{m_\pi^2}{m_K^2} \frac{m_\eta^6}{m_K^6} \left(\frac{79}{630} + \log \left[\frac{m_\pi^2}{m_K^2}\right] \right) + \frac{1}{21} \frac{m_\pi^4}{m_K^4} \frac{m_\eta^4}{m_K^4}  \left(\frac{223}{315}-\log \left[\frac{m_\pi^2}{m_K^2}\right] \right) \nonumber \\
	& \quad - \frac{2}{35} \frac{m_\pi^2}{m_K^2} \frac{m_\eta^4}{m_K^4} \left(\frac{9}{140} + \log \left[\frac{m_\pi^2}{m_K^2}\right] \right) + \frac{3}{35} \frac{m_\eta^2}{m_K^2} \frac{m_\pi^4}{m_K^4} \left(\frac{533}{420}-\log \left[\frac{m_\pi^2}{m_K^2}\right] \right) + \frac{\pi ^2}{6} - 3 \Bigg\}
\label{H2pkkApprox}
\end{align}

\begin{align}
	& \overline{H}^\chi_{\pi 2K K} \approx \frac{1}{512\pi^4} \Bigg\{ 1 + \frac{\pi ^2}{6} - \frac{1}{10} \frac{m_\eta^2}{m_K^2} \frac{m_\pi^4}{m_K^4} \left(\frac{11}{15}-\log \left[ \frac{m_\pi^2}{m_K^2} \right]\right) + \frac{1}{60} \frac{m_\pi^6}{m_K^6} {}_2F_1 \bigg[ \begin{array}{c}
		2,2 \\
		\frac{7}{2} \\
	\end{array}	\bigg| \frac{m_\pi^2}{4 m_K^2} \bigg] \log \left[\frac{m_\pi^2}{4 m_K^2}\right] \nonumber \\
	& - \frac{3}{70} \frac{m_\pi^4}{m_K^8} \frac{m_\eta^4}{m_K^4} \left(\frac{43}{420}-\log \left[ \frac{m_\pi^2}{m_K^2}\right] \right) + \frac{1}{30} \frac{m_\pi^2}{m_K^2} \frac{m_\eta^4}{m_K^4} \left(\frac{23}{30} + \log \left[ \frac{m_\pi^2}{m_K^2}\right] \right) - \frac{2}{63} \frac{m_\eta^4}{m_K^4} \frac{m_\pi^6}{m_K^6} \left(\frac{223}{315}-\log \left[ \frac{m_\pi^2}{m_K^2} \right] \right) \nonumber \\
	&  - \frac{3}{70} \frac{m_\eta^2}{m_K^2} \frac{m_\pi^6}{m_K^6} \left(\frac{533}{420}-\log \left[ \frac{m_\pi^2}{m_K^2} \right] \right) - \frac{1}{6} \frac{m_\pi^2}{m_K^2}\frac{m_\eta^2}{m_K^2} \left(\frac{1}{6} - \log \left[ \frac{m_\pi^2}{m_K^2} \right] \right) + \frac{1}{2} \frac{m_\eta^2}{m_K^2}  {}_3F_2 \bigg[ \begin{array}{c}
		1,1,1 \\
		\frac{3}{2},3 \\
	\end{array}	\bigg| \frac{m_\eta^2}{4 m_K^2} \bigg]  \nonumber \\
	& - \frac{1}{6} \frac{m_\pi^4}{m_K^4} \Bigg( \frac{8}{3} - \log [4] - \left( 1 + \log \left[\frac{m_\pi^2}{4 m_K^2}\right] \right) {}_2F_1 \bigg[ \begin{array}{c}
		1,1 \\
		\frac{5}{2} \\
	\end{array}	\bigg| \frac{m_\pi^2}{4 m_K^2} \bigg]  \Bigg) - \frac{m_\pi^2}{m_K^2} \left(2-\log \left[ \frac{m_\pi^2}{m_K^2} \right] \right) \Bigg\}
\label{Hp2kkApprox} 
\end{align}

\section{Numerical analysis \label{Sec:NumAnalysis}}

Several numerical analyses are perfomed in this section. We first perform a study to determine the relative contribution of the various classes of terms making up the NNLO piece of $m_K$, $F_K$, $m_\eta$ and $F_\eta$, while also examining the difference that arises due to use of the GMO simplified, as contrasted to the physical, expressions. Next, by means of numerical tests, we justify the use of the truncated sunset expressions of Section~\ref{Sec:ApproxSunsets} instead of the exact expressions of Appendix~\ref{Sec:SunsetResults} in potential studies involing fits with lattice data. In both these studies, we do not provide uncertainties for the values provided, as the numerics are comparative rather than absolute in nature. In the last part of this section, we compute values for  $m_K$, $F_K$, $m_\eta$ and $F_\eta$ using our expressions, and physical meson mass values as inputs. Since our aim in this last part is to provide numbers that can be used to check our expressions, rather than to present new and carefully recalculated values of the $m_P$ and $F_P$, and in keeping with the convention used in \cite{Bijnens:2014lea}, with whose values our own are compared, we give only central values for the calculated quantities.

\subsection{Breakup of the contributions}

We begin by giving a numerical breakup of the various terms that make up the  masses and decay constants to show their relative contributions. As the expressions given earlier in this paper are `renormalized' ones, we can directly substitute physical values for the meson masses and the pion decay constant in them, and the error is of $\mathcal{O}(p^8)$.

Table \ref{TableContrib} gives numerical values for the various components of the two loop contributions to the kaon and eta masses and decay constants for two different sets of values of the LECs, i.e. the free fit and the BE14 fit, obtained from continuum fits at NNLO, and the results of which are summarized in Ref.\cite{Bijnens:2014lea}. These numbers have been obtained by using the full $m_{\eta}^2$ dependent expressions (i.e. that have not been simplified by use of the GMO relation), and by summing the first 1000 terms of the single series, and the first 10000 terms of the double series, of the expressions given in Appendix~\ref{Sec:SunsetResults} for the three mass scale sunsets.
\begin{table}
\centering
\begin{tabular}{| c | | c | c c c c c c | c | }
\hline
 & \multirow{2}{*}{Fit} & \multirow{2}{*}{$sunset$} & \multirow{2}{*}{$log \times log$} & \multirow{2}{*}{$log$} & \multirow{2}{*}{$log \times L_i$} & \multirow{2}{*}{${L}_i$} & \multirow{2}{*}{${L}_i \times {L}_j$} & \multirow{2}{*}{Sum} \\[2ex]
\hline
\hline
\multirow{2}{*}{$m^2_K$} & Free Fit & \multirow{2}{*}{$2.4100$} & \multirow{2}{*}{$0.9420$} & \multirow{2}{*}{$3.0586$} & $-2.8763$ & $0.1178$ & $-0.3124$ & $3.3396$  \\ 
					& BE14 & & & & $-4.3794$ & $0.2768$ & $0.0665$ &	$2.3745$ \\
\hline
\multirow{2}{*}{$F_K$} & Free Fit & \multirow{2}{*}{$-1.2220$} & \multirow{2}{*}{$1.7648$} & \multirow{2}{*}{$-7.3042$} & $18.3342$ & $-0.2398$ & $3.1301$ & $14.4631$  \\ 
					& BE14 & & & & $15.0591$ & $-0.5637$ & $1.2018$ & $8.9358$	 \\
\hline
\hline
\multirow{2}{*}{$m^2_\eta$} & Free Fit & \multirow{2}{*}{$4.1105$} & \multirow{2}{*}{$1.5896$} & \multirow{2}{*}{$5.9059$} & $-7.1642$ & $0.2018$ & $-1.1207$ & $3.5228$  \\ 
					& BE14 & & & & $-10.2093$ & $0.3845$ & $-0.6144$ &	$1.1668$ \\
\hline
\multirow{2}{*}{$F_\eta$} & Free Fit & \multirow{2}{*}{$-1.2220$} & \multirow{2}{*}{$1.7648$} & \multirow{2}{*}{$-7.3042$} & $18.3342$ & $-0.2398$ & $3.1301$ & $14.4631$  \\ 
					& BE14 & & & & $15.0591$ & $-0.5637$ & $1.2018$ & $8.9358$	 \\
\hline

\end{tabular}
\caption{Contribution (in units of $10^{-6}$) of NNLO component terms to  $m^2_P$ and $F_P$. The inputs are $m_{\pi} = m_{\pi^0} = 0.1350$, $m_K = m_K^{\text{avg}} = 0.4955$, $m_{\eta} = 0.5479$ and $F_{\pi} = F_{\pi\text{ phys}} = 0.0922$, all in GeV. The renormalization scale $\mu = 0.77$ GeV.}
\label{TableContrib}
\end{table}

For the kaon mass, we see that the largest contribution arises from the pure log term and the pure sunset contributions. The contribution from the terms involving both the chiral logs as well as the low energy constants is also large, but its negative sign serves to reduce the contribution rather than augment it. The contribution of the bilinear log terms is also substantial. The large uncertainty due to the $L_i$, however, means that the contribution to full loop contribution of both the $L_i \times L_j$ term and the $\log \times L_i$ term may be significantly different from what their central values suggest.
\begin{table}
\centering
\begin{tabular}{| c | | c | c c c c c c | c | }
\hline
 & \multirow{2}{*}{Input Masses} & \multirow{2}{*}{$sunset$} & \multirow{2}{*}{$log \times log$} & \multirow{2}{*}{$log$} & \multirow{2}{*}{$log \times L_i$} & \multirow{2}{*}{${L}_i$} & \multirow{2}{*}{${L}_i \times {L}_j$} & \multirow{2}{*}{Sum} \\[2ex]
\hline
\hline
\multirow{2}{*}{$m^2_K$} & Physical & $2.4100$ & $0.9420$ & $3.0586$ & $-4.3794$ & $0.2768$ & \multirow{2}{*}{$0.0665$} & $2.3745$  \\ 
					& GMO & $2.5102$ & $0.9289$ & $3.0225$ & $-4.0554$ & $0.0587$ & & $2.5313$ \\
\hline
\multirow{2}{*}{$F_K$} & Physical & $-1.2220$ & $1.7648$ & $-7.3042$ & $15.0591$ & $-0.5637$ & \multirow{2}{*}{$1.2018$} & $8.9358$  \\ 
					& GMO & $-1.2939$ & $1.7698$ & $-7.2988$ & $14.3140$ & $0.4358$ & & $9.1287$ \\
\hline
\hline
\multirow{2}{*}{$m^2_\eta$} & Physical & $4.1105$ & $1.5896$ & $5.9059$ & $-10.2093$ & $0.3845$ & $-0.6144$ & $1.1668$  \\ 
					& GMO & $4.8962$ & $1.3989$ & $5.9110$ & $-9.7738$ & $0.9473$ & $0.9980$ & $4.3775$ \\
\hline
\multirow{2}{*}{$F_\eta$} & Physical & $-1.2220$ & $1.7648$ & $-7.3042$ & $15.0591$ & $-0.5637$ & \multirow{2}{*}{$1.2018$} & $8.9358$  \\ 
					& GMO & $-1.2939$ & $1.7698$ & $-7.2988$ & $14.3140$ & $0.4358$ & & $9.1287$ \\
\hline
\end{tabular}
\caption{Contribution (in units of $10^{-6}$) of NNLO component terms to  $m^2_P$ and $F_P$ for physical and GMO-calculated eta masses. The inputs are the same as those used for Table~\ref{TableContrib}. The $L_i$ used are the BE14 fit values.}
\label{TablePhysVsGMOContrib}

\vspace*{12mm}

\centering
\begin{tabular}{|c||c|c|c|}
\hline 
 & \multirow{2}{*}{Physical} & \multirow{2}{*}{GMO} & \multirow{2}{*}{Lattice}  \\[2ex]
\hline 
\hline
$\overline{H}^{\chi}_{K \pi \eta}$ & $50.1058$ & $52.3059$ & $52.6996$ \\ 
$\overline{H}^{\chi}_{2K \pi \eta}$ & $47.1145$ & $43.9569$ & $25.3240$ \\ 
$\overline{H}^{\chi}_{K 2\pi \eta}$ & $-258.6990$ & $-264.8280$ & $25.3240$ \\ 
$\overline{H}^{\chi}_{K \pi 2\eta}$ & $63.0648$ & $65.3259$ & $38.1248$ \\ 
\hline
$c^K_{K \pi \eta}$ & $3.0345$ & $3.1439$ & $3.7614$ \\ 
$d^K_{K \pi \eta}$ & $-2.3367$ & $-2.2692$ & $6.3472$ \\ 
\hline
$c^K_{sunsets}$ & $2.4100$ & $2.5102$ & $4.1692$ \\ 
$d^K_{sunsets}$ & $-1.2220$ & $-1.2939$ & $-1.1516$ \\ 
\hline
\hline
$\overline{H}^{\chi}_{\pi K K}$ & $44.7862$ & $44.7750$ & $49.4563$ \\ 
$\overline{H}^{\chi}_{2\pi K K}$ & $-236.5110$ & $-234.5361$ & $29.5042$ \\ 
$\overline{H}^{\chi}_{\pi 2K K}$ & $58.2355$ & $59.1524$ & $32.2094$ \\ 
\hline
$c^\eta_{\pi K K}$ & $4.0771$ & $4.0273$ & $4.7771$ \\ 
$d^\eta_{\pi K K}$ & $0.1336$ & $0.3386$ & $11.6803$ \\ 
\hline
$c^\eta_{sunsets}$ & $4.1105$ & $4.8962$ & $6.0683$ \\ 
$d^\eta_{sunsets}$ & $-1.1654$ & $-1.6868$ & $-1.8894$ \\ 
\hline 
\end{tabular}
\caption{Contribution (in units of $10^{-6}$) of various components to  $m^2_P$ and $F_P$.} \label{TableA}

\vspace*{12mm}

\begin{tabular}{|c||c|c|c|}
\hline 
 & \multirow{2}{*}{Physical} & \multirow{2}{*}{GMO} & \multirow{2}{*}{Lattice}  \\[2ex]
\hline
\hline
$(m_K^2)^{(6)}_{loop}$ & $0.0329$ & $0.0350$ & $0.0656$ \\ 
$(F_K^2)^{(6)}_{loop}$ & $0.1237$ & $0.1263$ & $0.3305$ \\ 
\hline
$(m_K^2)^{(6)}_{CT}$ & $-0.0437$ & $-0.0437$ & $-0.0276$ \\ 
$(F_K^2)^{(6)}_{CT}$ & $0.0238$ & $0.0238$ & $-0.0097$ \\ 
\hline
\hline
$(m_\eta^2)^{(6)}_{loop}$ & $0.0161$ & $0.0606$ & $0.0779$ \\ 
$(F_\eta^2)^{(6)}_{loop}$ & $0.1888$ & $0.1856$ & $0.3678$ \\ 
\hline
$(m_\eta^2)^{(6)}_{CT}$ & $-0.0115$ & $-0.0115$ & $0.0035$ \\ 
$(F_\eta^2)^{(6)}_{CT}$ & $0.0009$ & $0.0009$ & $-0.0302$ \\ 
\hline
\end{tabular}
\caption{Contribution of various components to  $m^2_P$ and $F_P$.} \label{TableB}

\caption*{Tables \ref{TableA} and \ref{TableB}: The inputs for the physical and GMO case are the same as for Table~\ref{TableContrib}. The inputs for the lattice column are $m_{\pi} = 0.4023$ and $m_{K} = 0.5574$, both in GeV.}
\end{table}

In the case of the kaon decay constant, the largest contribution is from the $log \times L_i$ terms, and is of an order of magnitude higher than the next highest positive contributions, coming from the bilinear LEC and bilinear log terms. The linear chiral log terms and the pure sunset terms reduce the two-loop contribution due to their negative sign. As in the case of the kaon mass, the contribution of the $L_i \times L_j$ term may be significantly different due to the large uncertainty of its central value.

Similarly, with both the eta mass and decay constant, the largest contribution in absolute terms comes from the $log \times L_i$ terms. For the eta mass, though, the negative sign of this term serves to reduce the contributions from the $log$ and $sunset$ contributions that have the next largest values. In the eta decay constant, however, the $log \times L_i$ dominates the overall value of the $\mathcal{O}(p^6)$ contribution.

In Tables \ref{TablePhysVsGMOContrib}, \ref{TableA} and \ref{TableB}, we justify the use of the GMO relation to obtain simplified expressions for the  masses and decay constants. In all three tables, we see that the difference between quantities calculated using GMO masses varies from those using the physical masses by a maximum of around 4\% in most calculated quantities, exceptions being $\overline{H}_{2K \pi \eta}$, $c^\eta_{L_i}$, $c^\eta_{L_i \times L_j}$ (and consequently in $(m_\eta^2)_{loop}^{(6)}$), and $d^\eta_{\pi K K}$ (thus also in $(d^\eta)_{sunsets}$). However, at the level of the total NNLO contribution, the difference is negligible for the kaon mass and small for the kaon and eta decay constants. 

The column labelled `lattice' in Tables \ref{TableA} and \ref{TableB} gives values for the sunset integrals and various components making up the NNLO contributions to $m^2_K$ and $\overline{F}_K$ using as input a particular set of meson mass values used in the lattice simulations of \cite{Durr:2016ulb}. The large divergence between the numbers obtained using the physical and GMO mass inputs on one hand, and the lattice mass inputs on the other hand, demonstrate the necessity to use lattice results carefully when comparing with the expressions presented in this paper.

\subsection{Simplified expressions for three mass scale sunset results \label{Sec:NumApproxSunsets}}

We show here that the approximate expressions for the sunset integrals presented in Section~\ref{Sec:ApproxSunsets}, made by truncating the infinite series at suitable points, is sufficiently precise for purposes of data fitting against the results of the lattice simulations presented in \cite{Durr:2016ulb, Durr:2010hr}.

In Tables \ref{TableApprox1} and \ref{TableApprox2} are shown the results for three sets of mass inputs, all taken from \cite{Durr:2016ulb}. The `Lattice Low' columns have as inputs: $m_{\pi} = 0.1830$ GeV, $m_{K} = 0.4964$ GeV, which are values representative of the lower end of the range of values of masses used in \cite{Durr:2016ulb}. The `Lattice Mid' columns have as inputs: $m_{\pi} = 0.3010$ GeV, $m_{K} = 0.5625$ GeV, and the `Lattice High' columns have as inputs: $m_{\pi} = 0.4023$ GeV, $m_{K} = 0.5574$ GeV, which are values representative of the middle and upper end, respectively, of the range of masses used in the aforementioned lattice study. For each of these three sets of masses, the values of various quantities are calculated in two ways- using the exact values of the sunsets (as given by the results of Appendix~\ref{Sec:SunsetResults}, and using the approximate expressions for the sunsets (as given by Eq.(\ref{HkpeApprox})-(\ref{Hkp2eApprox}).

It can be seen from the results of these tables that the deviation from the exact results is less than $1\%$ in all cases apart from $(m_K^2)^{(6)}_{loop}$ calculated using the `Lattice Low' values. Indeed, the truncations were performed on the full expressions of the sunsets in such a manner, that the numerical deviation of the approximations from the exact values was less than $1\%$ for the majority of the meson masses used in \cite{Durr:2016ulb}. More specifically, for $\overline{H}^{\chi}_{K \pi \eta}$ Eq.(\ref{HkpeApprox}) differs from Eq.(\ref{Eq:Hkpe}) by less than $0.5\%$ for all 47 sets of masses used in \cite{Durr:2016ulb}. For $\overline{H}^{\chi}_{2K \pi \eta}$ Eq.(\ref{H2kpeApprox}) differs from Eq.(\ref{Eq:H2kpe}) by more than $1\%$ for seven of these sets of masses, and by less than $0.4\%$ for 38 sets. And for $\overline{H}^{\chi}_{K 2\pi \eta}$ and $\overline{H}^{\chi}_{K \pi 2\eta}$ both, the truncated results differ from the exact ones by more than $1\%$ for the same 3 sets of masses. Similarly, for the eta sunsets, $\overline{H}^{\chi}_{\pi K K}$ differs from Eq.(\ref{HpkkApprox}) by less than 1\% for all sets of masses, and $\overline{H}^{\chi}_{2\pi K K}$ and $\overline{H}^{\chi}_{\pi 2K K}$ differ from Eq.(\ref{H2pkkApprox}) and Eq.(\ref{Hp2kkApprox}) by less than 1\% for all but (the same) six sets of masses.

\begin{table}
\centering
\begin{tabular}{|c||c|c||c|c||c|c||}
\hline
	& \multicolumn{2}{|c||}{\multirow{2}{*}{Lattice Low}} & \multicolumn{2}{|c||}{\multirow{2}{*}{Lattice Mid}} & \multicolumn{2}{|c||}{\multirow{2}{*}{Lattice High}} \\[2ex]
\hline
	& \multirow{2}{*}{Approx} & \multirow{2}{*}{Exact} & \multirow{2}{*}{Approx} & \multirow{2}{*}{Exact} & \multirow{2}{*}{Approx} & \multirow{2}{*}{Exact} 	\\[2ex]
\hline 
\hline
$\overline{H}^{\chi}_{K \pi \eta}$ & $49.1972$ & $49.2763$ & $57.3564$ & $57.4264$ & $52.6594$ & $52.6996$ \\ 
$\overline{H}^{\chi}_{2K \pi \eta}$ & $40.3584$ & $40.3898$ & $33.4005$ & $33.5287$ & $25.3936$ & $25.3240$ \\
$\overline{H}^{\chi}_{K 2\pi \eta}$ & $-181.192$ & $-180.8920$ & $-94.4140$ & $-94.5730$ & $-37.4788$ & $-37.7974$ \\
$\overline{H}^{\chi}_{K \pi 2\eta}$ & $60.6167$ & $60.8187$ & $51.0392$ & $51.2868$ & $37.9694$ & $38.1248$ \\
\hline
$c^K_{K \pi \eta}$ & $2.9267$ & $2.9300$ & $5.1472$ & $5.1522$ & $3.7574$ & 3.7614 \\ 
$d^K_{K \pi \eta}$ & $-1.2676$ & $-1.2730$ & $1.6939$ & $1.6774$ & $6.3404$ & $6.3472$ \\
\hline
$c^K_{sunsets}$ & $2.4126$ & $2.4158$ & $4.6864$ & $4.6914$ & $4.1651$ & 4.1692 \\
$d^K_{sunsets}$ & $-1.2508$ & $-1.2562$ & $-1.6999$ & $-1.7164$ & $-1.1584$ & $-1.1516$ \\
\hline
\hline
$\overline{H}^{\chi}_{\pi K K}$ & $42.5595$ & $42.6486$ & $51.1414$ & $51.3158$ & $49.1902$ & $49.4563$ \\ 
$\overline{H}^{\chi}_{2\pi K K}$ & $-157.3080$  & $-157.1500$ & $-79.1677$ & $-79.2012$ & $-29.5237$ & $-29.5042$ \\
$\overline{H}^{\chi}_{\pi 2K K}$ & $54.2419$ & $54.1775$ & $44.4467$ & $44.3589$ & $32.3957$ & $32.2094$ \\
\hline
$c^\eta_{\pi K K}$ & $3.7206$ & $3.7247$ & $6.4170$ & $6.4305$ & $4.7598$ & $4.7771$ \\ 
$d^\eta_{\pi K K}$ & $1.0047$ & $1.0522$ & $5.3347$ & $5.4545$ & $11.4664$ & $11.6803$ \\
\hline
$c^\eta_{sunsets}$ & $4.5926$ & $4.5967$ & $8.1678$ & $8.1813$ & $6.0509$ & $6.0683$ \\
$d^\eta_{sunsets}$ & $-1.8788$ & $-1.8313$ & $-2.9205$ & $-2.8007$ & $-2.1033$ & $-1.8894$  \\
\hline 
\end{tabular}
\caption{Contribution (in units of $10^{-6}$) of various components to  $m^2_P$ and $F_P$ for three sets of meson mass inputs from lattice simulations. For `Lattice Low', $m_\pi=0.1830$ and $m_K=0.4964$; for `Lattice Mid', $m_\pi=0.3010$ and $m_K=0.5625$; for `Lattice High', $m_\pi=0.4023$ and $m_K=0.5574$; all in GeV.}
\label{TableApprox1}
\end{table} 

\begin{figure}
\centering
\begin{minipage}{0.45\textwidth}
\includegraphics[width=0.98\textwidth]{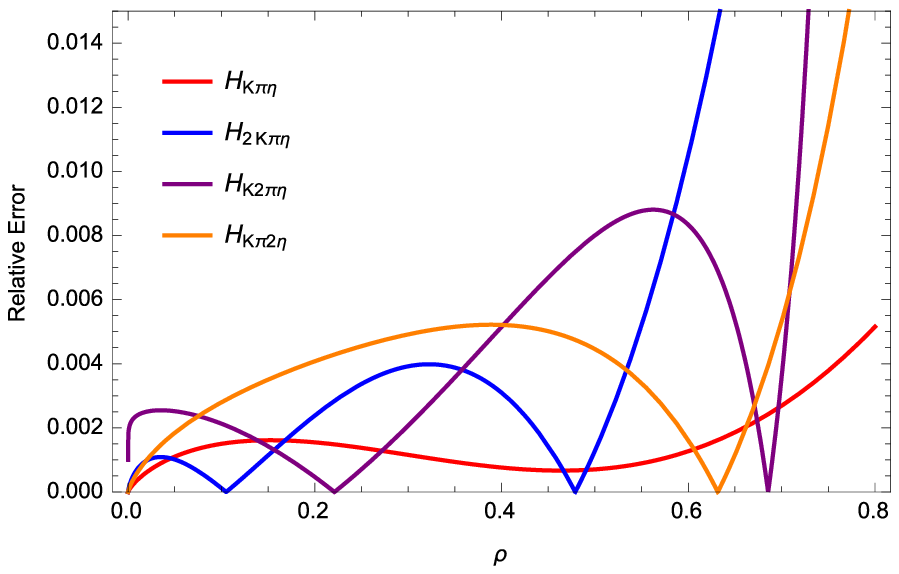} 
\end{minipage}
~~
\begin{minipage}{0.45\textwidth}
\includegraphics[width=0.98\textwidth]{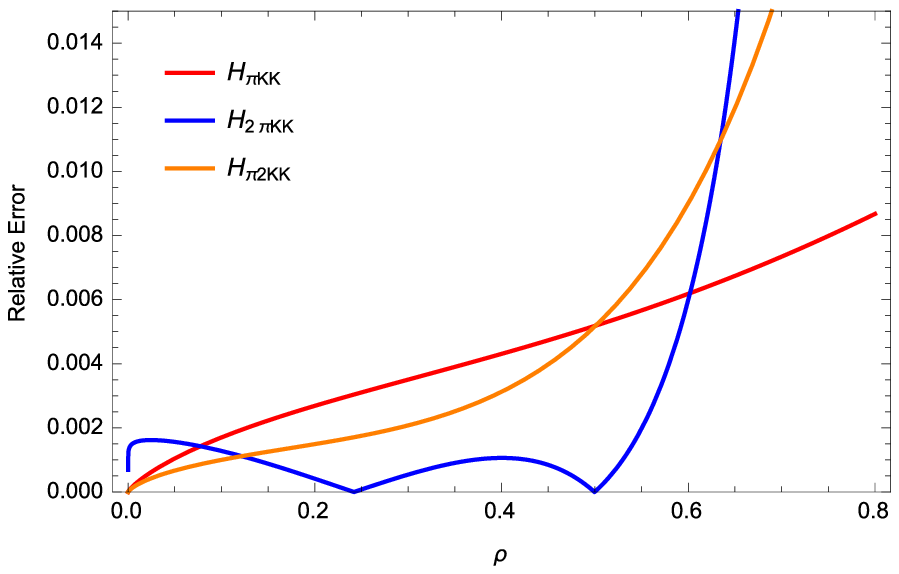}
\end{minipage}
\caption{Relative errors of the truncated sunset kaon (left) and eta (right) integrals.}
\label{FigHApproxError}
\end{figure}

Figure~\ref{FigHApproxError} gives a graphical representation of the relative errors of the truncated sunset expressions over a range of values of $\rho$. The points on the curves are the specific mass points used in \cite{Durr:2016ulb}. It is seen that for values of $\rho \lesssim 0.5$, which constitutes the majority of the mass values in the simulation of \cite{Durr:2016ulb}, the relative error is less than $1\%$.

\vspace*{15mm}

\begin{table}
\centering
\begin{tabular}{|c||c|c||c|c||c|c||}
\hline
	& \multicolumn{2}{|c||}{\multirow{2}{*}{Lattice Low}} & \multicolumn{2}{|c||}{\multirow{2}{*}{Lattice Mid}} & \multicolumn{2}{|c||}{\multirow{2}{*}{Lattice High}} \\[2ex]
\hline
	& \multirow{2}{*}{Approx} & \multirow{2}{*}{Exact} & \multirow{2}{*}{Approx} & \multirow{2}{*}{Exact} & \multirow{2}{*}{Approx} & \multirow{2}{*}{Exact} 	\\[2ex]
\hline 
\hline
$(m_K^2)^{(6)}_{loop}$ & 0.0353 & 0.0353 & 0.0710 & 0.0711 & 0.0655 & 0.0656 \\ 
$(\overline{F}_K^2)^{(6)}_{loop}$ & 0.1536 & 0.1536 & 0.2559 & 0.2557 & 0.3304 & 0.3305 \\ 
\hline
$(m_K^2)^{(6)}_{CT}$ & -0.0384 & -0.0384 & -0.0560 & -0.0560 & -0.0276 & -0.0276  \\ 
$(\overline{F}_K^2)^{(6)}_{CT}$ & 0.0183 & 0.0183 & 0.0108 & 0.0108 & -0.0097 & -0.0097 \\ 
\hline
\hline
$(m_\eta^2)^{(6)}_{loop}$ & $0.0217$ & $0.0218$ & $0.0544$ & $0.0544$ & $0.0776$ & $0.0779$ \\ 
$(\overline{F}_\eta^2)^{(6)}_{loop}$ & $0.2102$ & $0.2109$ & $0.3152$ & $0.3169$ & $0.3649$ & $0.3678$ \\ 
\hline
$(m_\eta^2)^{(6)}_{CT}$ & $-0.0076$ & $-0.0076$ & $-0.0023$ & $-0.0023$ & $0.0035$ & $0.0035$  \\ 
$(\overline{F}_\eta^2)^{(6)}_{CT}$ & $-0.0034$ & $-0.0034$ & $-0.0197$ & $-0.0197$ & $-0.0302$ & $-0.0302$ \\ 
\hline
\end{tabular}
\caption{Contribution (in units of $10^0$) of various components to  $m^2_P$ and $F_P$ for three sets of meson mass inputs from lattice simulations. For `Lattice Low', $m_\pi=0.1830$ and $m_K=0.4964$; for `Lattice Mid', $m_\pi=0.3010$ and $m_K=0.5625$; for `Lattice High', $m_\pi=0.4023$ and $m_K=0.5574$; all in GeV.}
\label{TableApprox2}
\end{table}

\subsection{Comparison with prior determinations}

In this section, we give numerical values for the quantities discussed in this paper in the form of LO + NLO + NNLO for both the BE14 and free fits. These have been calculated with the input parameters given under the tables of the previous section. We give both the values calculated using our GMO-simplified expressions, as well as the full ones.

\subsubsection{$m^2_K$}

Using the full expressions of Section~\ref{Sec:KaonMass} and the BE14 (free fit) LECs, we get the following values:
\begin{align}
	\frac{m_{K}^2}{m_{K,\text{phys}}^2} & = \frac{1}{m_{K,\text{phys}}^2} + \frac{\left(m_{K} \right)^{(4)}}{m_{K,\text{phys}}^2} + \frac{\left(m_{K}\right)^{(6)}_{\text{loop}}}{m_{K,\text{phys}}^2} + \frac{\left(m_{K}\right)^{(6)}_{\text{CT}}}{m_{K, \text{phys}}^2} \nonumber \\
	& =	1 -0.0690 (+0.0229) + 0.1338 (0.1882) -0.1779 (-0.2049)
\end{align}
and using the GMO-simplified expressions:
\begin{align}
		\frac{m_{K}^2}{m_{K,\text{phys}}^2} = 1 -0.0704 (+0.0215) + 0.1427(0.1959) -0.1779 (-0.2049)
\end{align}
These numbers are close to the literature values \cite{Ecker:2013pba}:
\begin{align}
	\left( \frac{m_{K}^2}{m_{K,\text{phys}}^2} \right)_{lit} & = 1.112(0.0994) -0.069(+0.022) -0.043(-0.016)
\end{align}
for the BE14 case, and although less so for the free fit case, are still compatible with them.

\subsubsection{$F_K$}

For the kaon decay constant, using the BE14 (free fit) low energy constants and the expressions of Section \ref{Sec:KaonDecay}, we obtain:
\begin{align}
	\frac{F_{K}}{F_{0}} &= 1 + F^{(4)}_{K} + \left(F_{K} \right)^{(6)}_{\text{loop}} + \left(F_{K} \right)^{(6)}_{\text{CT}} \nonumber\\
	& = 1 + 0.3849(0.4355) + 0.1237(0.2001) + 0.0238(0.0422)
\end{align}
Using the using the BE14 (free fit) low energy constants and GMO-simplified expressions, we get:
\begin{align}
	\frac{F_{K}}{F_{0}} = 1 + 0.3828 (0.4334) + 0.1263 (0.2012) + 0.0238 (0.0423)
\end{align}

To obtain $F_{K}/F_{\pi}$, we use the expansion presented in \cite{Bijnens:2011tb}:
\begin{align}
	\frac{F_K}{F_{\pi}} = 1 + \left( \frac{F_K}{F_0} \bigg|_{p^4} - \frac{F_{\pi}}{F_0} \bigg|_{p^4} \right)_{\text{NLO}} + \left( \frac{F_K}{F_0} \bigg|_{p^6} - \frac{F_{\pi}}{F_0} \bigg|_{p^6} - \frac{F_K}{F_0} \bigg|_{p^4} \frac{F_{\pi}}{F_0} \bigg|_{p^4} + \frac{F_{\pi}}{F_0} \bigg|^2_{p^4} \right)_{\text{NNLO}}
\end{align}
and values for the $F_{\pi}/F_0$ calculated in \cite{Ananthanarayan:2017yhz}. We get:
\begin{align}
	\frac{F_{K}}{F_{\pi}} = 1 + 0.1764 (0.1208) + 0.0226 (0.0769)
\end{align}
using the full expressions and the BE14 (free fit) LEC values. These values agree well with the numbers presented in \cite{Ecker:2013pba}:
\begin{align}
	\left( \frac{F_{K}}{F_{\pi}} \right)_{lit} = 1 + 0.176(0.121) + 0.023(0.077)
\end{align}

\subsubsection{$m^2_\eta$}

Using the full expressions of Section \ref{Sec:EtaMass} and the BE14 (free fit) LECs, we get the following values:
\begin{align}
	\frac{m_\eta^2}{m_{\eta,\text{phys}}^2} & = \frac{1}{m_{\eta,\text{phys}}^2} + \frac{\left(m_\eta \right)^{(4)}}{m_{\eta,\text{phys}}^2} + \frac{\left(m_\eta \right)^{(6)}_{\text{loop}}}{m_{\eta,\text{phys}}^2} + \frac{\left(m_\eta \right)^{(6)}_{\text{CT}}}{m_{\eta, \text{phys}}^2} \nonumber \\
	& =	1 -0.2126(-0.0736) +0.0538(0.1624) -0.0383(-0.1498)
\end{align}
and using the GMO-simplified expressions:
\begin{align}
	\frac{m_\eta^2}{m_{\eta,\text{phys}}^2} = 1 -0.2595(-0.1250) + 0.2018(0.2919) -0.0383(-0.1498)
\end{align}
As with the kaon, the BE14 numbers are close to the literature values \cite{Ecker:2013pba}, while the free fit numbers only mildly agree.
\begin{align}
	\left( \frac{m_{\eta}^2}{m_{\eta,\text{phys}}^2} \right)_{lit} & = 1.197(0.938) -0.214(-0.076) +0.017(0.014)
\end{align}

\subsubsection{$F_\eta$}

Using the full expressions of Section \ref{Sec:EtaDecay} and the BE14 (free fit) LECs, we get the following values:
\begin{align}
	\frac{F_\eta}{F_0} & = 1 + \left(F_\eta \right)^{(4)} + \left(F_\eta\right)^{(6)}_{\text{loop}} + \left(F_\eta\right)^{(6)}_{\text{CT}} \nonumber \\
	& =	1 + 0.4672(0.4996) + 0.1888(0.2597) + 0.0009(0.0254)
\end{align}
and using the GMO-simplified expressions:
\begin{align}
		m_\eta^2 = 1 +0.4672(0.4996) + 0.1797(0.2508) + 0.0009(0.0254)
\end{align}

\section{Lattice Fittings \label{Sec:LatticeFits}}

We present in this section a simplified form of the expressions for $m_K$, $F_K$, $m_\eta$ and $F_\eta$ that may conveniently be used in fits with lattice data. For this purpose, we used the simplified expressions of the sunset master integrals of Section~$\ref{Sec:ApproxSunsets}$, and expanded the $c^P_{sunset}$ and $d^P_{sunset}$ terms around the mass ratio $m_\pi^2/m_K^2 = 0$. Though the integrals $\overline{H}^\chi_{K 2\pi \eta}$ and $\overline{H}^\chi_{2\pi K K}$ diverge in the $m_\pi^2 \rightarrow 0$ limit, that they are multiplied by factors of $m_\pi^2$ ensures analyticity of the expressions in this limit.

\subsection{$m^2_K$}

The GMO expressions for the kaon mass can be written as:
\begin{align}
m_K^2 =& m_{K0}^2 + m_K^2 \left\{ \left(\frac{4}{9}\xi_K-\frac{1}{9}\xi_\pi\right) \lambda_\eta +\xi_K \hat L_{1M}^r + \xi_\pi \hat L_{2M}^r \right\} \nonumber\\
& \qquad + m_K^2\Bigg\{  \hat K_{1M}^r \lambda_\pi^2 + \hat K_{2M}^r \lambda_\pi\lambda_K
+ \hat K_{3M}^r \lambda_\pi\lambda_\eta
+ \hat K_{4M}^r \lambda_K^2
+ \hat K_{5M}^r \lambda_K\lambda_\eta
+ \hat K_{6M}^r \lambda_\eta^2 \nonumber\\
& \qquad \qquad \quad + \xi_K^2 F_M \left[\frac{m_\pi^2}{m_K^2}\right]
+ \hat C_{1M} \lambda_\pi+\hat C_{2M}\lambda_K+\hat C_{3M}\lambda_\eta
+ \hat C_{4M} \Bigg\}
\end{align}
where $\xi_\pi=m_\pi^2/(16\pi^2 F_\pi^2)$, $\xi_K= m_K^2/(16\pi^2 F_\pi^2)$ and $\lambda_i = \log(m_i^2/\mu^2)$. The coefficients $\hat L^r_{iM}$ are functions of the NLO LECs $L_i^r$. Each of the $\hat K_{iM}^r,\hat C_{iM}^r$ has three terms proportional to $\xi_\pi^2,\xi_\pi\xi_K,\xi_K^2$ respectively.
The $\hat K_{iM}$ and $F_M$ are fully determined, the $\hat C_{iM}^r, i=1,2,3$ depend linearly on the NLO LECs and $\hat C_{4M}$ depends up to quadratically on the NLO LECS and linearly on the NNLO LECs. There is some ambiguity in dividing the terms not depending on LECs between the various terms since $\log(m_i^2/m_K^2)=\lambda_i-\lambda_K$ for $i=\pi,\eta$.

The $F_I$ can be subdivided as:
\begin{align}
	F_I [ \rho ] =& a_{1I} + \bigg( a_{2I} + a_{3I} \log[\rho] + a_{4I} \log^2[\rho] \bigg) \rho  + \bigg( a_{5I} + a_{6I} \log[\rho] + a_{7I} \log^2[\rho] \bigg) \rho^2 \nonumber \\
	& + \bigg( a_{8I} + a_{9I} \log[\rho] + a_{10I} \log^2[\rho] \bigg) \rho^3  + \bigg( a_{11I} + a_{12I} \log[\rho] + a_{13I} \log^2[\rho] \bigg) \rho^4  + \mathcal{O} \left( \rho^5 \right) 
\label{Eq:FI}
\end{align}
where $\rho = m_\pi^2/m_K^2$, and for the kaon mass, $I=M$.

\begin{figure}

\centering

\begin{minipage}{0.45\textwidth}
\includegraphics[width=0.98\textwidth]{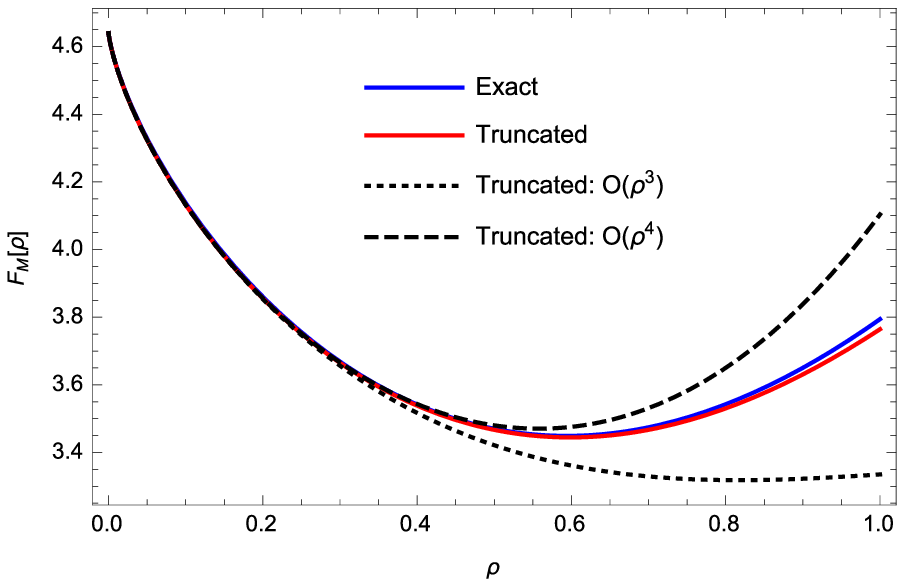} 
\end{minipage}
~~
\begin{minipage}{0.45\textwidth}
\includegraphics[width=0.98\textwidth]{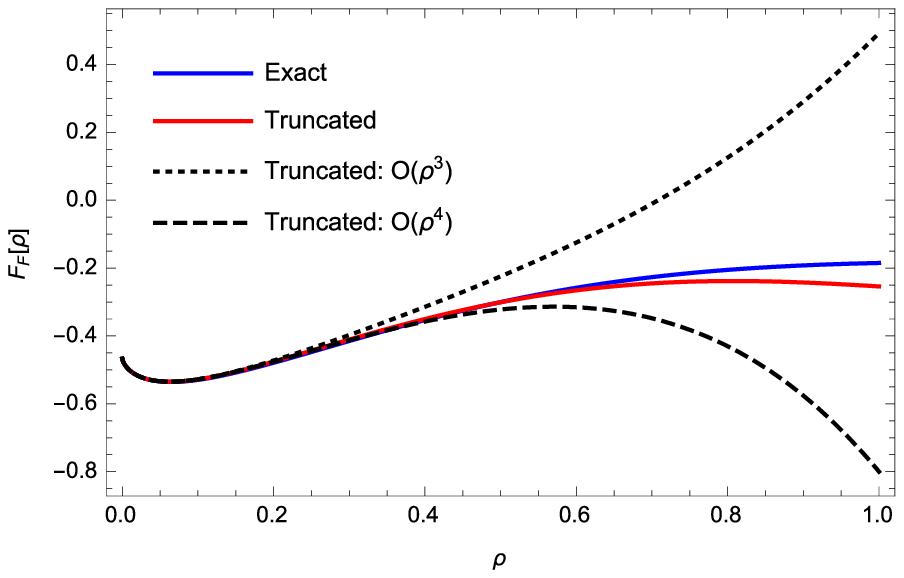}
\end{minipage}

\caption{$F_M$ (left) and $F_F$ (right) plotted against $\rho$ using exact and truncated sunset integral values, as well as expansions of the latter upto $\mathcal{O}(\rho^3)$ and $\mathcal{O}(\rho^4)$.}

\label{FigLatticeFit}
\end{figure}

For a more detailed discussion of the various possible ways in which the above expressions may be expressed for fitting with lattice data, see \cite{Ananthanarayan:2017yhz}. Note that unlike in \cite{Ananthanarayan:2017yhz} where $F_I [ \rho ]$ was truncated after $\mathcal{O} \left( \rho^3 \right)$, here we retain up to $\mathcal{O} \left( \rho^4 \right)$ terms. Our justification for doing so is that only at $\mathcal{O} \left( \rho^4 \right)$ does the expansion converge to the desired level of accuracy. This is shown graphically in Figure~\ref{FigLatticeFit}, where the expression $F_I$, which contains the terms $c_{\text{sunsets}}$ or $d_{\text{sunsets}}$ and terms from the bilinear chiral logs that are proportional to powers of $\rho$, are plotted with four different inputs. The blue plot was calculated using the exact values of the sunset integrals, the red using the approximate expressions of Section~\ref{Sec:ApproxSunsets}, and the dotted and dashed plots using the truncated sunset expressions expanded in $\rho$ up to $\mathcal{O}(\rho^3)$ and $\mathcal{O}(\rho^4)$ respectively. It is seen that only at $\mathcal{O}(\rho^4)$ do the expansions converge reasonably well to the exact ones over the entire range of interest of $\rho$, i.e. for $\rho \lesssim 0.5$.

Explicitly, for $m_K$, we have:
\begin{align}
	& \hat{L}^r_{1M} = -8 (4 \pi )^2 (2 L^r_{4}+L^r_{5}-4 L^r_{6}-2 L^r_{8}), \quad \hat{L}^r_{2M} = -8 (4 \pi )^2 (L^r_{4}-2 L^r_{6})
\end{align}
\begin{align}
	& \hat{K}^r_{1M} = \frac{1}{8} \xi _{\pi } \xi _K + \frac{169}{192} \xi _{\pi }^2, \quad \hat{K}^r_{2M} = \frac{1}{16} \xi _{\pi }^2 -\frac{3}{8} \xi _{\pi } \xi _K , \quad \hat{K}^r_{6M} = -\frac{11}{324} \xi_K^2 - \frac{47}{324} \xi _{\pi } \xi _K + \frac{1279}{5184} \xi _{\pi }^2 \nonumber \\
	& \hat{K}^r_{4M} = \frac{43}{24} \xi _K^2 + \frac{1}{9} \xi _{\pi } \xi _K, \quad \hat{K}^r_{5M} = \frac{7}{18} \xi _K^2 + \frac{19}{72} \xi _{\pi } \xi _K - \frac{1}{16} \xi _{\pi }^2, \quad \hat{K}^r_{3M} = -\frac{55}{72} \xi _{\pi } \xi _K - \frac{97}{288} \xi _{\pi }^2 
\end{align}

\begin{align}
	\hat{C}^r_{1M} =& \left(16 (4 \pi )^2 (2 L^r_{4}+L^r_{5}-4 L^r_{6}-2 L^r_{8})-\frac{11}{8}\right) \xi _{\pi } \xi _K \nonumber \\
	& - \left((4 \pi )^2 (48 L^r_{1}+12 L^r_{2}+15 L^r_{3}-68 L^r_{4}-12 L^r_{5}+88 L^r_{6}+24 L^r_{8})+\frac{455}{288}\right) \xi _{\pi }^2 
\end{align}

\begin{align}
	\hat{C}^r_{2M} =& \left(8 (4 \pi )^2 (L^r_{4}-2 L^r_{6})-\frac{41}{36}\right) \xi _{\pi } \xi _K \nonumber \\
	& - \left(2 (4 \pi )^2 (36 L^r_{1}+18 L^r_{2}+15 L^r_{3}-40 L^r_{4}-16 L^r_{5}+64 L^r_{6}+32 L^r_{8})+\frac{487}{144}\right) \xi _K^2
\end{align}

\begin{align}
	\hat{C}^r_{3M} =& \left(\frac{8}{9} (4 \pi )^2 (16 L^r_{1}+4 L^r_{2}+7 L^r_{3}-18 L^r_{4}-L^r_{5}+20 L^r_{6}-12 L^r_{7}+6 L^r_{8})+\frac{5}{8}\right) \xi _{\pi } \xi _K \nonumber \\
	& -\left(\frac{16}{9} (4 \pi )^2 (16 L^r_{1}+4 L^r_{2}+7 L^r_{3}-24 L^r_{4}-8 L^r_{5}+32 L^r_{6}+16 L^r_{8})+\frac{74}{27}\right) \xi _K^2  \nonumber \\
	& + \left(\frac{13}{864}-\frac{1}{9} (4 \pi )^2 (16 L^r_{1}+4 L^r_{2}+7 L^r_{3}-12 L^r_{4}+12 L^r_{5}+8 L^r_{6}-96 L^r_{7}-40 L^r_{8})\right) \xi _{\pi }^2 
\end{align}

\begin{align}
	\hat{C}^r_{4M} &= \frac{1}{27} (4 \pi )^2 \bigg\{ \bigg( 108 L^r_{1}+366 L^r_{2}+89 L^r_{3}-32 L^r_{5}+384 L^r_{7}+192 L^r_{8} \bigg) \xi _K^2  \nonumber \\
	& \quad -\bigg( 48 L^r_{2} + 4 L^r_{3} - 64 L^r_{5} + 768 L^r_{7} + 384 L^r_{8} \bigg) \xi _{\pi } \xi _K +  \bigg( 168 L^r_{2}+41 L^r_{3}-32 L^r_{5}+384 L^r_{7}+192 L^r_{8} \bigg) \xi _{\pi }^2 \bigg\} \nonumber \\
	& -16 \left( 16 \pi ^2\right)^2 \bigg\{ 2 \bigg (C^r_{12}+2 C^r_{13}+C^r_{14}+C^r_{15}+2 C^r_{16}-3 C^r_{19}-4 C^r_{20}-6 C^r_{21}-C^r_{31}-2 C^r_{32} + 16 (L^r_{4})^2  \nonumber \\
	& \qquad + 12 L^r_{4} L^r_{5} - 64 L^r_{4} L^r_{6} - 32 L^r_{4} L^r_{8} + 2 (L^r_{5})^2 - 24 L^r_{5} L^r_{6} - 12 L^r_{5} L^r_{8} + 64 (L^r_{6})^2 + 64 L^r_{6} L^r_{8} + 16 (L^r_{8})^2 \bigg) \xi _K^2  \nonumber \\
	& \quad +  \bigg( 2 C^r_{13}-2 C^r_{14}+C^r_{15}-4 C^r_{16}+2 C^r_{17}+6 C^r_{19}+2 C^r_{20}-12 C^r_{21}-2 C^r_{32} + 32 (L^r_{4})^2 + 16 L^r_{4} L^r_{5} \nonumber \\
	& \qquad - 128 L^r_{4} L^r_{6} - 24 L^r_{4} L^r_{8} +  4 (L^r_{5})^2 - 32 L^r_{5} L^r_{6} - 8 L^r_{5} L^r_{8} + 128 (L^r_{6})^2 + 48 L^r_{6} L^r_{8}  \bigg) \xi _{\pi } \xi _K \nonumber \\
	& \quad + \bigg( C^r_{14}+3 C^r_{16}-C^r_{17}-3 C^r_{19}-3 C^r_{20}-3 C^r_{21}+8 (L^r_{4}-2 L^r_{6}) (L^r_{4}+L^r_{5}-2 L^r_{6}-L^r_{8}) \bigg) \xi _{\pi }^2  \bigg\}
\end{align}

\begin{align}
	a_{1M} =& \frac{1165}{2592} \left(\text{Li}_2\left[ \frac{3}{4} \right] +\log [4] \log \left[ \frac{4}{3}\right] \right) +\frac{25 \pi ^2}{288}+\frac{2665}{3456}+\frac{23 \pi }{12 \sqrt{2}}-\frac{103}{192} \log ^2\left[\frac{4}{3}\right] - \frac{163}{216} \log \left[ \frac{4}{3} \right] \nonumber \\
	& - \frac{1}{24} \text{arccosec}^2\left[\sqrt{3}\right] + \left(\frac{\pi}{24}-\frac{23}{6 \sqrt{2}}\right) \text{arccosec} \left[ \sqrt{3} \right]
\end{align}
\begin{align}
	a_{2M} =& -\frac{689}{648} \left(\text{Li}_2\left[\frac{3}{4}\right]+\log [4] \log \left[\frac{4}{3}\right] \right)+\frac{11 \pi ^2}{72}-\frac{386 \gamma }{135}+\frac{71687}{16200}-\frac{221 \pi }{108 \sqrt{2}}-\frac{3277 \pi }{4320 \sqrt{3}}+\frac{5 \sqrt{2} \pi }{27} \nonumber \\
	& +\frac{53}{144} \log ^2\left[\frac{4}{3}\right] + \frac{55}{54} \log \left[ \frac{4}{3} \right] -\frac{1}{90}\log [4] - \frac{7 \pi}{288 \sqrt{3}}  \log \left[\frac{64}{3}\right] + \frac{19}{24} \text{arccosec}^2 \left[ \sqrt{3} \right] \nonumber \\
	& +\left(\frac{43 \sqrt{2}}{27}+\frac{17 \gamma }{3 \sqrt{2}}-\frac{19 \pi }{24}\right) \text{arccosec}\left[ \sqrt{3} \right] - \frac{1}{54} \psi \left[\frac{5}{2}\right]
\end{align}
\begin{align}
	a_{3M} = \frac{11}{8}+\frac{7 \pi}{288 \sqrt{3}}-\frac{1}{8} \log \left[ \frac{4}{3} \right], \quad a_{4M} = -\frac{1}{8}, \quad a_{7M} = -\frac{169}{192}, \quad a_{10M} = \frac{3}{16}, \quad a_{13M} = \frac{9}{64}
\end{align}
\begin{align}
	a_{5M} =& \frac{1031}{1296} \left(\text{Li}_2 \left[ \frac{3}{4}\right] +\log [4] \log \left[ \frac{4}{3} \right] \right) -\frac{23 \pi ^2}{48}+\frac{479393}{388800}+\frac{65 \pi }{72 \sqrt{2}}+\frac{706841 \pi }{331776 \sqrt{3}}+\frac{21737 \gamma }{6480} \nonumber \\
	& -\frac{55}{192} \log ^2\left[\frac{4}{3}\right]-\frac{151}{90} \log [4] - \frac{551}{1728}  \log \left[\frac{4}{3}\right] - \frac{62437 \pi}{55296 \sqrt{3}}  \log \left[\frac{64}{3}\right] - \frac{23}{24}  \text{arccosec}^2 \left[ \sqrt{3} \right]  \nonumber \\
	& -\frac{251}{648} \psi \left[ \frac{5}{2} \right] + \left(-\frac{173}{96 \sqrt{2}}-\frac{1009 \gamma }{144 \sqrt{2}}+\frac{23 \pi }{24}+\frac{23}{16 \sqrt{2}} \log [12] \right) \text{arccosec} \left[ \sqrt{3}\right]
\end{align}
\begin{align}
	& a_{6M} = -\frac{79}{48}+\frac{62437 \pi }{55296 \sqrt{3}}+\frac{43}{96} \log \left[ \frac{4}{3}\right] -\frac{23}{16 \sqrt{2}} \text{arccosec} \left[ \sqrt{3} \right]
\end{align}
\begin{align}
	a_{8M} &= -\frac{43}{216} \left(\text{Li}_2\left[\frac{3}{4}\right]+\log [4] \log \left[\frac{4}{3}\right] \right)+\frac{11 \pi ^2}{72}-\frac{199933 \gamma }{207360}-\frac{9347509}{6220800}-\frac{563 \pi }{2304 \sqrt{2}}-\frac{8967451 \pi }{13271040 \sqrt{3}} \nonumber \\
	& +\frac{30889}{51840} \log [4] + \frac{9653}{34560} \log \left[ \frac{4}{3}\right] + \frac{284179 \pi}{442368 \sqrt{3}} \log \left[ \frac{64}{3}\right] + \frac{5}{24} \text{arccosec}^2 \left[ \sqrt{3} \right] + \frac{47}{144} \psi\left[\frac{5}{2}\right] \nonumber \\
	& + \left(-\frac{5 \pi }{24}+\frac{1015}{1024 \sqrt{2}}+\frac{6313 \gamma }{4608 \sqrt{2}}-\frac{175}{768 \sqrt{2}}\log [12] \right) \text{arccosec} \left[ \sqrt{3} \right]
\end{align}
\begin{align}
	& a_{9M} = -\frac{5681}{17280}-\frac{284179 \pi }{442368 \sqrt{3}}+\frac{1}{24} \log \left[ \frac{4}{3}\right] + \frac{175}{768 \sqrt{2}}  \text{arccosec} \left[ \sqrt{3}\right]
\end{align}
\begin{align}
	a_{11M} &= \frac{5}{288} \left(\text{Li}_2 \left[ \frac{3}{4} \right] + \log [4] \log \left[\frac{4}{3}\right] \right)+\frac{25 \pi ^2}{288}+\frac{21213943}{33177600}+\frac{1981 \pi }{110592 \sqrt{2}}+\frac{331627 \pi }{42467328 \sqrt{3}}+\frac{166979 \gamma }{1105920} \nonumber \\
	& -\frac{61451}{1548288} \log \left[\frac{4}{3}\right] - \frac{737789}{3870720} \log [4] - \frac{708911 \pi}{7077888 \sqrt{3}}  \log \left[\frac{64}{3}\right] -\frac{7}{72} \psi\left[\frac{5}{2}\right] \nonumber \\
	& + \left(-\frac{2309 \gamma }{73728 \sqrt{2}}-\frac{8057}{442368 \sqrt{2}}+\frac{527}{24576 \sqrt{2}}  \log [12] \right) \text{arccosec} \left[ \sqrt{3}\right]
\end{align}
\begin{align}
	& a_{12M} = -\frac{499231}{1548288}+\frac{708911 \pi }{7077888 \sqrt{3}}-\frac{1}{96} \log \left[\frac{4}{3}\right]-\frac{527}{24576 \sqrt{2}}  \text{arccosec} \left[ \sqrt{3}\right]
\end{align}

\subsection{$F_K$}

We can fit $F_K$ in a similar manner as follows:
\begin{align}
\frac{F_K}{F} &= 1 + \left\{ -\frac{3}{8} \xi_{\pi} \lambda_{\pi} + \left(\frac{1}{8}\xi_{\pi}-\frac{1}{2}\xi_K \right) \lambda _{\eta } -\frac{3}{4}  \xi_K \lambda_K +\xi_K \hat L_{1F}^r + \xi_\pi \hat L_{2F}^r  \right\} \nonumber \\
& \qquad +\Bigg\{ \hat K_{1F}^r \lambda_\pi^2
+ \hat K_{2F}^r \lambda_\pi\lambda_K
+ \hat K_{3F}^r \lambda_\pi\lambda_\eta
+ \hat K_{4F}^r \lambda_K^2
+ \hat K_{5F}^r \lambda_K\lambda_\eta
+ \hat K_{6F}^r \lambda_\eta^2
\nonumber\\ 
& \qquad \quad
+ \xi_K^2 F_F\left[ \frac{m_\pi^2}{m_K^2} \right]
+ \hat C_{1F}\lambda_\pi+\hat C_{2F}\lambda_K+\hat C_{3F}\lambda_\eta
+ \hat C_{4F} \Bigg\}
\end{align}
where
\begin{align}
	\hat{L}^r_{1F} = 4 (4 \pi )^2 (2 L^r_{4}+L^r_{5}), \quad \hat{L}^r_{2F} = 4 (4 \pi )^2 L^r_{4}
\end{align}
\begin{align}
	& \hat{K}^r_{1F} = \frac{1}{6} \xi _{\pi } \xi _K - \frac{5}{192} \xi _{\pi }^2, \quad \hat{K}^r_{2F} = \frac{51}{32} \xi _{\pi } \xi _K - \frac{3}{32} \xi _{\pi }^2, \hat{K}^r_{6F} = \frac{31}{36} \xi _K^2 - \frac{11}{72} \xi _{\pi } \xi _K - \frac{21}{64} \xi _{\pi }^2 \nonumber \\
	& \hat{K}^r_{4F} = \frac{155}{288} \xi _K^2 + \frac{11}{144} \xi _{\pi } \xi _K, \quad \hat{K}^r_{5F} = -\frac{91}{72}  \xi _K^2 - \frac{53}{288} \xi _{\pi } \xi _K + \frac{3}{32} \xi _{\pi }^2, \quad \quad \hat{K}^r_{3F} = \frac{25}{24} \xi _{\pi } \xi _K + \frac{47}{96} \xi _{\pi }^2
\end{align}
\begin{align}
	\hat{C}^r_{1F} = \left(\frac{3}{16}-\frac{19}{2} (4 \pi )^2 (2 L^r_{4}+L^r_{5})\right) \xi _{\pi } \xi _K + \left(\frac{1}{2} (4 \pi )^2 (48 L^r_{1}+12 L^r_{2}+15 L^r_{3}-47 L^r_{4}-6 L^r_{5})+\frac{53}{64}\right) \xi _{\pi }^2
\end{align}
\begin{align}
	\hat{C}^r_{2F} =  \left(\frac{245}{96} + (4 \pi )^2 (36 L^r_{1}+18 L^r_{2}+15 L^r_{3}-30 L^r_{4}-7 L^r_{5}) \right) \xi _K^2 + \left(\frac{173}{144}-(4 \pi )^2 (7 L^r_{4}+6 L^r_{5})\right) \xi _{\pi } \xi _K 
\end{align}
\begin{align}
	\hat{C}^r_{3F} =& \left(\frac{19}{18} + \frac{2}{9} (4 \pi )^2 (64 L^r_{1}+16 L^r_{2}+28 L^r_{3}-66 L^r_{4}-3 L^r_{5}-72 L^r_{7}-36 L^r_{8})\right)  \xi _K^2  \nonumber \\
	& - \left(\frac{65}{144} + \frac{1}{18} (4 \pi )^2 (128 L^r_{1}+32 L^r_{2}+56 L^r_{3}-78 L^r_{4}+111 L^r_{5}-576 L^r_{7}-288 L^r_{8}) \right) \xi _{\pi } \xi _K \nonumber \\
	& + \left(\frac{3}{64} + \frac{1}{18} (4 \pi )^2 (16 L^r_{1}+4 L^r_{2}+7 L^r_{3}-3 L^r_{4}+42 L^r_{5}-288 L^r_{7}-144 L^r_{8}) \right) \xi _{\pi }^2
\end{align}
\begin{align}
	\hat{C}^r_{4F} &= 8 \left(16 \pi ^2\right)^2 \bigg\{ \bigg( -2 C^r_{14}+C^r_{15}-4 C^r_{16}+2 C^r_{17}+28 (L^r_{4})^2+14 L^r_{4} L^r_{5}-32 L^r_{4} L^r_{6}+4 (L^r_{5})^2-8 L^r_{5} L^r_{6} \bigg) \xi _{\pi } \xi _K \nonumber \\
	& \quad + \bigg( 2 C^r_{14}+2 C^r_{15}+4 C^r_{16}+(2 L^r_{4}+L^r_{5}) (14 L^r_{4}+3 L^r_{5}-16 L^r_{6}-8 L^r_{8} \bigg) \xi _K^2 \nonumber \\
	& \quad + \bigg( C^r_{14}+3 C^r_{16}-C^r_{17}+7 (L^r_{4})^2+8 L^r_{4} L^r_{5}-8 L^r_{4} L^r_{6}-8 L^r_{4} L^r_{8} \bigg) \xi _{\pi }^2  \bigg\} \nonumber \\
	& +\frac{2}{27} (4 \pi )^2 \bigg\{ \bigg( 12 L^r_{2}+L^r_{3}-36 L^r_{5}+432 L^r_{7}+216 L^r_{8} \bigg) \xi _{\pi } \xi _K - \left(42 L^r_{2}+\frac{41}{4} L^r_{3}-18 L^r_{5}+216 L^r_{7}+108 L^r_{8}\right) \xi _{\pi }^2 \nonumber \\
	& \quad -\left(27 L^r_{1}+\frac{183 L^r_{2}}{2}+\frac{89 L^r_{3}}{4}-18 L^r_{5}+216 L^r_{7}+108 L^r_{8}\right) \xi _K^2 \bigg\}
\end{align}

We subdivide $F_F$ as in Eq.(\ref{Eq:FI}) with $I=F$, and with the $a_{iF}$ given by:
\begin{align}
	a_{1F} =& -\frac{6337}{5184} \left(\text{Li}_2\left[\frac{3}{4}\right] + \log [4] \log \left[ \frac{4}{3} \right] \right) + \frac{41 \pi^2}{192} - \frac{11 \sqrt{2} \pi}{27} + \frac{10525}{6912} - \frac{119 \pi }{216 \sqrt{2}}-\frac{23}{1152} \log ^2\left[\frac{4}{3}\right] \nonumber \\
	& + \frac{127}{48} \log \left[\frac{4}{3}\right] + \frac{41}{48} \text{arccosec}^2 \left[ \sqrt{3}\right] + \left(\frac{295}{108 \sqrt{2}}-\frac{41 \pi }{48}\right) \text{arccosec} \left[\sqrt{3}\right]
\end{align}
\begin{align}
	a_{2F} =& \frac{5821}{2592}  \left(\text{Li}_2\left[\frac{3}{4}\right] + \log [4] \log \left[ \frac{4}{3} \right] \right) -\frac{25 \pi ^2}{96}-\frac{2050019}{388800}+\frac{145 \pi }{72 \sqrt{2}}+\frac{38693 \pi }{25920 \sqrt{3}}+\frac{82 \gamma }{405}-\frac{137}{576} \log ^2 \left[ \frac{4}{3} \right] \nonumber \\
	& - \frac{1687}{810} \log \left[ \frac{4}{3} \right] - \frac{281}{540} \log [4] - \frac{13 \pi}{1728 \sqrt{3}} \log \left[\frac{64}{3}\right] +\frac{11}{48} \text{arccosec} \left[ \sqrt{3} \right]^2 - \frac{29}{324} \psi \left[\frac{5}{2}\right] \nonumber \\
	& + \left(-\frac{11 \pi }{48}-\frac{13}{3 \sqrt{2}}-\frac{13 \gamma }{18 \sqrt{2}}+\frac{1}{6 \sqrt{2}} \log [1728] \right) \text{arccosec} \left[ \sqrt{3} \right]
\end{align}
\begin{align}
	a_{3F} = \frac{169}{6480} + \frac{13 \pi }{1728 \sqrt{3}}+\frac{7}{48} \log \left[ \frac{4}{3} \right] -\frac{1}{2 \sqrt{2}} \text{arccosec} \left[ \sqrt{3} \right]
\end{align}
\begin{align}
	a_{4F} = -\frac{1}{6}, \quad a_{7F} = \frac{325}{384}, \quad a_{10F} =  -\frac{9}{64}, \quad a_{13F} = - \frac{27}{128}
\end{align}
\begin{align}
	a_{5F} =& -\frac{845}{648} \left(\text{Li}_2\left[\frac{3}{4}\right]+\log [4] \log \left[ \frac{4}{3}\right] \right) +\frac{5 \pi ^2}{18} - \frac{1301 \sqrt{3} \pi }{512}-\frac{66191 \gamma }{12960}+\frac{25789}{155520}-\frac{145 \pi }{144 \sqrt{2}}+\frac{3572063 \pi }{663552 \sqrt{3}} \nonumber \\
	& + \frac{145}{384} \log^2 \left[ \frac{4}{3} \right] + \frac{15403}{6480} \log [4] + \frac{15941}{17280} \log \left[\frac{4}{3}\right] + \frac{176189 \pi}{110592 \sqrt{3}} \log \left[\frac{64}{3}\right] + \frac{59}{48} \text{arccosec}^2 \left[ \sqrt{3} \right]  \nonumber \\
	& + \frac{35}{144} \psi \left[ \frac{5}{2} \right] + \left(-\frac{59 \pi }{48}+\frac{323}{192 \sqrt{2}}+\frac{3167 \gamma }{288 \sqrt{2}}-\frac{115}{48 \sqrt{2}} \log [12] \right) \text{arccosec} \left[ \sqrt{3} \right]
\end{align}
\begin{align}
	a_{6F} = \frac{4427}{2160}-\frac{176189 \pi }{110592 \sqrt{3}}-\frac{155}{192} \log \left[ \frac{4}{3} \right] + \frac{115}{48 \sqrt{2}} \text{arccosec} \left[ \sqrt{3} \right] 
\end{align}
\begin{align}
	a_{8F} =& \frac{265}{864} \left( \text{Li}_2 \left[\frac{3}{4}\right] + \log [4] \log \left[ \frac{4}{3} \right] \right) - \frac{29 \pi ^2}{288} + \frac{11061169}{4147200} + \frac{4753 \pi }{13824 \sqrt{2}}+\frac{20910563 \pi }{26542080 \sqrt{3}}+\frac{199393 \gamma }{138240} \nonumber \\
	& - \frac{16337}{23040} \log [4] - \frac{10477}{27648} \log \left[\frac{4}{3}\right] -\frac{804611 \pi}{884736 \sqrt{3}} \log \left[ \frac{64}{3} \right] - \frac{5}{16} \text{arccosec}^2 \left[ \sqrt{3}\right] -\frac{119}{288} \psi \left[\frac{5}{2}\right] \nonumber \\
	& + \left(-\frac{19319 \gamma }{9216 \sqrt{2}}-\frac{84251}{55296 \sqrt{2}}+\frac{5 \pi }{16}+\frac{823}{3072 \sqrt{2}} \log [12] \right) \text{arccosec} \left[ \sqrt{3}\right]
\end{align}
\begin{align}
	a_{9F} = -\frac{2971}{27648}+\frac{804611 \pi }{884736 \sqrt{3}}-\frac{1}{96} \log \left[\frac{4}{3}\right] - \frac{823}{3072 \sqrt{2}} \text{arccosec} \left[ \sqrt{3} \right]
\end{align}
\begin{align}
	a_{11F} =& -\frac{5}{192} \left(\text{Li}_2 \left[ \frac{3}{4} \right] + \log [4] \log \left[ \frac{4}{3} \right] \right) -\frac{25 \pi ^2}{192}-\frac{4582831}{4423680}-\frac{1310311 \gamma }{6635520}-\frac{2135 \pi }{73728 \sqrt{2}}-\frac{13905571 \pi }{84934656 \sqrt{3}} \nonumber \\
	& +\frac{4453 \sqrt{3} \pi }{65536}+\frac{532067}{1935360} \log [4] + \frac{312911}{2903040} \log \left[ \frac{4}{3} \right] + \frac{1674775 \pi}{14155776 \sqrt{3}}  \log \left[ \frac{64}{3}\right] + \frac{97}{648} \psi \left[ \frac{5}{2} \right] \nonumber \\
	& + \left(-\frac{391 \gamma }{49152 \sqrt{2}}+\frac{9421}{294912 \sqrt{2}}-\frac{59}{4096 \sqrt{2}} \log [12] \right) \text{arccosec} \left[\sqrt{3}\right]
\end{align}
\begin{align}
	a_{12F} = \frac{5174549}{11612160}-\frac{1674775 \pi }{14155776 \sqrt{3}}+\frac{1}{64} \log \left[ \frac{4}{3} \right] + \frac{59}{4096 \sqrt{2}} \text{arccosec} \left[ \sqrt{3}\right]
\end{align}

The divergence of $F_F$ as given above from its exact value is shown in Figure~\ref{FigLatticeFit}.

\subsection{$m^2_\eta$}

The GMO expressions for the eta mass can similarly be expressed as:
\begin{align}
m_\eta^2 =& m_{\eta 0}^2 + \bigg\{ \frac{64 \pi^2}{3} \xi _K^2 \lambda _K - 8 \pi ^2 \xi _{\pi }^2 \lambda_\pi + \left( \frac{352 \pi^2}{27}  \xi_\pi \xi_K - \frac{512\pi^2}{27} \xi _K^2 - \frac{56\pi^2}{27} \xi_\pi^2 \right) \lambda_\eta \nonumber \\
	& \qquad \qquad - \frac{64}{9} \xi _K^2 \hat L_{1m}^r - \frac{16}{9} \xi_\pi \xi_K  \hat L_{2m}^r + \frac{8}{9} \xi_\pi^2 \hat L_{3m}^r \bigg\} \nonumber\\ &
 \qquad +\bigg\{  \hat K_{1m}^r \lambda_\pi^2 + \hat K_{2m}^r \lambda_\pi\lambda_K
+ \hat K_{3m}^r \lambda_\pi\lambda_\eta
+ \hat K_{4m}^r \lambda_K^2
+ \hat K_{5m}^r \lambda_K\lambda_\eta
+ \hat K_{6m}^r \lambda_\eta^2 \nonumber\\ & \hspace*{7ex}
+ m_K^2 \xi_K^2 F_m \left[\frac{m_\pi^2}{m_K^2}\right]
+ \hat C_{1m} \lambda_\pi+\hat C_{2m}\lambda_K+\hat C_{3m}\lambda_\eta
+ \hat C_{4m} \bigg\}
\end{align}

Note that in contrast with the kaon, there is an extra $m_K^2$ prefactor to $F_m$ aside from the $\xi_K^2$. Furthermore, each of the $\hat K_{im}^r,\hat C_{im}^r$ have six terms proportional to $\xi_\pi^2,\xi_\pi\xi_K,\xi_K^2$ and $m_\pi^2$ multiplied by either $m_\pi^2$ or $m_K^2$.

\begin{figure}

\centering

\begin{minipage}{0.45\textwidth}
\includegraphics[width=0.98\textwidth]{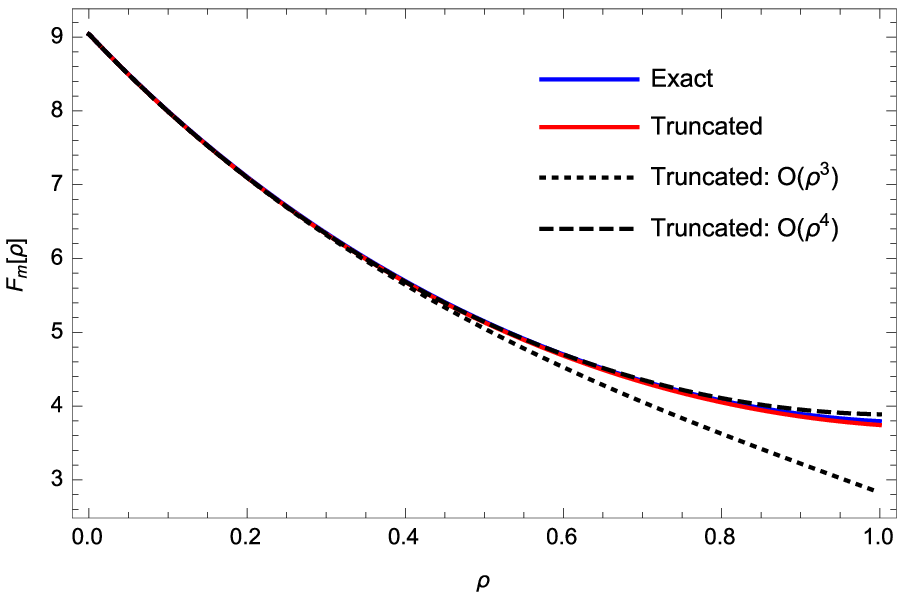} 
\end{minipage}
~~
\begin{minipage}{0.45\textwidth}
\includegraphics[width=0.98\textwidth]{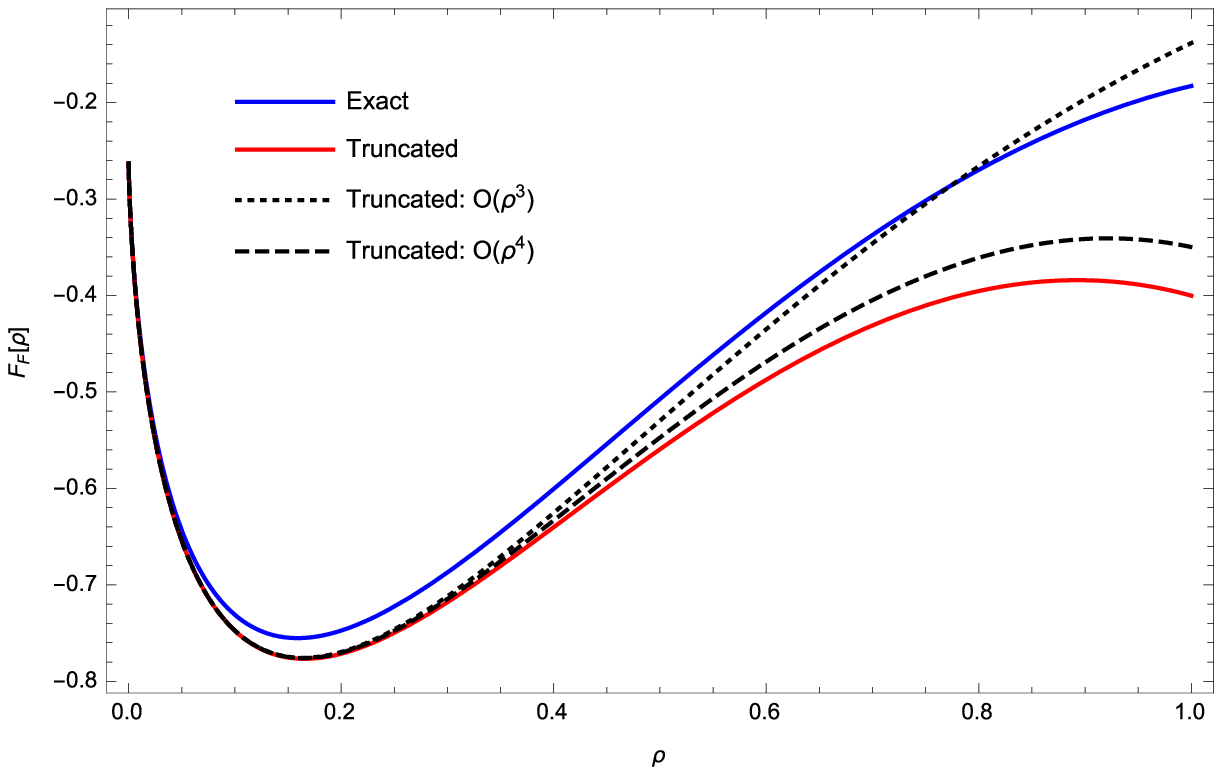}
\end{minipage}

\caption{$F_m$ (left) and $F_f$ (right) plotted against $\rho$ using exact and truncated sunset integral values, as well as expansions of the latter upto $\mathcal{O}(\rho^3)$ and $\mathcal{O}(\rho^4)$.}

\label{FigLatticeFitEta}
\end{figure}

Explicitly, for $m_\eta$, we have:

\begin{align}
	& \hat{L}^r_{1M} = (4 \pi )^4 ( 3 L^r_{4}+2 L^r_{5}-6 L^r_{6}-6 L^r_{7}-6 L^r_{8} ) \nonumber \\
	& \hat{L}^r_{2M} = (4 \pi )^4 ( 3 L^r_{4}-4 L^r_{5}-6 L^r_{6}+48 L^r_{7}+24 L^r_{8} ) \nonumber \\
	& \hat{L}^r_{3M} = (4 \pi )^4 ( 3 L^r_{4}-L^r_{5}-6 L^r_{6}+48 L^r_{7}+18 L^r_{8} )
\end{align}
\begin{align}
	& \hat{K}^r_{1M} = \left(\frac{5}{8} \xi _{\pi } \xi _K + \frac{65}{48} \xi _{\pi }^2 \right) m_\pi^2 - \left( \frac{3}{16} \xi_\pi \xi_K \right) m_K^2  \nonumber \\
	& \hat{K}^r_{2M} = \left( \frac{3}{4}  \xi_\pi \xi_K \right) m_\pi^2 + \left( \frac{55}{24} \xi_\pi \xi_K \right) m_K^2 \nonumber \\
	& \hat{K}^r_{3M} = \left(\frac{7}{36} \xi_\pi^2 - \frac{43}{27} \xi_\pi \xi_K \right) m_\pi^2 +  \left( \frac{64}{27} \xi_\pi \xi_K \right) m_K^2 \nonumber \\
	& \hat{K}^r_{4M} = \left( \frac{5}{6} \xi_\pi \xi_K \right) m_\pi^2 + \left( \frac{103}{36} \xi_K^2 -\frac{133}{72} \xi_\pi \xi_K \right) m_K^2 \nonumber \\
	& \hat{K}^r_{5M} = - \left( \frac{31}{108} \xi_\pi \xi_K \right) m_\pi^2 + \left(\frac{473}{216} \xi_\pi \xi_K - \frac{59}{18} \xi_K^2 \right) m_K^2 \nonumber \\
	& \hat{K}^r_{6M} = \left(\frac{1367}{648} \xi_\pi \xi_K - \frac{911}{3888} \xi_\pi^2 \right) m_\pi^2 +  \left(\frac{6185}{972} \xi_K^2 - \frac{2713}{432} \xi_\pi \xi_K \right) m_K^2
\end{align}

\begin{align}
	\hat{C}^r_{1M} =& \left(\frac{61}{54} \xi_\pi \xi_K - \frac{931}{864} \xi_\pi^2 \right) m_\pi^2 - \frac{3}{2} \xi_\pi \xi_K m_K^2 \nonumber \\
	&  + (16 \pi^2 ) \bigg\{ \left(\frac{128}{3} L^r_{4} + \frac{256}{9} L^r_{5} - \frac{256}{3} L^r_{6} - \frac{256}{3} L^r_{7} - \frac{256}{3} L^r_{8} \right) \xi_\pi \xi_K m_K^2 \nonumber \\
	& \qquad \qquad - \left( 64 L^r_{1}+16 L^r_{2}+16 L^r_{3}-\frac{232}{3} L^r_{4} + \frac{64}{9} L^r_{5} + \frac{272}{3} L^r_{6} - \frac{832}{3} L^r_{7} - \frac{320}{3} L^r_{8} \right) \xi_\pi \xi_K m_\pi^2  \nonumber \\
	& \qquad \qquad + \bigg( 16 L^r_{1}+4 L^r_{2}+4 L^r_{3}-24 L^r_{4}+32 L^r_{6}-192 L^r_{7}-72 L^r_{8} \bigg) \xi_\pi^2 m_\pi^2 \bigg\}
\end{align}
\begin{align}
	\hat{C}^r_{2M} =& -\frac{3}{2} \xi_\pi \xi_K m_\pi^2 + \left(\frac{961}{216} \xi_\pi \xi_K -\frac{577}{54} \xi_K^2 \right) m_K^2 \nonumber \\
	& + (16\pi^2) \bigg\{ \left(\frac{64}{3} L^r_{1} + \frac{16}{3} L^r_{2} + \frac{28}{3} L^r_{3} - 16 L^r_{4} - \frac{16}{3} L^r_{5} + \frac{32}{3} L^r_{6} + 128 L^r_{7} + \frac{224}{3} L^r_{8} \right) \xi_\pi \xi_K m_K^2 \nonumber \\
	& \qquad \qquad -\left(\frac{256}{3} L^r_{1} + \frac{64}{3} L^r_{2} + \frac{112}{3} L^r_{3} - \frac{320}{3} L^r_{4} - \frac{352}{9} L^r_{5} + 128 L^r_{6} + \frac{256}{3} L^r_{7} + \frac{320}{3} L^r_{8} \right) \xi_K^2 m_K^2  \nonumber \\
	& \qquad \qquad + \left(-\frac{8}{3} L^r_{4} + \frac{8}{9} L^r_{5} + \frac{16}{3} L^r_{6} - \frac{128}{3} L^r_{7} - 16 L^r_{8} \right) \xi_\pi \xi_K  m_\pi^2  \bigg\}
\end{align}
\begin{align}
	\hat{C}^r_{3M} =& \left(\frac{371}{972} \xi_\pi \xi_K -\frac{2045}{23328} \xi_\pi^2 \right) m_\pi^2 + \left(\frac{41}{648} \xi_\pi \xi_K - \frac{1093}{1458} \xi_K^2 \right) m_K^2 + \nonumber \\
	& + \left(16 \pi ^2\right) \bigg\{ \left(\frac{128}{3} L^r_{1} + \frac{128}{3} L^r_{2} + \frac{64}{3} L^r_{3} - 32 L^r_{4} - \frac{1808}{27} L^r_{5} + \frac{832}{9} L^r_{6} +\frac{1216}{3} L^r_{7} + \frac{6752}{27} L^r_{8} \right) \xi_\pi \xi_K m_K^2 \nonumber \\
	& \qquad \qquad -\left(\frac{512}{9} L^r_{1} + \frac{512}{9} L^r_{2} + \frac{256}{9} L^r_{3} - \frac{512}{9} L^r_{4} - \frac{512}{9} L^r_{5} + \frac{4096}{27} L^r_{6} + \frac{2048}{9} L^r_{7} + \frac{5120}{27} L^r_{8} \right) m_K^2 \xi _K^2 \nonumber \\
	& \qquad \qquad - \left(\frac{32}{3} L^r_{1} + \frac{32}{3} L^r_{2} + \frac{16}{3} L^r_{3} - 8 L^r_{4} - \frac{832}{27} L^r_{5} + \frac{208}{9} L^r_{6} + \frac{640}{3} L^r_{7} + \frac{3232}{27} L^r_{8} \right) \xi_\pi \xi_K m_\pi^2 \nonumber \\
	& \qquad \qquad + \left(\frac{8}{9} L^r_{1} + \frac{8}{9} L^r_{2} + \frac{4}{9} L^r_{3} - \frac{8}{9} L^r_{4} - \frac{128}{27} L^r_{5} + \frac{64}{27} L^r_{6} + \frac{320}{9} L^r_{7} + \frac{520}{27} L^r_{8} \right) \xi_\pi^2 m_\pi^2 \bigg\}
\end{align}
\begin{align}
	\hat{C}^r_{4M} &= \frac{2}{81} (16 \pi^2) \bigg\{ \bigg(384 L^r_{1}+816 L^r_{2}+228 L^r_{3}+128 L^r_{5}-1536 L^r_{7}-768 L^r_{8} \bigg) \xi _K^2 m_K^2 \nonumber \\
	& \qquad \qquad \qquad - \bigg( 288 L^r_{1}+396 L^r_{2}+153 L^r_{3}+312 L^r_{5}-3744 L^r_{7}-1872 L^r_{8} \bigg) \xi_\pi \xi_K m_K^2 \nonumber \\
	& \qquad \qquad \qquad + \bigg( 72 L^r_{1}+396 L^r_{2}+144 L^r_{3}+240 L^r_{5}-2880 L^r_{7}-1440 L^r_{8} \bigg) \xi_\pi \xi_K m_\pi^2  \nonumber \\
	& \qquad \qquad \qquad + \bigg( -6 L^r_{1}-87 L^r_{2}-30 L^r_{3}-56 L^r_{5}+672 L^r_{7}+336 L^r_{8} \bigg) \xi_\pi^2 m_\pi^2 \bigg\} \nonumber \\
	& + (16 \pi^2)^2 \bigg\{ \frac{128}{27} (3 L^r_{4}+5 L^r_{5}-6 L^r_{6}-6 L^r_{8}) \left( 3 L^r_{4}-L^r_{5}-6 L^r_{6}+48 L^r_{7}+18 L^r_{8} \right) \xi_\pi^2 m_\pi^2  \nonumber \\
	& \qquad \qquad - \frac{256}{27} \bigg( 8 C^r_{12}+12 C^r_{13}+6 C^r_{14}+6 C^r_{15}+9 C^r_{16}+6 C^r_{17}+6 C^r_{18}-27 C^r_{19}-27 C^r_{20}-27 C^r_{21} \nonumber \\
	& \qquad \qquad \qquad -18 C^r_{31}-18 C^r_{32}-18 C^r_{33} \bigg) \xi_K^2 m_K^2 \nonumber \\
	& \qquad \qquad - \frac{1024}{27} \bigg( 6 L^r_{4}+L^r_{5}-12 L^r_{6}-6 L^r_{8}) (3 L^r_{4}+2 L^r_{5}-6 L^r_{6}-6 L^r_{7}-6 L^r_{8} \bigg) \xi_K^2 m_K^2 \nonumber \\
	& \qquad \qquad + \frac{16}{27} \bigg( 2 C^r_{12}-6 C^r_{13}+9 C^r_{14}-3 C^r_{15}+27 C^r_{16}+9 C^r_{17}+24 C^r_{18}-27 C^r_{19}+27 C^r_{20}-27 C^r_{21} \nonumber \\
	& \qquad \qquad \qquad -18 C^r_{31}+54 C^r_{32} \bigg) \xi_\pi^2 m_\pi^2 \nonumber \\
	& \qquad \qquad -\frac{32}{9} \bigg( 4 C^r_{12}-6 C^r_{13}+10 C^r_{14}-3 C^r_{15}+24 C^r_{16}+10 C^r_{17}+24 C^r_{18}-54 C^r_{19}-18 C^r_{20}-36 C^r_{31} \nonumber \\
	& \qquad \qquad \qquad + 6 C^r_{32}-48 C^r_{33} \bigg) \xi _\pi \xi_K m_\pi^2  \nonumber \\
	& \qquad \qquad + \frac{64}{9} \bigg( 8 C^r_{12}+10 C^r_{14}+15 C^r_{16}+10 C^r_{17}+18 C^r_{18}-54 C^r_{19}-27 C^r_{20}+27 C^r_{21}-36 C^r_{31} \nonumber \\
	& \qquad \qquad \qquad -12 C^r_{32}-48 C^r_{33} \bigg) \xi_\pi \xi_K m_K^2 \nonumber \\
	&  \qquad \qquad - \frac{128}{9} \bigg( 36 (L^r_{4})^2+15 L^r_{4} L^r_{5}-144 L^r_{4} L^r_{6}+144 L^r_{4} L^r_{7}+42 L^r_{4} L^r_{8}+12 (L^r_{5})^2-30 L^r_{5} L^r_{6} -48 L^r_{5} L^r_{7} \nonumber \\
	& \qquad \qquad \qquad -32 L^r_{5} L^r_{8}+144 (L^r_{6})^2-288 L^r_{6} L^r_{7}-84 L^r_{6} L^r_{8}-96 L^r_{7} L^r_{8}-48 (L^r_{8})^2 \bigg) \xi_\pi \xi_K m_K^2 \nonumber \\
	& \qquad \qquad -\frac{128}{9} \bigg(3 L^r_{4} L^r_{5}+6 L^r_{4} L^r_{8}-10 (L^r_{5})^2-6 L^r_{5} L^r_{6}+144 L^r_{5} L^r_{7}+76 L^r_{5} L^r_{8}-12 L^r_{6} L^r_{8}-96 L^r_{7} L^r_{8} \nonumber \\
	& \qquad \qquad \qquad -48 (L^r_{8})^2 \bigg) \xi_\pi \xi_K m_\pi^2 \bigg\}
\end{align}

The $F_m$ can be subdivided as:
\begin{align}
	F_m^\eta [ \rho ] =& a_{1m} + \bigg( a_{2m} + a_{3m} \log[\rho] + a_{4m} \log^2[\rho] \bigg) \rho  + \bigg( a_{5m} + a_{6m} \log[\rho] + a_{7m} \log^2[\rho] \bigg) \rho^2 \nonumber \\
	& \quad + \bigg( a_{8m} + a_{9m} \log[\rho] + a_{10m} \log^2[\rho] \bigg) \rho^3 + \bigg( a_{11m} + a_{12m} \log[\rho] + a_{13m} \log^2[\rho] \bigg) \rho^4 + \mathcal{O} \left( \rho^5 \right) 
\end{align}

Note that we omit the factor of $1/(16\pi^2)^2$ in this definition in contrast to Eq.(\ref{Eq:FI}). In Figure~\ref{FigLatticeFitEta}, we see that the $\mathcal{O}(\rho^4)$ expansion version of $F_m$ agrees with the exact valued $F_m$ well within our desired range of $\rho$.

\begin{align}
	a_{1m} &= \frac{1165}{864} \left( \frac{\pi^2}{3} + \log ^2 \left[ 2 \sqrt{3}-3\right] - \log \left[ \frac{4}{3} \right] \log \left[ 3+2\sqrt{3} \right] - 2 \text{Li}_2 \left[ \frac{\sqrt{3}}{2}\right] + 2 \text{Li}_2 \left[ 2\sqrt{3}-3 \right] \right) \nonumber \\
	& + \frac{875}{486}-\frac{1157}{384} \log^2\left[ \frac{4}{3} \right] - \frac{19}{24} \log \left[ \frac{4}{3} \right] + \frac{1}{8} \csc ^{-1}\left[ \sqrt{3} \right]^2 + \frac{23}{2\sqrt{2}} \csc ^{-1}\left[ \sqrt{3} \right] 
\end{align}
\begin{align}
	a_{2m} &= -\frac{9859}{3456}  \left( \frac{\pi^2}{3} + \log ^2 \left[ 2 \sqrt{3}-3 \right] - \log \left[ \frac{4}{3}\right] \log \left[ 3+2\sqrt{3} \right] -2 \text{Li}_2 \left[ \frac{\sqrt{3}}{2} \right] +2 \text{Li}_2 \left[ -3+2 \sqrt{3} \right] \right) \nonumber \\
	& + \frac{18889}{22680} + \frac{16865}{4608} \log^2 \left[\frac{4}{3}\right] + \frac{683}{288} \log \left[ \frac{4}{3} \right] - \frac{75}{32} \csc^{-1} \left[ \sqrt{3} \right]^2 - \frac{517}{72 \sqrt{2}} \csc ^{-1} \left[ \sqrt{3} \right]
\end{align}
\begin{align}
	a_{3m} = \frac{41}{27}, \quad a_{4m} = \frac{3}{16}, \quad a_{6m} = \frac{947}{3780}, \quad a_{7m} = -\frac{5}{8}
\end{align}
\begin{align}
	a_{5m} &= \frac{7711}{4608} \left( \log ^2\left[ 2 \sqrt{3}-3\right] - \log \left[ \frac{4}{3} \right] \log \left[ 3 + 2\sqrt{3} \right] -2 \text{Li}_2\left[ \frac{\sqrt{3}}{2}\right] + 2 \text{Li}_2 \left[2 \sqrt{3}-3 \right] \right) +\frac{8735 \pi ^2}{13824} \nonumber \\
	& -\frac{206795171}{57153600}-\frac{8629}{6144} \log^2 \left[\frac{4}{3}\right] - \frac{179}{1152} \log \left[\frac{4}{3}\right] + \frac{293}{128} \csc ^{-1} \left[ \sqrt{3} \right]^2 + \frac{1043}{288 \sqrt{2}} \csc ^{-1}\left[\sqrt{3}\right]
\end{align}
\begin{align}
	a_{8m} &= -\frac{1099}{6144}  \left( \log^2\left[2 \sqrt{3}-3\right] -\log \left[ \frac{4}{3} \right]  \log \left[ 3+2 \sqrt{3}\right] -2 \text{Li}_2\left[\frac{\sqrt{3}}{2}\right] +2 \text{Li}_2 \left[ 2 \sqrt{3}-3 \right] \right) -\frac{10465 \pi^2}{55296} \nonumber \\
	& + \frac{27092374721}{31120135200} - \frac{437}{24576} \log^2 \left[ \frac{4}{3} \right] + \frac{\log [3]}{480} - \frac{13681}{69120} \log \left[\frac{4}{3}\right]-\frac{27}{512} \csc^{-1} \left[\sqrt{3}\right]^2 - \frac{323}{576 \sqrt{2}} \csc^{-1} \left[\sqrt{3}\right]
\end{align}
\begin{align}
	a_{9m} = \frac{1}{3} \log \left[\frac{4}{3}\right]-\frac{52837}{561330}, \quad a_{10m} =  -\frac{11}{48}, \quad 	a_{12m} = -\frac{1}{8} \log \left[\frac{4}{3}\right] - \frac{4327283}{8981280}, \quad a_{13m} = \frac{1}{16}
\end{align}
\begin{align}
	a_{11m} &= \frac{181}{24576} \left( \log ^2\left[2 \sqrt{3}-3\right]-\log \left[\frac{4}{3}\right] \log \left[3+2 \sqrt{3}\right] -2 \text{Li}_2\left[ \frac{\sqrt{3}}{2} \right] + 2 \text{Li}_2\left[ 2 \sqrt{3}-3 \right] \right) +\frac{356603876663}{569053900800} \nonumber \\
	& + \frac{3253\pi^2}{73728} + \frac{5963}{98304} \log^2\left[\frac{4}{3}\right] + \frac{177301}{967680} \log \left[\frac{4}{3}\right] -\frac{31 \log [3]}{1120} - \frac{27}{2048} \csc ^{-1} \left[ \sqrt{3} \right]^2 - \frac{67}{2048 \sqrt{2}} \csc^{-1} \left[\sqrt{3}\right]
\end{align}

\subsection{$F_\eta$}

The expression for $F_\eta$ can be written as:
\begin{align}
\frac{F_\eta}{F} &= 1 + \left\{ \frac{8}{3} \xi_K \hat{L}^r_{1f} + \frac{4}{3} \xi_\pi \hat{L}^r_{2f} - \frac{3}{2} \xi_K \lambda_K \right\}
\nonumber\\
& \qquad + \Bigg\{ \hat K_{1F}^r \lambda_\pi^2
+ \hat K_{2f}^r \lambda_\pi\lambda_K
+ \hat K_{3f}^r \lambda_\pi\lambda_\eta
+ \hat K_{4f}^r \lambda_K^2
+ \hat K_{5f}^r \lambda_K\lambda_\eta
+ \hat K_{6f}^r \lambda_\eta^2
\nonumber\\ & \hspace*{7ex}
+ m^2_K \xi_K^2 F_f\left[ \frac{m_\pi^2}{m_K^2} \right]
+ \hat C_{1f}\lambda_\pi+\hat C_{2f}\lambda_K+\hat C_{3f}\lambda_\eta
+ \hat C_{4f} \Bigg\}
\end{align}
where

\begin{align}
	\hat{L}^r_{1f} = (4 \pi)^2 (3 L^r_{4} + 2 L^r_{5}), \quad \hat{L}^r_{2f} = (4 \pi)^2 (3 L^r_{4} -  L^r_{5})
\end{align}
\begin{align}
	& \hat{K}^r_{1f} = \frac{99}{128} \rho - \frac{141}{512} \rho^2 + \frac{3}{2048} \rho^3 + \frac{3}{8192} \rho^4 + \mathcal{O}(\rho)^5, \quad \hat{K}^r_{3f} = 0 \nonumber \\
	& \hat{K}^r_{2f} = \frac{93}{64} \rho - \frac{3}{256} \rho^2 - \frac{3}{1024} \rho^3 - \frac{3}{4096} \rho^4 + \mathcal{O}(\rho)^5, \quad \hat{K}^r_{5f} = -\frac{119}{48} + \frac{1}{4} \rho \nonumber \\
	& \hat{K}^r_{4f} = \frac{191}{96} + \frac{35}{128} \rho + \frac{3}{512} \rho^2 + \frac{3}{2048} \rho^3 + \frac{3}{8192} \rho^4 + \mathcal{O}(\rho)^5, \quad \hat{K}^r_{6f} = \frac{71}{96} + \frac{1}{8} \rho - \frac{1}{32} \rho^2
\end{align}
\begin{align}
	\hat{C}^r_{1f} =& \left(\frac{3}{16}-\frac{16}{3} (4 \pi)^2 (3 L^r_{4}+2 L^r_{5})\right) \xi_\pi \xi_K + \left(\frac{2}{3} (4 \pi )^2 (36 L^r_{1}+9 L^r_{2}+9 L^r_{3}-33 L^r_{4}+2 L^r_{5})+\frac{47}{64}\right) \xi_\pi^2
\end{align}
\begin{align}
	\hat{C}^r_{2F} =& \left(2 (4 \pi)^2 (16 L^r_{1}+4 L^r_{2}+7 L^r_{3}-18 L^r_{4}-4 L^r_{5}) + \frac{17}{48}\right) \xi_K^2 + \left(\frac{3}{4}-\frac{2}{3} (4 \pi )^2 (15 L^r_{4}+13 L^r_{5})\right) \xi_\pi \xi_K 
\end{align}
\begin{align}
	\hat{C}^r_{3F} =& \left(\frac{32}{9} (4 \pi )^2 (6 L^r_{1}+6 L^r_{2}+3 L^r_{3}-3 L^r_{4}-2 L^r_{5})+\frac{16631}{3888}\right) \xi_K^2 \nonumber \\
	& - \left(\frac{16}{9} (4 \pi )^2 (6 L^r_{1}+6 L^r_{2}+3 L^r_{3}-3 L^r_{4}-2 L^r_{5})+\frac{4363}{3888}\right)\xi_\pi \xi_K \nonumber \\
	& + \left(\frac{2}{9} (4 \pi )^2 (6 L^r_{1}+6 L^r_{2}+3 L^r_{3}-3 L^r_{4}-2 L^r_{5})+\frac{3713}{15552}\right) \xi_\pi^2
\end{align}
\begin{align}
	\hat{C}^r_{4F} &= \frac{1}{9} (4 \pi )^2 \bigg\{ 8 ( 2 L^r_{1}+2 L^r_{2}+L^r_{3}) \xi_\pi \xi_K - (32 L^r_{1}+68 L^r_{2}+19 L^r_{3}) \xi_K^2 - (2 L^r_{1}+29 L^r_{2}+10 L^r_{3})  \xi_\pi^2 \bigg\} \nonumber \\
	& + \frac{8}{9} (4\pi)^4 \bigg\{ 4 \bigg( 6 C^r_{14}+6 C^r_{15}+9 C^r_{16}+6 C^r_{17}+6 C^r_{18}+(3 L^r_{4}+2 L^r_{5}) (21 L^r_{4}+4 L^r_{5}-24 L^r_{6}-12 L^r_{8}) \bigg) \xi_K^2 \nonumber \\
	&  - \bigg( 24 C^r_{14} - 6 C^r_{15} + 36 C^r_{16} + 24 C^r_{17} + 48 C^r_{18} - 252 (L^r_{4})^2 - 108 L^r_{4} L^r_{5} + 288 L^r_{4} L^r_{6} - 56 (L^r_{5})^2 + 48 L^r_{5} L^r_{6} \bigg) \xi_\pi \xi_K \nonumber \\
	&  + \bigg( 9 C^r_{14}-3 C^r_{15}+27 C^r_{16}+9 C^r_{17}+24 C^r_{18}+(3 L^r_{4}-L^r_{5}) (21 L^r_{4}+25 L^r_{5}-24 L^r_{6}-24 L^r_{8} ) \bigg) \xi_\pi^2 \bigg\}
\end{align}

Due to the numerically large prefactors of the masses in the expression for $d^\eta_{\pi K K}$, the errors that arise due to the use of the truncated sunset expressions get magnified significantly, resulting in a poorly converging expression if these approximate results for the sunsets are used. This can be seen in Figure~\ref{FigLatticeFitEta}, where the divergence between the truncated and exact values is significant even for small values of $\rho$. Therefore, in this case of $F_f$, we present an expansion in $\rho$ taken from the sunset integral series evaluated to a high order (and which therefore results in rapid convergence to the exact result), but an expansion which is numerical.

\begin{align}
	F_f [ \rho ] = 9.03816 & + \bigg( -7.82805 + 1.51852 \log (\rho) + 0.1875 \log ^2(\rho)  \bigg) \rho \nonumber \\
	&  + \bigg( 2.69955 + 0.250529 \log (\rho) - 0.625 \log ^2(\rho) \bigg) \rho^2 \nonumber \\
	& + \bigg( -1.08218 + 0.00176579 \log (\rho) - 0.229167 \log^2(\rho) \bigg) \rho^3 \nonumber \\
	& + \bigg( 0.722228 - 0.306794 \log (\rho) + 0.0625 \log ^2(\rho) \bigg) \rho^4 + \mathcal{O}(\rho^5)
\end{align}

\section{Summary and Conclusion}

$SU(3)$ ChPT is the effective theory of the strong interactions at low-energies, and describes the pseudo-scalar octet degrees of freedom and their interactions. Of the many properties associated with this sector, the masses and decay constants are amongst the most fundamental. The predictions for these from the effective theory and from the lattice constitute some of the most important tests of this part of the standard model and the standard picture of spontaneous symmetry breaking of the axial-vector symmetries associated with the massless limit of the theory. In the limit of isospin invariance, there are three masses in the theory, namely $m_\pi$, $m_K$ and $m_\eta$.

At two-loop order, the meson mass expressions involve the computation of the sunset diagrams, while the decay constants also require calculation of the energy derivative of the sunsets, all evaluated on-shell. Sunset integrals have been investigated in great detail independently of ChPT, and much is known about them. In the most general mass configuration, it has been shown that any sunset can be expressed in terms of at most four MI. If some of the masses are equal, then the number of MI reduces. On the other hand, if any of the masses is set to zero, the sunset is known to suffer from infra-red problems. All the above features contribute to the complexity of analyzing the masses and decay constants in ChPT.

Analytic treatments of the pion mass and decay constant have been performed in \cite{Ananthanarayan:2017yhz, Kaiser:2007kf}, where due to strangeness conservation, the only configuration not corresponding to a pseudo-threshold is a sunset with a kaon pair and an $\eta$ in the propagators. Since the pion mass is the smallest parameter in the theory, it is possible in this case to provide an expansion in this parameter to get the corresponding analytic expression. On the other hand, for the eta, in which a similar configuration appears, except with the pion mass in the propagator and the eta mass in the external momentum, it is not possible to expand in the small parameter without encountering IR divergences. In the case of the kaon, the sole configuration not of the pseudo-threshold type is one in which all the three particles are present in the propagators. For the quantities of interest, there are then two mass ratios present and one may wish to provide a double series representation in these mass ratios. In this work we have carried out precisely this exercise, by introducing MB representation for the sunset diagrams at hand (see \cite{Ananthanarayan:2018} for details). Whereas for problems with a single MB parameter, a simple approach exists which allows one to carry out the evaluation and summation of residues of poles, closing the contour in the complex plane to the left or to the right, and then using simple ratio tests to figure out the regions of convergence in the single parameter, a more sophisticated analysis is required when two or (especially) more MB parameters appear. The case at hand is a concrete realization of this scenario with two parameters. Our work follows several steps which will be summarized now.

\begin{enumerate}

\item We decompose the vector, tensor and derivative sunsets integrals appearing in the expressions for the $m_P$ and $F_P$ by applying integration by parts to express then in terms of the MI.

\item The resulting MI are of various mass configurations, and appear with up to three distinct mass scales. Each of these MI are then evaluated by using MB representations. The solutions of the one and two mass scale MI appearing in this analysis can all be written in closed form. The solutions of the three mass scale master integrals, however, are expressed as linear combinations of single and double infinite series. The full results are given in Appendix~\ref{Sec:SunsetResults}; see also \cite{Ananthanarayan:2017qmx} for an equivalent rewriting of these results in terms of Kamp\' e de F\' eriet series. We show in Appendix \ref{Sec:PionSunsets} how to get analytic results valid for the pion case.

\item We substitute the sunset integrals results into the expressions of the $m_P$ and $F_P$.

\item The GMO relation is then applied to these expressions. As the GMO is a tree-level relation, and we wish to express the $m_P$ and $F_P$ in terms of physical meson masses, this involves calculating and including contributions from lower $\mathcal{O}(p^4)$ terms to the higher orders ones $\mathcal{O}(p^6)$. The motivation for applying the GMO relation stems partly from the desire to provide simple expressions that can be compared against lattice simulations, in which the eta mass is generally calculated using the GMO relation and is not an independent parameter.

\item We isolate the contributions to the $m_P$ and $F_P$ from different terms, e.g. linear chiral log terms, bilinear chiral logs, terms involving the $\mathcal{O}(p^4)$ LEC, etc., to determine their relative weight in the final expressions of the masses and decay constants.

\item The $m_P$ and $F_P$ expressions for $P=K,\eta$ without application of the GMO relation, but separated into terms of different classes, is given in Appendix~\ref{Sec:NonGMOExpr}.

\item A set of results are given for the three mass scale sunsets that are truncations of the exact results, but which are numerically close to the latter for the lattice input sets of \cite{Durr:2010hr}. The approximate results for the the sunsets appearing in the kaon and eta expressions is given in Section~\ref{Sec:NumApproxSunsets}, and those for the pion are given in Appendix~\ref{Sec:PionSunsets}. The numerical justification for some of these approximations is presented in Section~\ref{Sec:NumAnalysis}.

\item Also presented as an ancillary tool to this paper is a \texttt{Mathematica} based code that allows one to obtain a truncated expression for the three mass master sunset integrals when the level of precision and values of the input meson masses are provided. This allows lattice practitioners, amongst others, to obtain analytic approximations for the sunsets for any given set of lattice inputs. These can then be used to construct relatively compact analytic expressions for easy comparison with lattice or experimental data.

\item A numerical study is done in Section~\ref{Sec:NumAnalysis} for $m_K$, $F_K$, $m_\eta$ and $F_\eta$ to provide a breakup of the relative numerical contributions to the NNLO part of their different constituents. This shows that the sunset integral contribution is significant.

\item In Section~\ref{Sec:NumAnalysis}, we also numerically justify the use of our GMO-simplified expressions by showing that the error on various components constituting the NNLO contribution due to the use of the GMO relation does not exceed 5\% in most cases, and that the final error on the NNLO contribution is effectively zero for the kaon mass, and very small for the kaon and eta decay constants.

\item We provide in Section~\ref{Sec:LatticeFits} a set of expressions for $m_K$, $m_\eta$, $F_K$ and $F_\eta$ that can be easily fit with lattice data, and in which the term that depends on the approximation of the loop integrals may easily be substituted by other approximations (calculated, for example, using tools such as the aforementioned supplementary \texttt{Mathematica} files).

\item We also calculate values of $m_K$, $F_K$, $m_\eta$ and $F_\eta$, and see that the comparison of our results with prior determinations shows good agreement when the BE14 LEC values are used. When the free fit LEC values are used, our results show some divergence with some  literature values.

\end{enumerate}

In this paper, we adopt a phenomenology practitioners perspective, and provide principally the final results that are of relevance in this respect. The results given in Appendix~\ref{Sec:SunsetResults}, for example, are for the $\mathcal{O}(\epsilon^0)$ term, and are only convergent for the values of mass ratios shown in Figure~\ref{Fig:RegOfConv}. In a forthcoming publication \cite{Ananthanarayan:2018}, we describe the calculation of the three mass scale sunset integrals in detail, and give the complete $\epsilon$-expansion for all possible values of the meson masses.

An important field where analytic expressions may be of use, and one which we have emphasised strongly in this work, is in lattice QCD. In \cite{Ananthanarayan:2017qmx}, the use of analytic expressions to determine values of ChPT parameters was demonstrated. Data from recent lattice simulations for $m_K$, $m_\eta$, $F_K$ and $F_\eta$ is not publically available, but we hope that the expressions and tools provided in this work will encourage and assist lattice practioners to perform such a cross-disciplinary study.

\section*{Acknowledgements}
SF thanks David Greynat for helpful discussions and correspondence. JB is supported in part by the Swedish Research Council grants contract numbers 2015-04089 and 2016-05996 and by the European Research Council under the European Union’s Horizon 2020 research and innovation programme (grant agreement No 668679). BA is partly supported by the MSIL Chair of the Division of Physical and Mathematical Sciences, Indian Institute of Science.

\appendix

\section{Expressions without the use of GMO \label{Sec:NonGMOExpr}}

We present here the expressions for the masses and decay constants in which the physical eta mass has been retained and not simplified by use of the GMO relation. We give only those terms that change when the GMO relation is used. 

\subsection{Kaon Mass}

The expression for the kaon mass, not simplified using the GMO relation, is:
\begin{align}
	M^2_{K} = m^{2}_{K} + \left( m^{2}_{K} \right)^{(4)} + \left( m^{2}_{K} \right)^{(6)}_{CT} + \left( m^{2}_{K} \right)^{(6)}_{loop} + \mathcal{O}(p^8) 
\end{align}
where $m^{2}_{K}$ is given by Eq.(\ref{cTreeMass}), $\left( m^{2}_{K} \right)^{(6)}_{CT}$ is given by Eq.(\ref{cCT}),
\begin{align}
	\frac{F_{\pi}^2}{m_K^2}	\left( m^{2}_{K} \right)^{(4)} = 8(m_{\pi}^2 + 2 m_K^2)(2 L_6^r - L_4^r) + 8 m_K^2 (2L_8^r - L_5^r) + \frac{m_{\eta}^4}{2 m_{K}^2} l^r_{\eta} +  \frac{m_{\pi}^2 m_{\eta}^2}{6 m_{K}^2} l^r_{\eta} 
\end{align}
and
\begin{align}
	F_{\pi}^4 \left( m^{2}_{K} \right)^{(6)}_{loop} = c^{K}_{L_i} + c^{K}_{L_i \times L_j} + c^{K}_{log \times L_i} + c^{K}_{log} + c^{K}_{log \times log} + c^{K}_{sunset}
\end{align}

where:

\begin{align}
	27 (16 \pi^2) c^{K}_{L_i} &= 108 m_K^6 L_1^r + 3 \left( 122 m_K^6 - 16 m_K^4 m_{\pi}^2 + 56 m_K^2 m_{\pi}^4 \right)L_2^r \nonumber \\
	& + \left(89 m_k^6 - 4 m_K^4 m_{\pi}^2 + 41 m_K^2 m_{\pi}^4 \right) L_3^r 
\end{align}

\begin{align}
	c^{K}_{log \times L_i} =& -2 m_K^2 m_{\pi}^2 \bigg( 48 m_{\pi}^2 L_1^r + 12 m_{\pi}^2 L_2^r + 15 m_{\pi}^2 L_3^r - 4 \left(8 m_K^2+17 m_{\pi}^2\right) L_4^r - 4  \left(4 m_K^2+3 m_{\pi}^2\right) L_5^r \nonumber \\
	& \quad + 8 \left(8 m_K^2+11 m_{\pi}^2\right) L_6^r + 8 \left(4 m_K^2+3 m_{\pi}^2\right) L_8^r \bigg) l_{\pi}^r \nonumber \\
	& - 4 m_K^4 \bigg( 36 m_K^2 L_1^r + 18 m_K^2 L_2^r + 15 m_K^2 L_3^r - 4 \left(10 m_K^2 + m_{\pi}^2 \right) L_4^r - 16 m_K^2 L_5^r \nonumber \\
	& \quad + 8 \left(8 m_K^2+m_{\pi}^2 \right) L_6^r + 32 m_K^2 L_8^r \bigg) l_K^r \nonumber \\
	& - \frac{2}{9} m_{\eta}^2 \bigg( 48 m_{K}^2 \left(4 m_{K}^2-m_{\pi}^2\right) L_1^r + 12 m_{K}^2 \left(4 m_{K}^2 - m_{\pi}^2\right) L_2^r + 21 m_{K}^2 \left(4 m_{K}^2-m_{\pi}^2\right) L_3^r \nonumber \\
	& \quad + 36 m_{K}^2 \left(m_{\pi}^2-8 m_{K}^2\right) L_4^r - 4 \left(28 m_{K}^4-3 m_{K}^2 m_{\pi}^2+2 m_{\pi}^4\right) L_5^r + 24 m_{K}^2 \left(16 m_{K}^2-m_{\pi}^2\right) L_6^r \nonumber \\
	& \quad + 96 \left(2 m_{K}^4-3 m_{K}^2 m_{\pi}^2+m_{\pi}^4\right) L_7^r + 24 \left(12 m_{K}^4-7 m_{K}^2 m_{\pi}^2+2 m_{\pi}^4\right) L_8^r \bigg) l_{\eta}^r
\end{align}

\begin{align}
	\left( 16 \pi^2 \right) c_{log}^{K} & =  \left(\frac{3}{8} m_{\eta}^2 m_{K}^2 m_{\pi}^2 - \frac{13}{4} m_{K}^4 m_{\pi}^2 - \frac{45}{16} m_{K}^2 m_{\pi}^4 \right) l_{\pi}^r  \nonumber \\
	& - \bigg( \frac{487}{72} m_{K}^6 + \frac{9}{16} m_{\eta}^4 m_{K}^2 - \frac{1}{12} m_{\eta}^2 m_{K}^4 + \frac{3}{4} m_{\eta}^2 m_{K}^2 m_{\pi}^2 + \frac{7}{4} m_{K}^4 m_{\pi}^2 + \frac{3}{16} m_{K}^2 m_{\pi}^4 \bigg) l_K^r \nonumber \\
	& + \left(\frac{9}{16} m_{\eta}^4 m_{K}^2 - \frac{143}{36} m_{\eta}^2 m_{K}^4 - \frac{1}{8} m_{\eta}^2 m_{K}^2 m_{\pi}^2 \right) l_{\eta}^r
\end{align}

\begin{align}
	c_{log \times log}^{K} &=  \left(\frac{15}{16} m_{\eta}^4 m_{\pi}^2 - \frac{41}{16} m_{\eta}^2 m_{K}^2 m_{\pi}^2 + \frac{15}{16} m_{\eta}^2 m_{\pi}^4 + \frac{9}{4} m_{K}^4 m_{\pi}^2 + \frac{9}{4} m_{K}^2 m_{\pi}^4 \right) (l^r_{\pi})^2 \nonumber \\
	 & + \left(-\frac{9}{8} m_{\eta}^4 m_{K}^2 + \frac{11}{12} m_{\eta}^2 m_{K}^4 - \frac{3}{2} m_{\eta}^2 m_{K}^2 m_{\pi}^2 + \frac{143}{18} m_{K}^6 + \frac{7}{4} m_{K}^4 m_{\pi}^2 - \frac{3}{8} m_{K}^2 m_{\pi}^4 \right) (l^r_{K})^2  \nonumber \\
	& + \left(-\frac{121}{72} m_{K}^2 m_{\eta}^4 + \frac{5}{32} m_{\eta}^4 m_{\pi}^2 + \frac{205}{36} m_{\eta}^2 m_{K}^4 - \frac{41}{16} m_{\eta}^2 m_{K}^2 m_{\pi}^2 + \frac{227}{288} m_{\eta}^2 m_{\pi}^4 \right) (l^r_{\eta})^2 \nonumber \\
	& + \left(\frac{3}{2} m_{\eta}^2 m_{K}^2 m_{\pi}^2 - \frac{7}{2} m_{K}^4 m_{\pi}^2 + \frac{3}{4} m_{K}^2 m_{\pi}^4 \right)  l^r_{\pi} l^r_{K} + \left(\frac{9}{4} m_{\eta}^4 m_{K}^2 - \frac{9}{2} m_{\eta}^2 m_{K}^4 + \frac{3}{2} m_{\eta}^2 m_{K}^2 m_{\pi}^2\right)  l^r_{K} l^r_{\eta} \nonumber \\
 & + \left(-\frac{15}{8} m_{\eta}^4 m_{\pi}^2 + \frac{5}{24} m_{\eta}^2 m_{K}^2 m_{\pi}^2 - \frac{37}{24} 
 m_{\eta}^2 m_{\pi}^4 \right) l^r_{\pi} l^r_{\eta} 
\end{align}

\begin{align}
	c^K_{sunset} &= \frac{1}{(16\pi^2)^2} \Bigg\{ \left(\frac{413}{128}+\frac{97 \pi ^2}{384}\right) m_{\eta}^4 m_{K}^2 -\left(\frac{33}{8}+\frac{15 \pi ^2}{64}\right) m_{\eta}^6 - \left(\frac{427}{3456}+\frac{179 \pi ^2}{648}\right) m_{K}^6 \nonumber \\
	&  + \left(\frac{29}{64}-\frac{\pi ^2}{32}\right) m_{\eta}^2 m_{K}^4 - \left(\frac{3}{4}+\frac{3 \pi ^2}{64}\right) m_{\pi}^6 + \left(\frac{3}{8}-\frac{5 \pi ^2}{64}\right) m_{\eta}^2 m_{\pi}^4 + \left(\frac{3}{8}-\frac{5 \pi ^2}{64}\right) m_{\eta}^4 m_{\pi}^2  \nonumber \\
	&  +\left(\frac{209}{1728}-\frac{265 \pi ^2}{2592}\right) m_{K}^4 m_{\pi}^2-\left(\frac{67}{384}+\frac{9 \pi ^2}{128}\right) m_{K}^2 m_{\pi}^4 + \left(\frac{3}{2}+\frac{53 \pi ^2}{192}\right) m_{\eta}^2 m_{K}^2 m_{\pi}^2 \Bigg\} \nonumber \\
	&  + c^K_{K \pi \pi} + c^K_{K \eta \eta}  + c^K_{K \pi \eta}
\end{align}

\begin{align}
	c^K_{K \eta \eta} &= \left(\frac{45}{32} m_{\eta}^4 - \frac{19}{16} m_{\eta}^2 m_{K}^2 + \frac{25}{288} m_{K}^4 \right) \overline{H}^{\chi}_{K \eta \eta} - \left(\frac{15}{8} m_{\eta}^4 m_{K}^2 - \frac{4}{3} m_{\eta}^2 m_{K}^4 - \frac{13}{24} m_{K}^6 \right) \overline{H}^{\chi}_{2K \eta \eta}
\end{align}

\begin{align}
	c^K_{K \pi \eta} &= \left(-\frac{3}{32} m_{\eta}^4 + \frac{5}{16} m_{\eta}^2 m_{K}^2 - \frac{3}{4} m_{\eta}^2 m_{\pi}^2 + \frac{13}{16} m_{K}^4 + \frac{5}{16} m_{K}^2 m_{\pi}^2 - \frac{3}{32} m_{\pi}^4 \right) \overline{H}^{\chi}_{K \pi \eta} \nonumber \\
	& + \left(\frac{3}{32} m_{\eta}^6 - \frac{13}{32} m_{\eta}^4 m_{K}^2 + \frac{39}{32} m_{\eta}^4 m_{\pi}^2 + \frac{1}{8}m_{\eta}^2 m_{K}^4 - \frac{51}{32} m_{\eta}^2 m_{K}^2 m_{\pi}^2 + \frac{9}{16} m_{\eta}^2 m_{\pi}^4 \right) \overline{H}^{\chi}_{K \pi 2\eta} \nonumber \\
	& + \left(\frac{9}{16} m_{\eta}^4 m_{\pi}^2 - \frac{51}{32} m_{\eta}^2 m_{K}^2 m_{\pi}^2 + \frac{39}{32} m_{\eta}^2 m_{\pi}^4 + \frac{1}{8} m_{K}^4 m_{\pi}^2 -\frac{13}{32} m_{K}^2 m_{\pi}^4 + \frac{3}{32} m_{\pi}^6 \right) \overline{H}^{\chi}_{K 2\pi \eta}
\end{align}
where $c^{K}_{L_i \times L_j}$ is given by Eq.(\ref{cLiLj}) and $c^K_{K \pi \pi}$ by Eq.(\ref{cBarkpp}).

\subsection{Kaon Decay Constant}

The expression for the kaon decay constant, not simplified using the GMO relation, is:
\begin{align}
	\frac{F_K}{F_0} = 1 + F_K^{(4)} + \left( F_K \right)^{(6)}_{CT} + \left( F_K \right)^{(6)}_{loop} + \mathcal{O}(p^8) 
\end{align}
where:
\begin{align}
	F_{\pi}^2 F_K^{(4)} = 4\left(2 m_{K}^2+m_{\pi}^2\right) L_4^r  + 4 m_{K}^2 L_5^r - \frac{3}{4} m_{\pi}^2 l_{\pi}^r - \frac{3}{2} m_{K}^2 l_{K} - \frac{3}{4} m_{\eta}^2 l_{\eta}^r
\end{align}
and $ \left( F_K \right)^{(6)}_{CT}$ is given by Eq.(\ref{dCT}). We also have:
\begin{align}
	F_{\pi}^4 \left( F_K \right)^{(6)}_{loop} = d^{K}_{L_i} + d^{K}_{L_i \times L_j} + d^{K}_{log \times L_i} + d^{K}_{log} +  d^{K}_{log \times log} + d^{K}_{sunset}
\end{align}

where:

\begin{align}
	-54(16\pi^2) d^{K}_{L_i} = (108 m_{K}^4) L_1^r + 6 \left(61 m_{K}^4 - 8 m_{K}^2 m_{\pi}^2 + 28 m_{\pi}^4 \right) L_2^r + \left(89 m_{K}^4 - 4 m_{K}^2 m_{\pi}^2 + 41 m_{\pi}^4 \right) L_3^r
\end{align}

\begin{align}
	d^{K}_{log \times L_i} =& \left(48 m_{\pi}^2 L_1^r + 12 m_{\pi}^2  L_2^r + 15 m_{\pi}^2 L_3^r - \left(38 m_{K}^2+47 m_{\pi}^2\right) L_4^r - \left(19 m_{K}^2+6 m_{\pi}^2\right) L_5^r \right) m_{\pi}^2 l_{\pi}^r \nonumber \\
	& + 2 \left( 36 m_{K}^2 L_1^r + 18 m_{K}^2 L_2^r + 15 m_{K}^2 L_3^r - \left(30 m_{K}^2 + 7 m_{\pi}^2\right) L_4^r - \left(7 m_{K}^2 + 6 m_{\pi}^2\right) L_5^r \right) m_{K}^2 l_K^r \nonumber \\
	& + \left(\frac{1}{3} \left(4 m_{K}^2-m_{\pi}^2\right) (16 L_1^r +4 L_2^r + 7 L_3^r ) - \left(22 m_{K}^2-m_{\pi}^2\right) L_4^r - 3 \left(m_{K}^2 + 2 m_{\pi}^2 \right) L_5^r \right) m_{\eta}^2 l_{\eta}  
\end{align}

\begin{align}
 	\left(16\pi^2\right) d_{log}^{K} =& \left( - \frac{9}{16} m_{\eta}^2 m_{\pi}^2 + \frac{9}{8} m_{K}^2 m_{\pi}^2 + \frac{39}{32} m_{\pi}^4 \right) l_{\pi}^r + \left( - \frac{27}{32} m_{\eta}^4 + \frac{41}{24} m_{\eta}^2 m_{K}^2 - \frac{5}{16} m_{\eta}^2 m_{\pi}^2 \right) l_{\eta}^r \nonumber \\
 	& + \left( \frac{27}{32} m_{\eta}^4 + \frac{5}{48} m_{\eta}^2 m_{K}^2 + \frac{9}{8} m_{\eta}^2 m_{\pi}^2 + \frac{643}{144} m_{K}^4 + \frac{27}{16} m_{K}^2 m_{\pi}^2 + \frac{9}{32} m_{\pi}^4 \right) l_K^r
\end{align}

\begin{align}
	d_{log \times log}^{K} =& \left(-\frac{45}{32}\frac{m_{\eta}^4 m_{\pi}^2}{m_{K}^2} - \frac{45}{32} \frac{m_{\eta}^2 m_{\pi}^4}{m_{K}^2} + \frac{103}{32} m_{\eta}^2 m_{\pi}^2 - \frac{9}{8} m_{K}^2 m_{\pi}^2 + \frac{51}{32} m_{\pi}^4 \right) (l^r_{\pi})^2 \nonumber \\
	& + \left(\frac{27}{16} m_{\eta}^4 - \frac{11}{12} m_{\eta}^2 m_{K}^2 + \frac{9}{4} m_{\eta}^2 m_{\pi}^2 + \frac{3}{8} m_{K}^4 - \frac{3}{2} m_{K}^2 m_{\pi}^2 + \frac{9}{16} m_{\pi}^4 \right) (l^r_{K})^2 \nonumber \\
	& + \left(-\frac{45}{32}\frac{m_{\eta}^4 m_{\pi}^2}{m_{K}^2} + \frac{45}{16} m_{\eta}^4 - \frac{45}{32} \frac{m_{\eta}^2 m_{\pi}^4}{m_{K}^2} - \frac{19}{6} m_{\eta}^2 m_{K}^2 + \frac{7}{2} m_{\eta}^2 m_{\pi}^2 \right) (l^r_{\eta})^2 \nonumber \\
	& + \left(-\frac{27}{8} m_{\eta}^4 + \frac{53}{24} m_{\eta}^2 m_{K}^2 - \frac{9}{4} m_{\eta}^2 m_{\pi}^2 \right) l^r_{K} l^r_{\eta} + \left( \frac{45}{16} \frac{m_{\eta}^4 m_{\pi}^2}{m_{K}^2} + \frac{45}{16} \frac{m_{\eta}^2 m_{\pi}^4}{m_{K}^2} -\frac{5}{8} m_{\eta}^2 m_{\pi}^2 \right) l^r_{\pi} l^r_{\eta} \nonumber \\
	& + \left(-\frac{9}{4} m_{\eta}^2 m_{\pi}^2 + \frac{75}{8} m_{K}^2 m_{\pi}^2 - \frac{9}{8} m_{\pi}^4 \right) l^r_{\pi} l^r_{K}
\end{align}

\begin{align}
	d^{K}_{sunset} &= \frac{1}{\left( 16 \pi ^2\right)^2} \Bigg\{ \left(\frac{99}{16}+\frac{45 \pi ^2}{128}\right)  \frac{m_{\eta}^6}{m_{K}^2} - \left(\frac{9}{16}-\frac{15 \pi ^2}{128}\right) \frac{m_{\eta}^4 m_{\pi}^2}{m_{K}^2} - \left(\frac{1111}{256}+\frac{263 \pi ^2}{768}\right) m_{\eta}^4 \nonumber \\
	&  - \left(\frac{9}{16}-\frac{15 \pi ^2}{128}\right) \frac{m_{\eta}^2 m_{\pi}^4}{m_{K}^2} + \left(\frac{67}{192}+\frac{185 \pi ^2}{1728}\right) m_{\eta}^2 m_{K}^2-\left(\frac{9}{4}+\frac{139 \pi ^2}{384}\right) m_{\eta}^2 m_{\pi}^2 \nonumber \\
	&  + \left(\frac{583}{2304}+\frac{3 \pi ^2}{32}\right) m_{K}^4 + \left(\frac{9}{8}+\frac{9 \pi ^2}{128}\right)  \frac{m_{\pi}^6}{m_{K}^2} - \left(\frac{5}{192}-\frac{19 \pi ^2}{192}\right) m_{K}^2 m_{\pi}^2 + \left(\frac{745}{768}+\frac{15 \pi ^2}{256}\right) m_{\pi}^4 \Bigg\}  \nonumber \\
	&  + d^{K}_{K \pi \pi} + d^{K}_{K \eta \eta} + d^{K}_{K \pi \eta} 
\end{align}

\begin{align}
	d^K_{K \eta \eta} &= \left(- \frac{135}{64} \frac{m_{\eta}^4}{m_{K}^2} + \frac{25}{16} m_{\eta}^2 - \frac{229}{576}m_{K}^2 \right) \overline{H}^\chi_{K \eta \eta} + \left(\frac{45}{16} m_{\eta}^4 - \frac{41}{24} m_{\eta}^2 m_{K}^2 + \frac{37}{144} m_{K}^4 \right) \overline{H}^\chi_{2K \eta \eta}
\end{align}

\begin{align}
	d^K_{K \pi \eta} &= \left(\frac{9}{64} \frac{m_{\eta}^4}{m_{K}^2}+ \frac{9}{8} \frac{m_{\eta}^2 m_{\pi}^2}{m_{K}^2} - \frac{5}{16} m_{\eta}^2 + \frac{9}{64} \frac{m_{\pi}^4}{m_{K}^2} + \frac{7}{32} m_{K}^2 - \frac{5}{16} m_{\pi}^2 \right) \overline{H}^\chi_{K \pi \eta} - \left( \frac{1}{2} m_{K}^4 \right) \overline{H}^\chi_{2K \pi \eta} \nonumber \\
	& + \left(- \frac{9}{64} \frac{m_{\eta}^6}{m_{K}^2} - \frac{117}{64} \frac{m_{\eta}^4 m_{\pi}^2}{m_{K}^2} + \frac{29}{64}m_{\eta}^4 - \frac{27}{32} \frac{m_{\eta}^2 m_{\pi}^4}{m_{K}^2} - \frac{13}{16} m_{\eta}^2 m_{K}^2 + \frac{123}{64} m_{\eta}^2 m_{\pi}^2 \right) \overline{H}^\chi_{K \pi 2\eta} \nonumber \\
	& + \left(- \frac{27}{32} \frac{m_{\eta}^4 m_{\pi}^2}{m_{K}^2} - \frac{117}{64} \frac{m_{\eta}^2 m_{\pi}^4}{m_{K}^2} + \frac{123}{64} m_{\eta}^2 m_{\pi}^2 - \frac{9}{64} \frac{m_{\pi}^6}{m_{K}^2} - \frac{13}{16} m_{K}^2 m_{\pi}^2 + \frac{29}{64} m_{\pi}^4 \right) \overline{H}^\chi_{K 2\pi \eta}
\end{align}
where $d^{K}_{L_i \times L_j}$ is given by Eq.(\ref{dBarLiLj}) and $d^K_{K \pi \pi} $ by Eq.(\ref{dBarkpp}).

\subsection{Eta Mass}

The expression for the eta mass, not simplified using the GMO relation, is:
\begin{align}
	M^2_{\eta} = m^{2}_{\eta} + \left( m^{2}_{\eta} \right)^{(4)} + \left( m^{2}_{\eta} \right)^{(6)}_{CT} + \left( m^{2}_{\eta} \right)^{(6)}_{loop} + \mathcal{O}(p^8) 
\end{align}
where $m^{2}_{\eta}$ is given by Eq.(\ref{cTreeMass}), $\left( m^{2}_{\eta} \right)^{(6)}_{CT}$ is given by Eq.(\ref{cCT}),
\begin{align}
		\frac{F_{\pi}^2}{m_\eta^2}	\left( m_{\eta}^{2} \right)^{(4)} =& -8 m_{\eta}^2 \left(2 m_{K}^2+m_{\pi}^2\right) L_4^r + \frac{8}{3} m_{\eta}^2 \left(m_{\pi}^2-4 m_{K}^2\right) L_5^r + \frac{16}{3} \left(8 m_{K}^4 + 2 m_{K}^2 m_{\pi}^2 - m_{\pi}^4 \right) L_6^r \nonumber \\
	& + \frac{128}{3} \left(m_{K}^2-m_{\pi}^2\right)^2 L_7^r + \frac{16}{3} L_8^r \left(8 m_{K}^4-8 m_{K}^2 m_{\pi}^2+3 m_{\pi}^4\right) - m_{\pi}^4 l^r_{\pi} \nonumber \\
	& + \frac{2}{3} m_{K}^2 \left(3 m_{\eta}^2 + m_{\pi}^2 \right) l^r_{K} + \frac{1}{9} m_{\eta}^2 \left(7 m_{\pi}^2 - 16 m_{K}^2 \right) l^r_{\eta}
\end{align}
and
\begin{align}
	F_{\pi}^4 \left( m^{2}_{\eta} \right)^{(6)}_{loop} = c^{\eta}_{L_i} + c^{\eta}_{L_i \times L_j} + c^{\eta}_{log \times L_i} + c^{\eta}_{log} + c^{\eta}_{log \times log} + c^{\eta}_{sunset}
\end{align}
where:

\begin{align}
 ( 16 \pi^2 ) c_{log}^{\eta} &= \left(-\frac{4}{3}\frac{m_K^6 m_{\pi}^2}{m_{\eta}^2} + \frac{5}{3}\frac{m_K^4 m_{\pi}^4}{m_{\eta}^2} - \frac{7}{12} \frac{m_K^2 m_{\pi}^6}{ m_{\eta}^2} + \frac{1}{16} \frac{m_{\pi}^8}{m_{\eta}^2} - 2 m_K^4 m_{\pi}^2 + \frac{2}{3}m_K^2 m_{\pi}^4 - \frac{41}{24} m_{\pi}^6 \right) l^r_{\pi} \nonumber \\
  & + \left(\frac{20}{3} \frac{m_K^8}{m_{\eta}^2} - \frac{10}{3} \frac{m_K^6 m_{\pi}^2}{m_{\eta}^2} + \frac{5}{12} \frac{m_K^4 m_{\pi}^4}{m_{\eta}^2} - 24 m_K^6 + \frac{82}{9}m_K^4 m_{\pi}^2 - 3 m_K^2 m_{\pi}^4\right) l^r_K \nonumber \\
  & + \bigg(-\frac{20}{3} \frac{m_K^8}{m_{\eta}^2} + \frac{14}{3} \frac{m_K^6 m_{\pi}^2}{m_{\eta}^2} - \frac{25}{12} \frac{m_K^4 m_{\pi}^4}{m_{\eta}^2} - \frac{262}{243} m_{\eta}^2 m_K^4 + \frac{7}{12} \frac{m_K^2 m_{\pi}^6}{m_{\eta}^2} + \frac{823}{486} m_{\eta}^2 m_K^2 m_{\pi}^2 \nonumber \\
  & \quad - \frac{1}{16} \frac{m_{\pi}^8}{m_{\eta}^2} - \frac{83}{486} m_{\eta}^2 m_{\pi}^4 + \frac{16}{9} m_K^6 - \frac{16}{9} m_K^4 m_{\pi}^2 + \frac{5}{3} m_K^2 m_{\pi}^4 - \frac{1}{3} m_{\pi}^6 \bigg) l^r_{\eta} 
\end{align}

\begin{align}
	c_{log \times log}^{\eta} &= \bigg( -\frac{20}{3} \frac{m_{K}^8 m_{\pi}^2}{m_{\eta}^4} + \frac{22}{3} \frac{m_{K}^6 m_{\pi}^4}{m_{\eta}^4} - \frac{29}{12} \frac{m_{K}^4 m_{\pi}^6}{m_{\eta}^4} + \frac{1}{4}\frac{m_{K}^2 m_{\pi}^8}{m_{\eta}^4} + 4 \frac{m_{K}^6 m_{\pi}^2}{m_{\eta}^2} - \frac{14}{3} \frac{m_{K}^4 m_{\pi}^4}{m_{\eta}^2} \nonumber \\
	& \quad + \frac{11}{12} \frac{m_{K}^2 m_{\pi}^6}{m_{\eta}^2} + 3 m_{K}^2 m_{\pi}^4 + \frac{65}{12} m_{\pi}^6 \bigg) (l_{\pi}^r)^2 \nonumber \\
	& + \bigg(- \frac{20}{3} \frac{m_{K}^8 m_{\pi}^2}{m_{\eta}^4} + \frac{22}{3} \frac{m_{K}^6 m_{\pi}^4}{m_{\eta}^4} - \frac{29}{12} \frac{m_{K}^4 m_{\pi}^6}{m_{\eta}^4} + \frac{1}{4} \frac{m_{K}^2 m_{\pi}^8}{m_{\eta}^4} - \frac{20}{3} \frac{m_{K}^8}{m_{\eta}^2 } + \frac{22}{3} \frac{m_{K}^6 m_{\pi}^2}{m_{\eta}^2}  \nonumber \\
	& \quad - \frac{61}{12} \frac{m_{K}^4 m_{\pi}^4}{m_{\eta}^2} + \frac{11}{12} \frac{m_{K}^2 m_{\pi}^6}{m_{\eta}^2} + \frac{100}{9} m_{K}^6 -\frac{71}{9} m_{K}^4 m_{\pi}^2 + \frac{23}{6} m_{K}^2 m_{\pi}^4 \bigg) (l_{K}^r)^2 \nonumber \\
	& + \bigg( -\frac{25}{3} m_{\eta}^4 m_{K}^2 +\frac{55}{12} m_{\eta}^4 m_{\pi}^2 - \frac{20}{3} \frac{m_{K}^8}{m_{\eta}^2} + \frac{10}{3} \frac{m_{K}^6 m_{\pi}^2}{m_{\eta}^2} - \frac{5}{12} \frac{m_{K}^4 m_{\pi}^4}{m_{\eta}^2} + \frac{2312}{81} m_{\eta}^2 m_{K}^4 \nonumber \\
	& \quad - \frac{1636}{81} m_{\eta}^2 m_{K}^2 m_{\pi}^2 + \frac{619}{162} m_{\eta}^2 m_{\pi}^4 + \frac{8}{9} m_{K}^6 + \frac{4}{9} m_{K}^4 m_{\pi}^2 - \frac{1}{6}m_{K}^2 m_{\pi}^4 \bigg) (l_{\eta}^r)^2 \nonumber \\
	& + \bigg( \frac{40}{3} \frac{m_{K}^8 m_{\pi}^2}{ m_{\eta}^4} - \frac{44}{3} \frac{m_{K}^6 m_{\pi}^4}{m_{\eta}^4} + \frac{29}{6} \frac{m_{K}^4 m_{\pi}^6}{ m_{\eta}^4} - \frac{1}{2} \frac{m_{K}^2 m_{\pi}^8}{m_{\eta}^4} - 8 \frac{m_{K}^6 m_{\pi}^2}{m_{\eta}^2} + \frac{28}{3} \frac{m_{K}^4 m_{\pi}^4}{m_{\eta}^2} \nonumber \\
	& \quad -\frac{11}{6} \frac{m_{K}^2 m_{\pi}^6}{m_{\eta}^2} - \frac{32}{3} m_{K}^4 m_{\pi}^2 - \frac{8}{3} m_{K}^2 m_{\pi}^4 \bigg) l_{\pi}^r l_{K}^r \nonumber \\
	& + \bigg( \frac{40}{3} \frac{m_{K}^8}{m_{\eta}^2} - \frac{20}{3} \frac{m_{K}^6 m_{\pi}^2}{m_{\eta}^2} + \frac{5}{6} \frac{m_{K}^4 m_{\pi}^4}{ m_{\eta}^2} - \frac{64}{9} m_{\eta}^2 m_{K}^4 + \frac{28}{9} m_{\eta}^2 m_{K}^2 m_{\pi}^2 - \frac{16}{9} m_{K}^6 - \frac{8}{9} m_{K}^4 m_{\pi}^2 \nonumber \\
	& \quad + \frac{1}{3} m_{K}^2 m_{\pi}^4 \bigg) l_{K}^r l_{\eta}^r + \bigg( \frac{64}{9} m_{\eta}^2 m_{K}^2 m_{\pi}^2 - \frac{35}{9}  m_{\eta}^2 m_{\pi}^4 \bigg) l_{\pi}^r l_{\eta}^r
\end{align}

\begin{align}
	c^{\eta}_{sunset} = \frac{1}{\left(16 \pi ^2\right)^2}  &\Bigg\{ -\left(\frac{80}{3}+\frac{20 \pi ^2}{9}\right) \frac{m_K^{10}}{m_{\eta}^4} + \left(\frac{58}{3}+2 \pi ^2\right) \frac{ m_K^8 m_{\pi}^2}{m_{\eta}^4} - \left(6+\frac{11 \pi ^2}{12}\right) \frac{m_K^6 m_{\pi}^4}{m_{\eta}^4} \nonumber \\
	& + \left(\frac{49}{24}+\frac{2 \pi ^2}{9}\right) \frac{m_K^4 m_{\pi}^6}{m_{\eta}^4} - \left(\frac{7}{12}+\frac{\pi ^2}{48}\right) \frac{m_K^2 m_{\pi}^8}{m_{\eta}^4} + \frac{1}{16} \frac{m_{\pi}^{10}}{m_{\eta}^4} + \left(\frac{91}{6}+\frac{17 \pi ^2}{9}\right) \frac{m_K^8}{m_{\eta}^2} \nonumber \\
	& - \left(\frac{77}{12}+\frac{23 \pi ^2}{18}\right) \frac{m_K^6 m_{\pi}^2}{m_{\eta}^2} + \left(\frac{127}{96}+\frac{73 \pi ^2}{144}\right) \frac{m_K^4 m_{\pi}^4}{m_{\eta}^2} + \left(\frac{455}{324}+\frac{32 \pi ^2}{243}\right) m_{\eta}^2 m_K^4 \nonumber \\
	& -\left(\frac{45}{32}+\frac{11 \pi ^2}{144}\right) \frac{m_K^2 m_{\pi}^6}{m_{\eta}^2}-\left(\frac{911}{648}+\frac{28 \pi ^2}{243}\right) m_{\eta}^2 m_K^2 m_{\pi}^2 + \frac{119}{384} \frac{m_{\pi}^8}{m_{\eta}^2} \nonumber \\
	& + \left(\frac{1547}{5184}+\frac{49 \pi ^2}{1944}\right) m_{\eta}^2 m_{\pi}^4+\left(\frac{6095}{972}-\frac{767 \pi ^2}{729}\right) m_K^6+\left(\frac{85 \pi ^2}{108}-\frac{451}{144}\right) m_K^4 m_{\pi}^2 \nonumber \\
	& -\left(\frac{2857}{2592}+\frac{457 \pi ^2}{972}\right) m_K^2 m_{\pi}^4-\left(\frac{1417}{7776}+\frac{91 \pi ^2}{11664}\right) m_{\pi}^6 	\Bigg\} + c_{\pi \pi \eta}^{\eta} + c_{K K \eta}^{\eta} + c_{\pi K K}^{\eta}
\end{align}

\begin{align}
	c_{K K \eta}^{\eta} &= \left(10 \frac{m_K^8}{m_{\eta}^4} - 5 \frac{m_K^6 m_{\pi}^2}{m_{\eta}^4} + \frac{5}{8} \frac{m_K^4 m_{\pi}^4}{m_{\eta}^4} - \frac{20}{3} \frac{m_K^6}{m_{\eta}^2} + 2 \frac{m_K^4 m_{\pi}^2}{m_{\eta}^2} - \frac{1}{12} \frac{m_K^2 m_{\pi}^4}{m_{\eta}^2} - \frac{2}{3} m_K^4 + \frac{11}{9} m_K^2 m_{\pi}^2 - \frac{5}{24} m_{\pi}^4 \right) \overline{H}^\chi_{\eta K K} \nonumber \\
	& + \left(- \frac{40}{3} \frac{m_K^8}{m_{\eta}^2} + \frac{20}{3} \frac{m_K^6 m_{\pi}^2}{m_{\eta}^2} - \frac{5}{6} \frac{m_K^4 m_{\pi}^4}{m_{\eta}^2} + \frac{56}{9} m_{\eta}^2 m_K^4 - \frac{44}{9} m_{\eta}^2 m_K^2 m_{\pi}^2 + \frac{5}{6} m_{\eta}^2 m_{\pi}^4 + \frac{64}{9} m_K^6 - \frac{16}{9} m_K^4 m_{\pi}^2 \right) \overline{H}^\chi_{2\eta K K}
\end{align}

\subsection{Eta Decay Constant}

The expression for the eta decay constant, not simplified using the GMO relation, is:
\begin{align}
	\frac{F_K}{F_0} = 1 + F_{\eta}^{(4)} + \left( F_{\eta} \right)^{(6)}_{CT} + \left( F_{\eta} \right)^{(6)}_{loop} + \mathcal{O}(p^8) 
\end{align}
where:
\begin{align}
	F_{\pi}^2 F_{\eta}^{(4)} = 4 \left(2 m_K^2+m_{\pi}^2\right) L^r_{4} + \frac{4}{3} \left(4 m_K^2-m_{\pi}^2\right) L^r_{5} - 3 m_K^2 l^r_K
\end{align}
and $ \left( F_K \right)^{(6)}_{CT}$ is given by Eq.(\ref{dCT}). We also have:
\begin{align}
	F_{\pi}^4 \left( F_K \right)^{(6)}_{loop} = d^{K}_{L_i} + d^{K}_{L_i \times L_j} + d^{K}_{log \times L_i} + d^{K}_{log} +  d^{K}_{log \times log} + d^{K}_{sunset}
\end{align}
where:

\begin{align}
 	\left(16\pi^2\right) d_{log}^{\eta} &= \left( 2 \frac{m_K^6 m_{\pi}^2}{m_{\eta}^4} - \frac{5}{2} \frac{m_K^4 m_{\pi}^4}{m_{\eta}^4} + \frac{7}{8} \frac{m_K^2 m_{\pi}^6}{m_{\eta}^4} - \frac{3}{32} \frac{m_{\pi}^8}{m_{\eta}^4} - \frac{m_K^4 m_{\pi}^2}{m_{\eta}^2} + \frac{11}{6} \frac{m_K^2 m_{\pi}^4}{m_{\eta}^2} - \frac{19}{48} \frac{m_{\pi}^6}{m_{\eta}^2} + \frac{9}{8} m_{\pi}^4 \right) l^r_{\pi} \nonumber \\
 	& + \left(-10 \frac{m_K^8}{m_{\eta}^4} + 5 \frac{m_K^6 m_{\pi}^2}{m_{\eta}^4} - \frac{5}{8} \frac{m_K^4 m_{\pi}^4}{m_{\eta}^4} + \frac{4}{9} \frac{m_K^6}{m_{\eta}^2} - \frac{10}{9} \frac{m_K^4 m_{\pi}^2}{m_{\eta}^2} + \frac{1}{4} \frac{m_K^2 m_{\pi}^4}{m_{\eta}^2} + 6 m_K^4 + \frac{9}{4} m_K^2 m_{\pi}^2 \right) l^r_K \nonumber \\
 	& + \bigg( 10 \frac{m_K^8}{m_{\eta}^4} - 7 \frac{m_K^6 m_{\pi}^2}{m_{\eta}^4} + \frac{25}{8} \frac{m_K^4 m_{\pi}^4}{m_{\eta}^4} - \frac{7}{8} \frac{m_K^2 m_{\pi}^6}{m_{\eta}^4} + \frac{3}{32} \frac{m_{\pi}^8}{m_{\eta}^4} - \frac{4}{9} \frac{m_K^6}{m_{\eta}^2} + \frac{19}{9} \frac{m_K^4 m_{\pi}^2}{m_{\eta}^2} - \frac{25}{12} \frac{m_K^2 m_{\pi}^4}{m_{\eta}^2} \nonumber \\
 	& \quad  + \frac{3}{4} m_{\eta}^2 m_K^2 + \frac{19}{48} \frac{m_{\pi}^6}{m_{\eta}^2} - \frac{3}{4} m_{\eta}^2 m_{\pi}^2 + \frac{550}{243} m_K^4 - \frac{1331}{972} m_K^2 m_{\pi}^2 + \frac{2221}{3888} m_{\pi}^4 \bigg) l^r_{\eta}
\end{align}

\begin{align}
	d_{log \times log}^{\eta} &= \left( 10 \frac{m_K^8}{m_{\eta}^4} - 5 \frac{m_K^6 m_{\pi}^2}{m_{\eta}^4} + \frac{5}{8} \frac{m_K^4 m_{\pi}^4}{m_{\eta}^4} + \frac{9}{8} m_{\eta}^4 - \frac{8}{9} \frac{m_K^6}{m_{\eta}^2} - \frac{4}{9} \frac{m_K^4 m_{\pi}^2}{m_{\eta}^2} + \frac{1}{6} \frac{m_K^2 m_{\pi}^4}{m_{\eta}^2} - 3 m_{\eta}^2 m_K^2 + \frac{3}{4}  m_{\eta}^2 m_{\pi}^2 \right) (l^r_{\eta})^2 \nonumber \\
	& + \left(10 \frac{m_K^8 m_{\pi}^2}{m_{\eta}^6} - 11 \frac{m_K^6 m_{\pi}^4}{m_{\eta}^6} + \frac{29}{8} \frac{m_K^4 m_{\pi}^6}{m_{\eta}^6} - \frac{3}{8} \frac{m_K^2 m_{\pi}^8}{m_{\eta}^6} - 2 \frac{m_K^6 m_{\pi}^2}{m_{\eta}^4} + \frac{11}{3} \frac{m_K^4 m_{\pi}^4}{m_{\eta}^4} - \frac{19}{24} \frac{m_K^2 m_{\pi}^6}{m_{\eta}^4} - \frac{9}{8} m_{\pi}^4 \right) (l^r_{\pi})^2 \nonumber \\
	& + \left(-20\frac{m_K^8}{m_{\eta}^4} + 10\frac{m_K^6 m_{\pi}^2}{m_{\eta}^4} - \frac{5}{4} \frac{m_K^4 m_{\pi}^4}{m_{\eta}^4} + \frac{16}{9} \frac{m_K^6}{m_{\eta}^2} + \frac{8}{9} \frac{m_K^4 m_{\pi}^2}{m_{\eta}^2} - \frac{1}{3} \frac{m_K^2 m_{\pi}^4}{m_{\eta}^2}\right) l^r_{\eta} l^r_K \nonumber \\
	& + \bigg( 10 \frac{m_K^8 m_{\pi}^2}{m_{\eta}^6} - 11 \frac{m_K^6 m_{\pi}^4}{m_{\eta}^6} + \frac{29}{8} \frac{m_K^4 m_{\pi}^6}{m_{\eta}^6} - \frac{3}{8} \frac{m_K^2 m_{\pi}^8}{m_{\eta}^6} + 10 \frac{m_K^8}{m_{\eta}^4} - 7 \frac{m_K^6 m_{\pi}^2}{m_{\eta}^4} + \frac{103}{24} \frac{m_K^4 m_{\pi}^4}{m_{\eta}^4} \nonumber \\
	& \quad - \frac{19}{24} \frac{m_K^2 m_{\pi}^6}{m_{\eta}^4} - \frac{8}{9} \frac{m_K^6}{m_{\eta}^2} - \frac{4}{9} \frac{m_K^4 m_{\pi}^2}{m_{\eta}^2} + \frac{1}{6} \frac{m_K^2 m_{\pi}^4}{m_{\eta}^2} + 3 m_K^4-\frac{3}{2} m_K^2 m_{\pi}^2 \bigg) (l^r_K)^2 \nonumber \\
	& + \bigg( -20 \frac{m_K^8 m_{\pi}^2}{m_{\eta}^6} + 22 \frac{m_K^6 m_{\pi}^4}{m_{\eta}^6} - \frac{29}{4} \frac{m_K^4 m_{\pi}^6}{m_{\eta}^6} + \frac{3}{4} \frac{m_K^2 m_{\pi}^8}{m_{\eta}^6} + 4 \frac{m_K^6 m_{\pi}^2}{m_{\eta}^4} - \frac{22}{3} \frac{m_K^4 m_{\pi}^4}{m_{\eta}^4} + \frac{19}{12} \frac{m_K^2 m_{\pi}^6}{m_{\eta}^4} \nonumber \\
	& \quad +12 m_K^2 m_{\pi}^2 \bigg) l^r_K l^r_{\pi}
\end{align}

\begin{align}
	d^{\eta}_{sunset} &= \frac{1}{\left( 16 \pi ^2\right)^2} \Bigg\{ \left(40+\frac{10 \pi ^2}{3}\right) \frac{m_K^{10}}{m_{\eta}^6} - \left(29+3 \pi ^2\right) \frac{m_K^8 m_{\pi}^2}{m_{\eta}^6} + \left(9+\frac{11 \pi ^2}{8}\right) \frac{ m_K^6 m_{\pi}^4}{m_{\eta}^6} -\left(\frac{49}{16}+\frac{\pi ^2}{3}\right) \frac{m_K^4 m_{\pi}^6}{m_{\eta}^6} \nonumber \\
	&  + \left(\frac{7}{8}+\frac{\pi ^2}{32}\right) \frac{m_K^2 m_{\pi}^8}{m_{\eta}^6} - \frac{3}{32} \frac{m_{\pi}^{10}}{m_{\eta}^6} - \left(\frac{91}{4}+\frac{43 \pi ^2}{18}\right)  \frac{m_K^8}{m_{\eta}^4} + \left(\frac{151}{24}+\frac{49 \pi^2}{36}\right) \frac{m_K^6 m_{\pi}^2}{m_{\eta}^4} \nonumber \\
	&  - \left(\frac{125}{192}+\frac{131 \pi ^2}{288}\right) \frac{m_K^4 m_{\pi}^4}{m_{\eta}^4} + \left(\frac{301}{192}+\frac{19 \pi ^2}{288}\right) \frac{ m_K^2 m_{\pi}^6 }{m_{\eta}^4} - \frac{277}{768} \frac{m_{\pi}^8}{m_{\eta}^4} + \left(\frac{5 \pi ^2}{27}-\frac{23}{12}\right) \frac{m_K^6}{m_{\eta}^2}  \nonumber \\
	& + \left(\frac{13}{4}+\frac{\pi ^2}{18}\right) \frac{m_K^4 m_{\pi}^2}{m_{\eta}^2} - \left(\frac{331}{192}+\frac{\pi ^2}{48}\right) \frac{m_K^2 m_{\pi}^4}{m_{\eta}^2} + \left(\frac{59}{384}+\frac{\pi ^2}{72}\right) \frac{m_{\pi}^6}{m_{\eta}^2} +\left(\frac{12349}{7776}+\frac{5 \pi ^2}{24}\right) m_K^4 \nonumber \\
	& +\left(\frac{\pi ^2}{48}-\frac{4133}{7776}\right) m_K^2 m_{\pi}^2+\left(\frac{6761}{15552}+\frac{5 \pi ^2}{96}\right) m_{\pi}^4  \Bigg\} + d^{\eta}_{\pi \pi \eta} + d^{\eta}_{K K \eta} + d^{\eta}_{\pi K K}
\end{align}

\begin{align}
	d^{\eta}_{\pi \pi \eta} &= \left( \frac{1}{12} m_{\pi}^4 \right) \overline{H}_{2\eta \pi \pi} - \left( \frac{1}{12} \frac{m_{\pi}^4}{m_{\eta}^2} \right) \overline{H}^\chi_{\eta \pi \pi}
\end{align}
\begin{align}
	d^{\eta}_{K K \eta} &= \bigg(-15 \frac{m_K^8}{m_{\eta}^6}+\frac{15}{2} \frac{m_K^6 m_{\pi}^2}{m_{\eta}^6} - \frac{15}{16} \frac{m_K^4 m_{\pi}^4}{m_{\eta}^6} + \frac{28}{3} \frac{m_K^6}{m_{\eta}^4} - \frac{10}{3} \frac{m_K^4 m_{\pi}^2}{m_{\eta}^4} + \frac{1}{4}\frac{m_K^2 m_{\pi}^4}{m_{\eta}^4} - \frac{8}{9} \frac{m_K^4}{m_{\eta}^2} \nonumber \\
	& \quad - \frac{2}{9} \frac{m_K^2 m_{\pi}^2}{m_{\eta}^2} + \frac{1}{12} \frac{m_{\pi}^4}{m_{\eta}^2} - \frac{3}{4} m_K^2 + \frac{3}{16} m_{\pi}^2 \bigg)  \overline{H}^\chi_{K K \eta}  \nonumber \\
	& + \bigg( 20 \frac{m_K^8}{m_{\eta}^4} - 10 \frac{m_K^6 m_{\pi}^2}{m_{\eta}^4} + \frac{5}{4} \frac{m_K^4 m_{\pi}^4}{m_{\eta}^4} - \frac{88}{9} \frac{m_K^6}{m_{\eta}^2} + \frac{28}{9} \frac{m_K^4 m_{\pi}^2}{m_{\eta}^2} - \frac{1}{6} \frac{m_K^2 m_{\pi}^4}{m_{\eta}^2} + \frac{8}{9} m_K^4  \nonumber \\
	& \quad + \frac{2}{9} m_K^2 m_{\pi}^2 - \frac{1}{12} m_{\pi}^4 \bigg) \overline{H}^\chi_{K K 2\eta}
\end{align}

\begin{align}
	d^{\eta}_{\pi K K} &= \bigg(5 \frac{m_K^8}{m_{\eta}^6} - \frac{9}{2} \frac{m_K^6 m_{\pi}^2}{m_{\eta}^6} + \frac{45}{16} \frac{m_K^4 m_{\pi}^4}{m_{\eta}^6} - \frac{7}{8} \frac{m_K^2 m_{\pi}^6}{m_{\eta}^6} + \frac{3}{32} \frac{m_{\pi}^8}{m_{\eta}^6} - 2 \frac{m_K^6}{m_{\eta}^4} + \frac{10}{3} \frac{m_K^4 m_{\pi}^2}{m_{\eta}^4} - \frac{55}{24} \frac{m_K^2 m_{\pi}^4}{m_{\eta}^4}  \nonumber \\
	& \quad + \frac{19}{48} \frac{m_{\pi}^6}{m_{\eta}^4} +\frac{1}{2} \frac{m_K^4}{m_{\eta}^2} - \frac{1}{6} \frac{m_K^2 m_{\pi}^2}{m_{\eta}^2} + \frac{25}{96} \frac{m_{\pi}^4}{m_{\eta}^2} - \frac{9}{16}  m_{\pi}^2 \bigg) \overline{H}^\chi_{\pi K K} \nonumber \\
	& + \bigg( - 20 \frac{m_K^{10}}{m_{\eta}^6} + 25 \frac{m_K^8 m_{\pi}^2}{m_{\eta}^6} - \frac{59}{4} \frac{m_K^6 m_{\pi}^4}{m_{\eta}^6} + \frac{63}{16} \frac{m_K^4 m_{\pi}^6}{m_{\eta}^6} - \frac{3}{8} \frac{m_K^2 m_{\pi}^8}{m_{\eta}^6} + \frac{9}{m_{\eta}^4}  m_K^8 - \frac{77}{6} \frac{m_K^6 m_{\pi}^2}{m_{\eta}^4} \nonumber \\
	& \quad + \frac{355}{48} \frac{m_K^4 m_{\pi}^4}{m_{\eta}^4} - \frac{19}{16} \frac{m_K^2 m_{\pi}^6}{m_{\eta}^4} - \frac{m_K^6}{m_{\eta}^2} + \frac{1}{3} \frac{m_K^4 m_{\pi}^2}{m_{\eta}^2} - \frac{25}{48} \frac{m_K^2 m_{\pi}^4}{m_{\eta}^2} \bigg) \overline{H}^\chi_{\pi 2K K} \nonumber \\
	& + \bigg(-\frac{m_K^6 m_{\pi}^4}{m_{\eta}^6}-\frac{m_K^4 m_{\pi}^6}{m_{\eta}^6} + \frac{11}{16} \frac{m_K^2 m_{\pi}^8}{m_{\eta}^6} - \frac{3}{32} \frac{m_{\pi}^{10}}{m_{\eta}^6} + \frac{m_K^6 m_{\pi}^2}{m_{\eta}^4} - \frac{3}{2} \frac{m_K^4 m_{\pi}^4}{m_{\eta}^4} + \frac{91}{48} \frac{m_K^2 m_{\pi}^6}{m_{\eta}^4}  \nonumber \\
	& \quad - \frac{19}{48} \frac{m_{\pi}^8}{m_{\eta}^4} - \frac{1}{2} \frac{m_K^4 m_{\pi}^2}{m_{\eta}^2} + \frac{1}{6} \frac{m_K^2 m_{\pi}^4}{m_{\eta}^2} - \frac{25}{96} \frac{m_{\pi}^6}{m_{\eta}^2} \bigg) \overline{H}^\chi_{2\pi K K} 
\end{align}
where $d^{\eta}_{L_i \times L_j}$ is given by Eq.(\ref{dBarLiLj}).

\section{Sunset Integral Results \label{Sec:SunsetResults}}

The results presented in this Appendix have been checked by doing the calculations analytically in two different ways (see \cite{Ananthanarayan:2018} for details). They have also been checked numerically using \texttt{AMBRE} \cite{Gluza:2007rt,Gluza:2010rn} and other related \texttt{Mathematica} packages, as well as using \textbf{CHIRON} \cite{Bijnens:2014gsa}. Equivalent expressions in terms of Kamp\'e de F\'eriet series may be found in \cite{Ananthanarayan:2017qmx}.

\subsection{Three mass scale kaon sunsets}

\begin{align}\label{Eq:Hkpe}
	& \overline{H}^{\chi}_{K \pi \eta} = \frac{m_{K}^2}{512\pi ^4} \Bigg\{ -\frac{1}{4}+\frac{5 \pi ^2}{6}-\frac{7}{4}\left(\frac{m_{\eta}^4}{m_{K}^4}+\frac{m_{\pi}^4}{m_{K}^4}\right) + \left(1-\frac{\pi^2}{2}\right)\left(\frac{m_{\eta}^2}{m_{K}^2}+\frac{m_{\pi}^2}{m_{K}^2}\right) +\frac{m_{\pi}^4}{2 m_{K}^4} \log\left[\frac{m_{\pi}^2}{m_{K}^2}\right] \nonumber \\
	& \quad +\frac{m_{\pi}^2}{m_{K}^2} \frac{m_{\eta}^2}{m_{K}^2} \left(7+\frac{2 \pi^2}{3}-2 \log\left[\frac{m_{\eta}^2}{m_{K}^2}\right]-2 \log\left[\frac{m_{\pi}^2}{m_{K}^2}\right]+\log\left[\frac{m_{\eta}^2}{m_{K}^2}\right] \log\left[\frac{m_{\pi}^2}{m_{K}^2}\right]\right) +\frac{m_{\eta}^4}{2 m_{K}^4} \log\left[\frac{m_{\eta}^2}{m_{K}^2}\right] \nonumber \\
	& \quad -\frac{m_{\pi}^2}{m_{K}^2} \log\left[\frac{m_{\pi}^2}{m_{K}^2}\right]^2-\frac{m_{\eta}^2}{m_{K}^2} \log\left[\frac{m_{\eta}^2}{m_{K}^2}\right]^2 +\frac{8 \pi }{3}\left(\frac{m_{\eta}^2}{m_{K}^2}\right)^{3/2} 
		{}_2F_1 \bigg[ \begin{array}{c}
		\frac{1}{2},-\frac{1}{2} \\
		\frac{5}{2} \\
	\end{array}	\bigg| \frac{m_\eta^2}{4m_K^2} \bigg] 
	+\frac{1}{36}\frac{m_{\eta}^6}{m_{K}^6}
		{}_3F_2 \bigg[ \begin{array}{c}
		1,1,2 \\
		\frac{5}{2},4 \\
	\end{array}	\bigg| \frac{m_\eta^2}{4m_K^2} \bigg] \nonumber \\
	& \quad
	 + \frac{1}{36} \frac{m_{\pi}^6}{m_{K}^6}
		{}_3F_2 \bigg[ \begin{array}{c}
		1,1,2 \\
		\frac{5}{2},4 \\
	\end{array}	\bigg| \frac{m_\pi^2}{4m_K^2} \bigg] + \frac{1}{6} \frac{m_{\eta}^4}{m_K^4} \frac{m_{\pi}^2}{m_K^2}
	 \left( 2\gamma_E - 1 + \log \left[\frac{m_{\eta}^2}{4 m_K^2}\right] + \log \left[\frac{m_{\pi}^2}{4 m_K^2}\right] \right) {}_2F_1 \bigg[ \begin{array}{c}
		1,1 \\
		\frac{5}{2} \\
	\end{array}	\bigg| \frac{m_\pi^2}{4m_K^2} \bigg] \nonumber \\
	& \quad + 2 \sqrt{\pi} \frac{m_{\eta}^2}{m_{K}^2} \sum_{m=0}^{\infty} \frac{\Gamma(1+m)}{ \Gamma(\frac{5}{2}+m)} \left(\frac{m_{\pi}^2}{4m_{K}^2}\right)^{m+2}  \bigg( 2 \psi(m+1)+\psi(m+2)+\psi(m+3)-2 \psi\left(m+\frac{5}{2}\right) \bigg) \nonumber \\
	& \quad + 8 \sqrt{\pi} \sum_{m,n=0}^{\infty} \frac{ \Gamma (m+n+1) \Gamma (m+n+2) \Gamma (m+n+3)}{\Gamma (m+2) \Gamma (m+3) \Gamma (n+1) \Gamma (n+2) \Gamma \left(m+n+\frac{5}{2}\right)}\left(\frac{m_{\eta}^2}{4 m_{K}^2}\right)^{m+2}\left(\frac{m_{\pi}^2}{4 m_{K}^2}\right)^{n+1} \nonumber \\
	& \qquad \times \bigg( \log \left[\frac{m_{\eta}^2}{4 m_{K}^2}\right]+\log \left[\frac{m_{\pi}^2}{4 m_{K}^2}\right] -\psi(m+2)-\psi(m+3)-\psi(n+1)-\psi(n+2)\nonumber \\
	& \qquad \quad + 2 \psi(m+n+1) +2 \psi(m+n+2)+2 \psi(m+n+3) -2 \psi\left(m+n+\frac{5}{2}\right) \bigg) \nonumber \\
	& \quad + \frac{32}{\pi^{3/2}} \left(\frac{m_{K}^2}{4m_{\eta}^2}\right)^{\frac{1}{2}} \sum_{m,n=0}^{\infty} \frac{\Gamma \left(m+\frac{1}{2}\right) \Gamma \left(m+\frac{3}{2}\right) \Gamma \left(n-\frac{1}{2}\right) \Gamma \left(n+\frac{1}{2}\right) \Gamma \left(n+\frac{3}{2}\right)}{\Gamma (n+1) \Gamma (m+n+2) \Gamma (m+n+3)} \left(\frac{m_{\pi}^2}{m_{\eta}^2}\right)^m\left(\frac{m_{\pi}^2}{4m_{K}^2}\right)^{n+2} \nonumber \\
	& \qquad \times \left( \log \left[\frac{m_\pi^2}{m_\eta^2}\right] + \psi\left(m+\frac{1}{2}\right) + \psi\left(m+\frac{3}{2}\right) - \psi(m+n+2) - \psi(m+n+3) \right) \nonumber \\
	& \quad - 8 \sqrt{\pi} \sum_{m,n=0}^{\infty} \frac{\Gamma \left(m+n-\frac{1}{2}\right) \Gamma \left(m+n+\frac{1}{2}\right) \Gamma \left(m+n+\frac{3}{2}\right)}{\Gamma \left(m+\frac{1}{2}\right) \Gamma \left(m+\frac{3}{2}\right) \Gamma (n+1) \Gamma (n+2) \Gamma (m+n+1)} \left(\frac{m_{\eta}^2}{4m_{K}^2}\right)^{\frac{1}{2}+m} \left(\frac{m_{\pi}^2}{4m_{K}^2}\right)^{1+n} \nonumber 
	\\
	& \qquad \times \left( \log \left[\frac{m_\pi^2}{m_\eta^2}\right] + \psi\left(m+\frac{1}{2}\right) + \psi\left(m+\frac{3}{2}\right) - \psi(n+1) - \psi(n+2) \right) \Bigg\}
\end{align}

\begin{align}\label{Eq:H2kpe}
	& \overline{H}^{\chi}_{2K \pi \eta} = \frac{1}{512\pi ^4} \Bigg\{ \frac{5 \pi^2}{6} -1 -\frac{m_{\eta}^2}{m_{K}^2} \bigg( 1 + \frac{\pi^2}{3} + \frac{1}{2} \log ^2 \left[\frac{m_{K}^2}{m_{\eta}^2}\right] + \log \left[\frac{m_{K}^2}{m_{\eta}^2}\right] + \text{Li}_2 \left[ 1-\frac{m_{\pi}^2}{m_{\eta}^2} \right] \bigg)  \nonumber \\
	& \quad -\frac{m_{\pi}^2}{m_{K}^2} \bigg( 1 + \frac{\pi^2}{3} - \log \left[\frac{m_{\pi}^2}{m_{K}^2}\right] - \frac{1}{2} \log^2 \left[ \frac{m_{K}^2}{m_{\eta}^2} \right] - \log \left[\frac{m_{K}^2}{m_{\eta}^2}\right] \log \left[\frac{m_{\pi}^2}{m_{K}^2}\right] - \text{Li}_2 \left[1-\frac{m_{\pi}^2}{m_{\eta}^2}\right] \bigg) \nonumber \\
	& \quad - \frac{m_{\pi}^4}{4 m_{K}^4}
	{}_3F_2 \bigg[ \begin{array}{c}
		1,1,1 \\
		\frac{3}{2},3 \\
	\end{array}	\bigg| \frac{m_\pi^2}{4m_K^2} \bigg]	
	 - \frac{m_{\eta}^4}{4 m_{K}^4} 
	 {}_3F_2 \bigg[ \begin{array}{c}
		1,1,1 \\
		\frac{3}{2},3 \\
	\end{array}	\bigg| \frac{m_\eta^2}{4m_K^2} \bigg] 
	+ \frac{2 \pi}{3} \left(\frac{m_{\eta}^2}{m_{K}^2}\right)^{\frac{3}{2}}
	{}_2F_1 \bigg[ \begin{array}{c}
		\frac{1}{2},\frac{1}{2} \\
		\frac{5}{2} \\
	\end{array}	\bigg| \frac{m_\eta^2}{4m_K^2} \bigg] \nonumber \\
	& \quad + 4 \sqrt{\pi} \sum_{m,n=0}^{\infty} \frac{\Gamma \left(m+n+\frac{1}{2}\right)^2 \Gamma \left(m+n+\frac{3}{2}\right)}{\Gamma \left(m+\frac{1}{2}\right) \Gamma \left(m+\frac{3}{2}\right) \Gamma (n+1) \Gamma (n+2) \Gamma (m+n+1)} \left(\frac{m_{\eta}^2}{4m_{K}^2}\right)^{\frac{1}{2}+m} \left(\frac{m_{\pi}^2}{4m_{K}^2}\right)^{1+n} \nonumber \\
	& \qquad \times \left( \log \left[\frac{m_\pi^2}{m_\eta^2}\right] + \psi\left(m+\frac{1}{2}\right) + \psi\left(m+\frac{3}{2}\right) - \psi(n+1) - \psi(n+2) \right) \nonumber \\
	& \quad - 4 \sqrt{\pi} \sum_{m,n=0}^{\infty}  \frac{ \Gamma (m+n+1)^2 \Gamma (m+n+2) }{\Gamma (m+1) \Gamma (m+2) \Gamma (n+1) \Gamma (n+2) \Gamma \left(m+n+\frac{3}{2} \right) } \left(\frac{m_{\eta}^2}{4m_{K}^2}\right)^{1+m} \left(\frac{m_{\pi}^2}{4m_{K}^2}\right)^{1+n} \nonumber \\
	& \qquad \times \bigg( \log \left[\frac{m_{\pi}^2}{4 m_{K}^2}\right] + \log \left[\frac{m_{\eta}^2}{4 m_{K}^2}\right] -\psi(m+1)-\psi(m+2) -\psi(n+1)-\psi(n+2) \nonumber \\
	& \qquad \quad + 4 \psi(m+n+1)+2 \psi(m+n+2)-2 \psi\left(m+n+\frac{3}{2}\right) \bigg) \nonumber \\
	& \quad - \frac{4}{\pi^{3/2}} \sum_{m,n=0}^{\infty} \frac{\Gamma \left(m+\frac{1}{2}\right) \Gamma \left(m+\frac{3}{2}\right) \Gamma \left(n+\frac{1}{2}\right)^2 \Gamma \left(n+\frac{3}{2}\right)}{\Gamma (n+1) \Gamma (m+n+2) \Gamma (m+n+3)} \left(\frac{m_{\pi}^2}{m_{\eta}^2}\right)^{m+\tfrac{1}{2}} \left(\frac{m_{\pi}^2}{4m_{K}^2}\right)^{n+\tfrac{3}{2}} \nonumber \\
	& \qquad \times \left( \log \left[\frac{m_\pi^2}{m_\eta^2}\right] + \psi\left(m+\frac{1}{2}\right) + \psi\left(m+\frac{3}{2}\right) - \psi(m+n+2) - \psi(m+n+3) \right) \Bigg\}
\end{align}

\begin{align}\label{Eq:Hk2pe}
	& \overline{H}^{\chi}_{K 2\pi \eta} = \frac{1}{512\pi ^4} \Bigg\{ \frac{1}{6} \frac{m_\eta^4}{m_K^4} \left( 2\gamma_E + \log \left[ \frac{m_{\eta}^2}{4 m_{K}^2} \right] + \log \left[ \frac{m_{\pi}^2}{4 m_{K}^2} \right] \right) 
	{}_2F_1 \bigg[ \begin{array}{c}
		1,1 \\
		\frac{5}{2} \\
	\end{array}	\bigg| \frac{m_\eta^2}{4m_K^2} \bigg] - 1 - \frac{\pi^2}{2} - 2\log \left[ \frac{m_{\pi}^2}{m_{K}^2} \right] \nonumber \\
	& + \frac{m_{\eta}^2}{m_{K}^2} \left( 5 + \frac{2 \pi^2}{3} - 2 \log \left[ \frac{m_{\pi}^2}{m_{K}^2} \right] + \log \left[ \frac{m_{K}^2}{m_{\eta}^2}\right] - \log \left[ \frac{m_{K}^2}{m_{\eta}^2}\right]  \log \left[ \frac{m_{\pi}^2}{m_{K}^2}\right] \right) + \frac{m_{\pi}^4}{12 m_{K}^4}
	{}_3F_2 \bigg[ \begin{array}{c}
		1,1,2 \\
		\frac{5}{2},3 \\
	\end{array}	\bigg| \frac{m_\pi^2}{4m_K^2} \bigg] \nonumber \\
	& + \frac{1}{3} \frac{m_{\eta}^2}{m_{K}^2} \frac{m_{\pi}^2}{m_{K}^2}\left( 2\gamma_E - 1 + \log \left[ \frac{m_{\eta}^2}{4 m_{K}^2} \right] + \log \left[ \frac{m_{\pi}^2}{4 m_{K}^2} \right] \right) 
	{}_3F_2 \bigg[ \begin{array}{c}
		1,1,3 \\
		2,\frac{5}{2} \\
	\end{array}	\bigg| \frac{m_\pi^2}{4m_K^2} \bigg]	-\frac{m_{\pi}^2}{m_{K}^2} \left(3-\log \left[ \frac{m_{\pi}^2}{m_{K}^2}\right] \right)  \nonumber \\	
		& -\log^2 \left[\frac{m_{\pi}^2}{m_{K}^2} \right] + 2 \sqrt{\pi} \sum_{m=0}^{\infty} \frac{\Gamma (m+1)}{\Gamma \left(m+\frac{5}{2}\right)} \left(\frac{m_{\eta}^2}{4 m_{K}^2}\right)^{m+2} \bigg( 2 \psi(m+1) +\psi(m+2) + \psi(m+3)-2 \psi\left(m+\frac{5}{2}\right) \bigg) \nonumber \\
		& + \frac{\sqrt{\pi}}{2} \frac{m_{\eta}^2}{m_{K}^2} \sum_{m=0}^{\infty} \frac{\Gamma (m+1) \Gamma (m+3)}{\Gamma (m+2) \Gamma \left(m+\frac{5}{2}\right)} \left(\frac{m_{\pi}^2}{4 m_{K}^2}\right)^{m+1} \bigg( 2 \psi(m+1)+2 \psi(m+3)-2 \psi\left(m+\frac{5}{2}\right) \bigg) \nonumber \\
		& + \frac{2}{\pi^{3/2}}  \sum_{m,n=0}^{\infty} \frac{\Gamma \left(m+\frac{1}{2}\right) \Gamma \left(m+\frac{3}{2}\right) \Gamma \left(n-\frac{1}{2}\right) \Gamma \left(n+\frac{1}{2}\right) \Gamma \left(n+\frac{3}{2}\right)}{\Gamma (n+1) \Gamma (m+n+2)^2} \left(\frac{m_{\pi}^2}{m_{\eta}^2}\right)^{m+\tfrac{1}{2}} \left(\frac{m_{\pi}^2}{4 m_{K}^2}\right)^{n+\tfrac{1}{2}} \nonumber \\
		& \qquad \times \left( \log \left[ \frac{m_{\pi}^2}{m_{\eta}^2} \right] + \psi\left(m+\frac{1}{2}\right) + \psi\left(m+\frac{3}{2}\right) -2 \psi(m+n+2) \right) \nonumber \\		
		& - 2 \sqrt{\pi} \sum_{m,n=0}^{\infty} \frac{\Gamma \left(m+n-\frac{1}{2}\right) \Gamma \left(m+n+\frac{1}{2}\right) \Gamma \left(m+n+\frac{3}{2}\right)}{\Gamma \left(m+\frac{1}{2}\right) \Gamma \left(m+\frac{3}{2}\right) \Gamma (n+1)^2 \Gamma (m+n+1)} \left(\frac{m_{\eta}^2}{4 m_{K}^2}\right)^{m+\frac{1}{2}} \left(\frac{m_{\pi}^2}{4 m_{K}^2}\right)^n \nonumber \\
		& \qquad \times \left( \log \left[ \frac{m_{\pi}^2}{m_{\eta}^2} \right] + \psi\left(m+\frac{1}{2}\right)+\psi\left(m+\frac{3}{2}\right)-2 \psi(n+1) \right) \nonumber \\
		& + 2 \sqrt{\pi} \sum_{m,n=0}^{\infty} \frac{\Gamma (m+n+2) \Gamma (m+n+3) \Gamma (m+n+4)}{\Gamma (m+2) \Gamma (m+3) \Gamma (n+2)^2 \Gamma \left(m+n+\frac{7}{2}\right)} \left(\frac{m_{\eta}^2}{4 m_{K}^2}\right)^{m+2} \left(\frac{m_{\pi}^2}{4 m_{K}^2}\right)^{n+1} \nonumber \\
		& \qquad \times \bigg( \log \left[ \frac{m_{\pi}^2}{4 m_{K}^2} \right] + \log \left[ \frac{m_{\eta}^2}{4 m_{K}^2} \right] - \psi(m+2) -\psi(m+3) -2 \psi(n+2) \nonumber \\
		& \qquad \quad +  2 \psi(m+n+2)+2 \psi(m+n+3)+2 \psi(m+n+4)-2 \psi\left(m+n+\frac{7}{2}\right) \bigg) \Bigg\}
\end{align}

\begin{align}
	& \overline{H}^{\chi}_{K \pi 2\eta} = \frac{1}{512\pi^4} \Bigg\{ 2 \log \left[ \frac{m_K^2}{m_{\eta}^2} \right] - \log^2 \left[\frac{m_{K}^2}{m_{\eta}^2}\right] + \pi \left(\frac{m_{\eta}^2}{m_{K}^2}\right)^{\tfrac{1}{2}} \left(4-\frac{m_{\eta}^2}{m_{K}^2}\right)^{\tfrac{1}{2}} - \frac{m_{\eta}^2}{m_{K}^2} \left(3 + \log \left[\frac{m_{K}^2}{m_{\eta}^2}\right] \right) \nonumber \\
	& \quad + \frac{m_{\pi}^2}{m_{K}^2} \left( 5 + \frac{2 \pi ^2}{3} + 2 \log \left[\frac{m_{K}^2}{m_{\eta}^2}\right] - \log \left[\frac{m_{\pi}^2}{m_{K}^2}\right] - \log \left[\frac{m_{\pi}^2}{m_{K}^2}\right] \log \left[ \frac{m_{K}^2}{m_{\eta}^2} \right] \right) + 2 \pi \frac{m_{\eta}}{m_K}
	{}_2F_1 \bigg[ \begin{array}{c}
		\frac{1}{2},\frac{1}{2} \\
		\frac{3}{2} \\
	\end{array}	\bigg| \frac{m_\eta^2}{4m_K^2} \bigg] \nonumber \\
	& \quad + \frac{m_{\eta}^4}{12 m_{K}^4}
	{}_3F_2 \bigg[ \begin{array}{c}
		1,1,2 \\
		\frac{5}{2},3 \\
	\end{array}	\bigg| \frac{m_\eta^2}{4m_K^2} \bigg]  + \frac{1}{6} \frac{m_\pi^4}{m_K^4} \left( 2\gamma_E + \log \left[ \frac{m_{\eta}^2}{4 m_{K}^2} \right] + \log \left[ \frac{m_{\pi}^2}{4 m_{K}^2} \right] \right)
	{}_2F_1 \bigg[ \begin{array}{c}
		1,1 \\
		\frac{5}{2} \\
	\end{array}	\bigg| \frac{m_\pi^2}{4m_K^2} \bigg]	\nonumber \\	
	& \quad + \frac{1}{3} \frac{m_{\eta}^2}{m_{K}^2} \frac{m_{\pi}^2}{m_{K}^2} \left( 2\gamma_E - 1 + \log \left[ \frac{m_{\eta}^2}{4 m_{K}^2} \right] + \log \left[ \frac{m_{\pi}^2}{4 m_{K}^2} \right] \right)
	{}_3F_2 \bigg[ \begin{array}{c}
		1,1,3 \\
		2,\frac{5}{2} \\
	\end{array}	\bigg| \frac{m_\eta^2}{4m_K^2} \bigg] - 1 -\frac{\pi^2}{2} \nonumber \\	
	& \quad + \frac{\sqrt{\pi}}{2} \frac{m_{\pi}^2}{m_{K}^2} \sum_{m=0}^{\infty} \frac{\Gamma (m+1) \Gamma (m+3)}{\Gamma (m+2) \Gamma \left(m+\frac{5}{2}\right)} \left(\frac{m_{\eta}^2}{4 m_{K}^2}\right)^{m+1} \bigg( 2 \psi(m+1)+2 \psi(m+3)-2 \psi\left(m+\frac{5}{2} \right) \bigg) \nonumber \\
	& \quad + 2 \sqrt{\pi} \sum_{m=0}^{\infty} \frac{\Gamma (m+1)}{\Gamma \left(m+\frac{5}{2}\right)} \left(\frac{m_{\pi}^2}{4 m_{K}^2}\right)^{m+2} \bigg( 2 \psi(m+1)+\psi(m+2)+\psi(m+3)-2 \psi\left(m+\frac{5}{2}\right) \bigg) \nonumber \\
	& \quad - 2 \sqrt{\pi} \sum_{m,n=0}^{\infty} \frac{\Gamma \left(m+n-\frac{1}{2}\right) \Gamma \left(m+n+\frac{1}{2}\right) \Gamma \left(m+n+\frac{3}{2}\right)}{\Gamma \left(m+\frac{1}{2}\right)^2 \Gamma (n+1) \Gamma (n+2) \Gamma (m+n+1)} \left(\frac{m_{\eta}^2}{4 m_{K}^2}\right)^{m-\tfrac{1}{2}} \left(\frac{m_{\pi}^2}{4 m_{K}^2}\right)^{n+1} \nonumber \\
	& \quad \qquad \times \left( \log \left[\frac{m_{\pi}^2}{m_{\eta}^2}\right] + 2 \psi\left(m+\frac{1}{2}\right) -\psi(n+1)-\psi(n+2) \right) \nonumber \\
	& \quad + 2 \sqrt{\pi} \sum_{m,n=0}^{\infty} \frac{\Gamma (m+n+2) \Gamma (m+n+3) \Gamma (m+n+4)}{\Gamma (m+2)^2 \Gamma (n+2) \Gamma (n+3) \Gamma \left(m+n+\frac{7}{2}\right)} \left(\frac{m_{\eta}^2}{4 m_{K}^2}\right)^{m+1} \left(\frac{m_{\pi}^2}{4 m_{K}^2}\right)^{n+2} \nonumber \\
	& \quad \qquad \times \bigg( \log \left[\frac{m_{\pi}^2}{4 m_{K}^2}\right] + \log \left[\frac{m_{\eta}^2}{4 m_{K}^2}\right] -2 \psi(m+2) - \psi(n+2)-\psi(n+3) \nonumber \\
	& \quad \qquad \quad + 2 \psi(m+n+2)+2 \psi(m+n+3)+2 \psi(m+n+4)-2 \psi\left(m+n+\frac{7}{2}\right) \bigg) \nonumber \\
	& \quad - \frac{2}{\pi^{3/2}} \sum_{m,n=0}^{\infty} \frac{\Gamma \left(m+\frac{3}{2}\right)^2 \Gamma \left(n-\frac{1}{2}\right) \Gamma \left(n+\frac{1}{2}\right) \Gamma \left(n+\frac{3}{2}\right)}{\Gamma (n+1) \Gamma (m+n+2) \Gamma (m+n+3)} \left(\frac{m_{\pi}^2}{m_{\eta}^2}\right)^{m+\tfrac{3}{2}} \left(\frac{m_{\pi}^2}{4 m_{K}^2}\right)^{n+\tfrac{1}{2}} \nonumber \\
	& \quad \qquad \times \left( \log \left[\frac{m_{\pi}^2}{m_{\eta}^2}\right] + 2 \psi\left(m+\frac{3}{2}\right) - \psi(m+n+2)-\psi(m+n+3) \right) \Bigg\}
\label{Eq:Hkp2e}
\end{align}

\subsection{Three mass scale eta sunsets}

\begin{align}
	& \overline{H}^\chi_{\pi K K} = \frac{m_{\pi}^2}{512 \pi ^4} \Bigg\{ \frac{\pi ^2}{6}-5 + 4 \log \left[\frac{m_{\pi}^2}{m_K^2}\right] - \log ^2 \left[\frac{m_{\pi}^2}{m_K^2}\right] + \frac{m_{\eta}^2}{m_{\pi}^2} \left( \log \left[ \frac{m_{K}^2}{m_{\eta}^2} \right] + \frac{5}{4} \right)  + \frac{m_{K}^2}{m_{\pi}^2} \left(6 + \frac{\pi ^2}{3}\right) \nonumber \\
	& \quad - \frac{1}{18} \frac{m_{\eta}^2}{m_{K}^2} \frac{m_{\eta}^2}{m_{\pi}^2}  {}_3F_2 \bigg[ \begin{array}{c}
		1,1,2 \\
		\frac{5}{2},4 \\
	\end{array}	\bigg| \frac{m_\eta^2}{4m_K^2} \bigg] - \frac{1}{3} \frac{m_{\pi}^2}{m_K^2} \log \left[ \frac{m_{\pi}^2}{4 m_K^2} \right] {}_2F_1 \bigg[ \begin{array}{c}
		1,1 \\
		\frac{5}{2} \\
	\end{array}	\bigg| \frac{m_{\pi}^2}{4 m_K^2} \bigg] \nonumber \\
	& \quad - \sqrt{\pi} \sum_{m,n=0}^{\infty} \frac{\Gamma (m+n+1) \Gamma (m+n+2) \Gamma (m+n+3)}{\Gamma (m+2) \Gamma (m+3) \Gamma (n+1) \Gamma (n+2) \Gamma \left(m+n+\frac{5}{2}\right)} \left(\frac{m_{\eta}^2}{4 m_{K}^2}\right)^{m+1} \left(\frac{m_{\pi}^2}{4 m_{K}^2}\right)^n \nonumber \\
	& \quad \times \Bigg( \log \left[\frac{m_{\pi}^2}{4 m_{K}^2}\right]-\psi(n+1)-\psi(n+2) + \psi(m+n+1) + \psi(m+n+2) +\psi(m+n+3) - \psi\left( m+n+\frac{5}{2} \right) \Bigg) \nonumber \\
	& \quad - \sqrt{\pi} \sum_{m=0}^{\infty} \frac{\Gamma (m+1)}{\Gamma \left(m+\frac{5}{2}\right)} \left(\frac{m_{\pi}^2}{4 m_{K}^2}\right)^{m+1} \Bigg( \psi(m+1) - \psi\left( m+\frac{5}{2} \right) \Bigg) \Bigg\}
\label{Eq:Hpkk}
\end{align}

\begin{align}
	& \overline{H}^\chi_{2\pi K K} = \frac{1}{512 \pi ^4} \Bigg\{ \frac{\pi ^2}{6} - 3 + 2 \log \left[ \frac{m_{\pi}^2}{m_{K}^2} \right] - \log^2 \left[ \frac{m_{\pi}^2}{m_{K}^2} \right] - \frac{1}{3} \frac{m_{\pi}^2}{m_K^2} \left( 1 + 2 \log \left[ \frac{m_{\pi}^2}{4 m_K^2} \right] \right)
	{}_2F_1 \bigg[ \begin{array}{c}
		1,1 \\
		\frac{5}{2} \\
	\end{array}	\bigg| \frac{m_{\pi}^2}{4 m_K^2} \bigg] \nonumber \\
	& \quad - \frac{1}{30} \frac{m_{\pi}^4}{m_K^4} \log \left[ \frac{m_{\pi}^2}{4 m_K^2} \right]  {}_2F_1 \bigg[ \begin{array}{c}
		2,2 \\
		\frac{7}{2} \\
	\end{array}	\bigg| \frac{m_{\pi}^2}{4 m_K^2} \bigg] - \sqrt{\pi} \sum_{m=0}^{\infty} \left(\frac{m_\pi^2}{4 m_K^2}\right)^{m+1} \frac{\Gamma (m+1) \Gamma (m+3)}{\Gamma (m+2) \Gamma \left(m+\frac{5}{2}\right)}  \Bigg(\psi(m+1)-\psi\left(m+\frac{5}{2}\right)\Bigg) \nonumber \\
	& \quad - \sqrt{\pi} \sum_{m,n=0}^{\infty} \frac{\Gamma (m+n+1) \Gamma (m+n+2) \Gamma (m+n+3)}{\Gamma (m+2) \Gamma (m+3) \Gamma (n+1)^2 \Gamma \left(m+n+\frac{5}{2}\right)} \left(\frac{m_{\eta}^2}{4 m_{K}^2}\right)^{m+1} \left(\frac{m_{\pi}^2}{4 m_{K}^2}\right)^n	\nonumber \\
	& \quad \times \Bigg( \log \left[\frac{m_{\pi}^2}{4m_{K}^2}\right] - \psi(n+1) - \psi(n+2) + \psi(m+n+1) + \psi(m+n+2) + \psi(m+n+3) - \psi \left( m+n+\frac{5}{2} \right) \Bigg) 	\nonumber \\
	& \quad - \sqrt{\pi} \sum_{m,n=0}^{\infty} \frac{\Gamma (m+n+1) \Gamma (m+n+2) \Gamma (m+n+3)}{\Gamma (m+2) \Gamma (m+3) \Gamma (n+1) \Gamma (n+2) \Gamma \left(m+n+\frac{5}{2}\right)} \left(\frac{m_{\eta}^2}{4 m_{K}^2}\right)^{m+1} \left(\frac{m_{\pi}^2}{4 m_{K}^2}\right)^n	\Bigg\}
\label{Eq:H2pkk}
\end{align}

\begin{align}
	& \overline{H}^\chi_{\pi 2K K} = \frac{1}{512 \pi ^4} \Bigg\{ 1 + \frac{\pi^2}{6} + \frac{1}{60} \frac{m_{\pi}^6}{m_K^6} \log \left[ \frac{m_{\pi}^2}{4 m_K^2} \right] {}_2F_1 \bigg[ \begin{array}{c}
		2,2 \\
		\frac{7}{2} \\
	\end{array}	\bigg| \frac{m_\pi^2}{4 m_K^2} \bigg] - \frac{m_{\pi}^2}{m_{K}^2} \left(2 - \log \left[ \frac{m_{\pi}^2}{m_{K}^2} \right] \right) \nonumber \\
	& \quad + \frac{1}{6} \frac{m_{\pi}^4}{m_K^4} \left( 1 + \log \left[ \frac{m_{\pi}^2}{4 m_K^2} \right] \right) {}_2F_1 \bigg[ \begin{array}{c}
		1,1 \\
		\frac{5}{2} \\
	\end{array}	\bigg| \frac{m_\pi^2}{4 m_K^2} \bigg] + \frac{1}{2} \frac{m_{\eta}^2}{ m_{K}^2}  {}_3F_2 \bigg[ \begin{array}{c}
		1,1,1 \\
		\frac{3}{2},3 \\
	\end{array}	\bigg| \frac{m_\eta^2}{4 m_K^2} \bigg] \nonumber \\		
	& \quad + 2 \sqrt{\pi } \sum_{m=0}^\infty \frac{\Gamma (m+2)}{\Gamma \left(m+\frac{5}{2}\right)} \left(\frac{m_\pi^2}{4 m_K^2}\right)^{m+2} \Bigg( \psi(m+1)-\psi\left(m+\frac{5}{2}\right) \Bigg) \nonumber \\
	& \quad+ 2 \sqrt{\pi} \sum_{m,n=0}^{\infty} \frac{\Gamma (m+n+2)^2 \Gamma (m+n+3)}{\Gamma (m+2) \Gamma (m+3) \Gamma (n+1) \Gamma (n+2) \Gamma \left(m+n+\frac{5}{2}\right)} \left(\frac{m_{\eta}^2}{4 m_{K}^2}\right)^{m+1} \left(\frac{m_{\pi}^2}{4 m_{K}^2}\right)^{n+1} \nonumber \\
	& \quad \times \Bigg( \log \left[\frac{m_{\pi}^2}{4m_{K}^2}\right] - \psi(n+1)-\psi(n+2) +\psi(m+n+1)+ \psi(m+n+2) +\psi(m+n+3) - \psi \left( m+n+\frac{5}{2} \right) \Bigg) \nonumber \\
	& \quad + 2 \sqrt{\pi} \sum_{m,n=0}^{\infty} \frac{\Gamma (m+n+1) \Gamma (m+n+2) \Gamma (m+n+3)}{\Gamma (m+2) \Gamma (m+3) \Gamma (n+1) \Gamma (n+2) \Gamma \left(m+n+\frac{5}{2}\right)} \left(\frac{m_{\eta}^2}{4 m_{K}^2}\right)^{m+1} \left(\frac{m_{\pi}^2}{4 m_{K}^2}\right)^{n+1}  \Bigg\}
\label{Eq:Hp2kk}
\end{align}

\subsection{One and two mass scale sunsets}

\begin{align}
\overline{H}^{\chi}_{K \eta \eta} &= \frac{m_{\eta}^2}{512\pi^4} \Bigg\{ 4 + \frac{\pi^2}{3} + \frac{m_{K}^2}{m_{\eta}^2} \left(\frac{\pi ^2}{6}-\frac{1}{4}\right) - \frac{m_{K}^2}{m_{\eta}^2} \log^2 \left[ \frac{m_{K}^2}{m_{\eta}^2} \right] + 2 \log \left[ \frac{m_K^2}{m_\eta^2} \right] \nonumber \\
& \qquad + 2 \left( \frac{m_\eta^2}{m_K^2} + \frac{m_K^2}{m_\eta^2}-2 \right) \left(\text{Li}_2\left[\frac{m_{K}^2}{m_{\eta}^2}\right] + \log \left[1-\frac{m_{K}}{m_{\eta}}\right] \log \left[\frac{m_K^2}{m_\eta^2}\right] \right) \Bigg\}
\end{align}

\begin{align}
\overline{H}^{\chi}_{2K \eta \eta} &= \frac{1}{512\pi^4} \Bigg\{ \frac{\pi^2}{6} - 1 - \log^2 \left[ \frac{m_{K}^2}{m_{\eta}^2} \right] + 2 \left(1-\frac{m_{\eta}^2}{m_{K}^2}\right) \left( \text{Li}_2 \left[ \frac{m_{K}^2}{m_{\eta}^2} \right] + \log^2 \left[ \frac{m_{K}^2}{m_{\eta}^2} \right] \log^2 \left[ 1 - \frac{m_{K}^2}{m_{\eta}^2} \right] \right) \Bigg\}
\end{align}

\begin{align}
\overline{H}^{\chi}_{K K K} &= \frac{m_K^2}{512 \pi^4} \left(\frac{15}{4}+\frac{\pi ^2}{2}\right)
\end{align}

\section{The Pion Mass and Decay Constant \label{Sec:PionSunsets}}

Analytic expressions for the pion mass and decay constant in ChPT at two loops are presented in \cite{Ananthanarayan:2017yhz}. The expressions given there are approximations obtained by taking an expansion of the three mass scale sunsets around zero external momentum $p^2=m_\pi^2=0$. To go beyond these approximations, one may use the exact result presented in Eq.(17) of \cite{Ananthanarayan:2017qmx}, as well as its derivatives, and substitute them into Eq.(28) and Eq.(46) of \cite{Ananthanarayan:2017yhz}. We would like to emphasize that it is proved in \cite{Ananthanarayan:2018} that although the expression in Eq.(\ref{Eq:Hpkk}) of the present paper and the one in Eq.(17) of \cite{Ananthanarayan:2017qmx} do not come from summing up the same sets of residues in the Mellin-Barnes intermediate computations, each may be derived from the other by the swap $m_\pi^2 \leftrightarrow m_\eta^2$. This is also true for Eqs.(\ref{Eq:H2pkk}) and (\ref{Eq:Hp2kk}) and their pion analogues. 

In \cite{Ananthanarayan:2017qmx}, some results of this paper and of \cite{Ananthanarayan:2017yhz} are used to obtain an expression for the quantity $F_K/F_\pi$. An approximate analytic expression that can be readily fit with lattice data is also presented there. The truncated three mass sunsets used to produce Eqs.(18)-(19) of \cite{Ananthanarayan:2017qmx} are:

\begin{align}
	& \overline{H}^\chi_{\eta K K} \approx \frac{m_\pi^2}{512 \pi^4} \Bigg\{ \frac{5}{4} + \log \left[\frac{m_K^2}{m_\pi^2}\right] + \frac{1}{30} \frac{m_\eta^6}{m_K^4 m_\pi^2}  \left(\gamma -1+\psi\left(\frac{7}{2}\right)\right) + \frac{1}{3} \frac{m_\eta^4}{m_K^2 m_\pi^2} \left(\gamma +\psi\left(\frac{5}{2}\right)\right) + \frac{m_K^2}{m_\pi^2} \left(6+\frac{\pi ^2}{3}\right) \nonumber \\
	& \quad  + \frac{m_\eta^2}{m_\pi^2} \left(\frac{\pi^2}{6}-5+\log (256) -\log ^2\left[\frac{m_\eta^2}{m_K^2}\right] \right) + \frac{ 4 m_\eta m_K}{m_\pi^2} \sqrt{4-\frac{m_\eta^2}{m_K^2}}  \log \left[\frac{m_\eta^2}{4 m_K^2}\right] \csc ^{-1}\left[\frac{2 m_K}{m_\eta}\right] \nonumber \\
	& \quad + \frac{1}{3} \frac{m_\eta^2}{m_K^2} \left(\frac{7}{6}-\log \left[\frac{m_\eta^2}{m_K^2}\right] \right) + \frac{1}{35} \frac{m_\eta^6}{m_K^6} \left(\frac{533}{420}-\log \left[\frac{m_\eta^2}{m_K^2}\right] \right) + \frac{1}{10} \frac{m_\eta^4}{m_K^4} \left(\frac{37}{30}-\log \left[\frac{m_\eta^2}{m_K^2}\right] \right) \Bigg\}
\end{align}

\begin{align}
	& \overline{H}^\chi_{2\eta K K} \approx \frac{1}{512 \pi^4} \Bigg\{ \frac{m_\eta^4}{10 m_K^4} \left( \frac{31}{15}-\log (4) - \frac{1}{3} \, _2F_1 \left[2,2;\frac{7}{2};\frac{m_\eta^2}{4 m_K^2}\right] \log \left[\frac{m_\eta^2}{4 m_K^2}\right] \right) + \frac{1}{105} \frac{m_\eta^6}{m_K^6} \left(2 \gamma -3+2 \psi\left(\frac{9}{2}\right)\right) \nonumber \\
	& \quad + \frac{1}{3} \frac{m_\pi^2}{m_K^2} \left(\frac{1}{6} - \log \left[\frac{m_\eta^2}{m_K^2}\right] \right) - \log^2 \left[\frac{m_K^2}{m_\eta^2}\right] - 2 \log \left[\frac{m_K^2}{m_\eta^2}\right] - 8 \log \left[ \frac{m_\eta^2}{4 m_K^2}\right] + \frac{3}{35}\frac{m_\eta^4 m_\pi^2 }{m_K^6} \left(\frac{131}{140} - \log \left[\frac{m_\eta^2}{m_K^2}\right] \right)  \nonumber \\
	& \quad + \frac{4 m_K}{m_\eta} \sqrt{4-\frac{m_\eta^2}{m_K^2}} \left(2 \log \left[ \frac{m_\eta^2}{4 m_K^2}\right] + 1 \right) \sin ^{-1} \left[\frac{m_\eta}{2 m_K}\right] + \frac{1}{5} \frac{m_\eta^2 m_\pi^2}{m_K^4} \left(\frac{11}{15}-\log \left[\frac{m_\eta^2}{m_K^2}\right] \right) + \frac{2}{3} \frac{m_\eta^2}{m_K^2} \left(\gamma +\psi\left(\frac{5}{2}\right)\right) \nonumber \\
	& \quad + \frac{5}{462} \frac{m_\eta^8 m_\pi^2}{m_K^{10}} \left(\frac{18107}{13860} - \log \left[\frac{m_\eta^2}{m_K^2}\right] \right) + \frac{2}{63} \frac{m_\eta^6 m_\pi^2}{m_K^8} \left(\frac{1627}{1260} - \log \left[\frac{m_\eta^2}{m_K^2}\right] \right) + \frac{\pi^2}{6}-7 \Bigg\}
\end{align}

\begin{align}
	& \overline{H}^\chi_{\eta 2K K} \approx \frac{1}{512 \pi^4} \Bigg\{ 1 + \frac{1}{2} \frac{m_\pi^2}{m_K^2}  \, _3F_2 \left[1,1,1;\frac{3}{2},3;\frac{m_\pi^2}{4 m_K^2}\right] + \frac{m_\eta^8}{m_K^8} \left(\frac{\log (4)}{140}-\frac{389}{29400}\right) - \frac{1}{6} \frac{m_\eta^4}{m_K^4} \left(\gamma +\psi\left(\frac{5}{2}\right)\right) \nonumber \\
& \quad + \frac{\pi^2}{6} + \frac{1}{60} \frac{m_\eta^6}{m_K^6}  \left(\, _2F_1 \left[ 2,2;\frac{7}{2};\frac{m_\eta^2}{4 m_K^2}\right] \log \left[\frac{m_\eta^2}{4 m_K^2}\right] - \frac{62}{15}+\log (16)\right) + \frac{m_\eta^2}{m_K^2} \left(3 \log \left[\frac{m_\eta^2}{4 m_K^2}\right] + \log (4)\right)  \nonumber \\
& \quad + \frac{1}{6} \frac{m_\eta^2 m_\pi^2}{m_K^4} \left(\log \left[\frac{m_\eta^2}{m_K^2}\right] - \frac{1}{6}\right) + \frac{1}{63} \frac{m_\eta^8 m_\pi^2}{m_K^{10}} \left(\log \left[\frac{m_\eta^2}{m_K^2}\right] - \frac{1627}{1260}\right) + \frac{3}{70} \frac{m_\eta^6 m_\pi^2}{m_K^8} \left(\log \left[\frac{m_\eta^2}{m_K^2}\right] - \frac{533}{420}\right) \nonumber \\
& \quad - \frac{2 m_\eta}{m_K} \sqrt{4-\frac{m_\eta^2}{m_K^2}} \left(\log \left[\frac{m_\eta^2}{4 m_K^2}\right] + 1 \right) \csc ^{-1} \left[\frac{2 m_K}{m_\eta}\right] + \frac{1}{10} \frac{m_\eta^4 m_\pi^2}{m_K^6} \left(\log \left[\frac{m_\eta^2}{m_K^2}\right]-\frac{11}{15}\right) \Bigg\}
\end{align}

\end{document}